\documentclass[aps,twocolumn,superscriptaddress]{revtex4-2}
\usepackage{amsmath, amssymb}
\usepackage{mathtools}
\usepackage[pdftex]{graphicx}
\usepackage[colorlinks]{hyperref}
 \usepackage{color}
\usepackage{setspace, enumitem}
\usepackage[normalem]{ulem}
\usepackage{enumitem}
\usepackage{physics}
\usepackage[dvipsnames]{xcolor}
\usepackage{simpler-wick}
\usepackage{float}
\usepackage{amsmath}
\usepackage{hyperref}
\usepackage[capitalise]{cleveref}
\usepackage{comment}
\usepackage{soul}

\definecolor{TTT}{RGB}{212,175,55} 
\definecolor{MPP}{RGB}{231, 84, 128} 

\newcommand{\mZ}{\mathcal{Z}}

\graphicspath{{Figures/}}

\begin{document}

\title{Generalized Lotka-Volterra model with sparse interactions: \\
non-Gaussian effects and topological multiple-equilibria phase}

\author{Tommaso Tonolo}
\affiliation{Gran Sasso Science Institute, Viale F. Crispi 7, 67100 L’Aquila, Italy} 
\affiliation{INFN-Laboratori Nazionali del Gran Sasso, Via G. Acitelli 22, 67100 Assergi (AQ), Italy}

\author{Maria Chiara Angelini}
\affiliation{Dipartimento di Fisica, Universit\`a di Roma ``Sapienza'', Piazzale A. Moro 2, I-00185, Roma, Italy}

\author{Sandro Azaele}
\affiliation{Dipartimento di Fisica, Universit\`a di Padova, Via F. Marzolo, 8, Padova, Italy}

\author{Amos Maritan}
\affiliation{Dipartimento di Fisica, Universit\`a di Padova, Via F. Marzolo, 8, Padova, Italy}

\author{Giacomo Gradenigo}
\affiliation{Gran Sasso Science Institute, Viale F. Crispi 7, 67100 L’Aquila, Italy}
\affiliation{INFN-Laboratori Nazionali del Gran Sasso, Via G. Acitelli 22, 67100 Assergi (AQ), Italy}

\begin{abstract}

We study the equilibrium phases of a generalized Lotka-Volterra model characterized by a species interaction matrix which is random, sparse and symmetric. Dynamical fluctuations are modeled by a demographic noise with amplitude proportional to the effective temperature $T$. The equilibrium distribution of species abundances is obtained by means of the cavity method and the Belief Propagation equations, which allow for an exact solution on sparse networks. Our results reveal a rich and non-trivial phenomenology that deviates significantly from the predictions of fully connected models. Consistently with data from real ecosystems, which are characterized by sparse rather than dense interaction networks, we find strong deviations from Gaussianity in the distribution of abundances. 
In addition to the study of these deviations from Gaussianity, which are not related to multiple-equilibria, we also identified a novel topological multiple-attractor phase, present at both finite temperature, as shown here, and at $T=0$, as previously suggested in the literature. The peculiarity of this phase, which differs from the multiple-equilibria phase of fully-connected networks, is its strong dependence on the presence of extinctions. These findings provide new insights into how network topology and disorder influence ecological networks, particularly emphasizing that sparsity is a crucial feature for accurately modeling real-world ecological phenomena.

\end{abstract}

\maketitle


\section{Introduction}

Since the pioneering work of May \cite{may1973qualitative}, recent
advances in statistical mechanics have opened new avenues for
understanding the stability and phase behaviour of large
ecosystems. In particular, the generalized Lotka-Volterra (gLV) model
has emerged as a powerful framework for studying the coexistence of
species in complex ecological networks.  In the presence of symmetric
and dense interactions it is possible to exactly solve the equilibrium
statistical mechanics, characterizing completely the phase diagram of
the system.  In general the equilibrium phases of the gLV model with
disordered interactions are controlled by three parameters: $\mu$, the
average interaction strength; $\sigma$, the width of the disordered
interactions distribution; $T$, the amplitude of demographic
fluctuations. Typically, in dense networks, at finite values of $\mu$
and $T$, a multiple-equilibria phase emerges for large values of
$\sigma$~\cite{altieri2021properties}. However, realistic ecological
networks are neither dense nor characterized by symmetric
interactions. While the effort to optimize ecosystems modeling is
still in progress~\cite{Barbier2018generic,Barbier2020complex}, it is legitimate to wonder which aspects of
the phenomenology of fully-connected networks represent features of
realistic ecosystems and which ones are simply due to model
approximations.  While several papers already considered the case of
asymmetric
interactions~\cite{bunin2017ecological,galla2018dynamically,ros2023generalized,
  ros2023quenched, lorenzana2024non}, here we focus on symmetric ones,
so that we can exploit the tools of equilibrium statistical
mechanics. We consider interactions which have quenched disorder, the
case of annealed one being discussed
elsewhere~\cite{suweis2024generalized}, and which are sparse, at
variance with most of the previous literature. In particular we will
focus on the study of species abundance distributions, which in sparse
networks may have very relevant fluctuations.  It is in fact well
known that species abundance distributions found from real ecosystems
data are typically log-normal or gamma-like
distributed~\cite{may2007theoretical, sole2002self,
  volkov2007patterns, azaele2016statistical,
  grilli2020macroecological,McGill2007SAD}; however, in the case of
normally distributed quenched interactions, truncated Gaussians are
the standard for dense networks~\cite{bunin2017ecological,
  galla2018dynamically,garcia2022well}. Whilst deviations from
Gaussianity in the marginal distributions of the single degrees of freedom can be obtained by assuming non-Gaussian quenched
disorder~\cite{azaele2024generalized}, and in particular long-tail distributions for the couplings \cite{neri2010phase}, the goal of this work is to
show that strong non-Gaussian effects can also emerge owing to sparsity,
while keeping Gaussian quenched disorder.

The main focus of our analysis will be to understand the behaviour of the gLV model on a Random Regular Graph, which we will also refer to as Bethe Lattice. Specifically, we consider graphs with \emph{small connectivity} $k$, meaning that each species interacts with $k$ other species selected at random from the $N$ possible ones (in the following, we will also refer to the connectivity of a species as its \emph{degree}).
We will study the properties of the gLV model on a sparse network varying three parameters: the mean value $\mu$ and the standard deviation $\sigma$ of the nonzero normally distributed symmetric interactions and the demographic noise strength $T$.
We find two main results: strong deviation from Gaussianity in abundance distributions already in the single equilibrium phase and, for large enough values of the parameter $\mu$, a topological multiple attractor phase at finite temperature and $\sigma=0$. This phase looks very different from the one of fully connected models; in particular, as first noticed in \cite{marcus2022local}, in this phase the multiple equilibria are linked to different interaction network topologies of
surviving species.
The gLV model on a random regular graph was in fact already considered in \cite{marcus2022local}, where a phase transition to multiple equilibria was studied at zero temperature, $T=0$, zero disorder, $\sigma=0$, and finite interaction strength $\mu$. Here we extend a similar analysis to finite temperatures.

In the last part of this introductory section we will discuss how our work fits within the existing literature and connects to previous studies.
Ref. \cite{valigi2024local} focused on sparse ecosystems stability using spectral analysis of sparse random graphs with various topologies. However, it did not address the analysis of species abundance distributions nor the study of transitions from single to multiple equilibria phases. 
In Ref.~\cite{poley2024interaction}, gLV equations are analyzed on non fully-connected networks. The authors propose that degree heterogeneities are essential to have non-gaussian distributions. In contrast, our results challenge this conclusion. However in~\cite{poley2024interaction} the discussion is limited to the case of dense interactions, as the average degree distribution is always of the order of the total number of species $N$. Their methods do not apply to the case of sparse networks where the connectivity remains finite in the limit $N\to\infty$, that is the case we are analyzing.
In Ref. \cite{azaele2024generalized}, realistic species abundance distributions were derived with non-Gaussian distributed interactions by extending Dynamical Mean Field Theory. Similarly, in \cite{suweis2024generalized} comparable results were obtained in a system with stochastically varying species interactions. Non-Gaussian species abundances distributions are thus usually obtained as a consequence of non-Gaussian and/or variable interactions.
Concluding, no exhaustive study of the equilibrium properties of a sparse system with finite connectivity at finite temperature, in particular regarding the features of abundance distributions, has ever been done. Our work fills this gap for the first time.

The work will be organized as follows:
In Section~\ref{sec:bethe-lattice} we will introduce the model and present the
so called cavity equations used to obtain the equilibrium abundance marginal distributions; in Section~\ref{sec:non-gaussianity} we will discuss in detail the strong deviations from Gaussianity which can be found in the abundance distributions on a sparse ecological network. We will then show that no multiple equilibria phase emerges at small values of $\mu$, at variance with fully-connected models. Sec.~\ref{sec:topological-glass} will be then devoted to study the equilibrium phase diagram of the model at zero disorder ($\sigma=0$), varying only $\mu$ and $T$. The equilibrium properties in this regime have been investigated using both the cavity equations and the Langevin dynamics, finding consistency between the two methods.
Conclusions and perspectives will then be reported in Sec.~\ref{sec:conclusions}.

\section{Generalized Lotka-Volterra model on the Bethe lattice}
\label{sec:bethe-lattice}

The generalized Lotka-Volterra model is defined by the equations:

\begin{equation}\label{LV-equations}
    \frac{d n_i(t)}{d t}=\frac{r_i}{K_i} n_i(t)\big[K_i-
      n_i(t)-\sum_{j\in\partial i}\alpha_{ij}n_j(t) \big]+\xi_i(t),
\end{equation}
where $n_i(t)$ is the abundance of species ($i=1, ..., N$) at time
$t$ and $\partial i$ represents the set of species which interact with species $i$. We will focus on random graphs for which the cardinality of $\partial i$ is fixed and equal to $k_i=3$ for all $i$'s, even if the methods we use could be applied with no modification also in the case of a poissonian distribution of the connectivities $k_i$'s. The
factors $r_i$ and $K_i$ are respectively the intrinsic growth rate and
the carrying capacity of species $i$, while $\xi_i(t)$ represents a
Gaussian multiplicative noise with zero mean and covariance

\begin{equation}
    \langle \xi_i(t)\xi_j(t')\rangle= 2T n_i(t)\delta_{ij}\delta(t-t').
\end{equation}

We consider a symmetric interaction matrix with elements $\alpha_{ij}=\alpha_{ji}$. Because of the symmetric interactions we know that the Langevin dynamics admits an equilibrium distribution of the form $P({\bf n})=\exp(-H_{\text{eff}}({\bf n})/T)$,
with ${\bf n}=\lbrace n_1, \ldots, n_N \rbrace$ where the Hamiltonian $H_{\text{eff}}({\bf n})$ has been firstly derived in ref. \cite{BiroliBuninCammarota2018} following the Itô convention. When the stochastic dynamics of Eq.~\eqref{LV-equations} is complemented with a reflecting wall condition $n_i^{\text{min}}=\lambda$ with $\lambda \ll 1$ for every species $i$, used to avoid unphysical negative values of abundances in numerical simulations, the effective
Hamiltonian $H_{\text{eff}}({\bf n})$ reads as:
\begin{align}\label{Ham-Altieri}
  & H_{\text{eff}}({\bf n}) =-\sum_{i=1}^N r_i \left({n}_i-\frac{{n}_i^2}{2 {K}_i}\right)+ \nonumber \\
  & \sum_{(ij)\in E}\frac{\alpha_{ij}}{2}\left(\frac{r_i}{{K}_i}+\frac{r_j}{{K}_j}\right){n}_i {n}_j+\sum_{i=1}^N \left[{T} \ln\, ({n}_i)-\ln \theta({n}_i-\lambda)\right],
\end{align}
where the term with the Heaviside function $\theta({n}_i-\lambda)$ accounts for the presence of the reflecting wall.
The symbol $E$ in the double sum of Eq.~\eqref{Ham-Altieri} denotes the set of pairs of interacting species, which in the present work are arranged in a sparse random graph. Without any lack of generality, we will consider identical carrying capacities, $K_i=K$, and identical intrinsic growth rates, $r_i=r$, for all species. In particular we choose $r=1$ and $K=280$.
We study a sparse network with finite connectivity $k=3$, where the variables $n_i$ take only integer non-negative values, consistently with the interpretation of $n_i$ as the number of individuals for the species $i$. In this particular case, the choice of discrete variables is also an algorithmic necessity for an efficient computational way to solve the Belief Propagation equations, which we will present in the following and in detail in Appendix \ref{app:BP-algorithm}. Please note, however, that the choice to take integer values for $n_i$ is not inevitable: indeed one can choose to discretize $n_i$ into small steps with a discretization step $dn<1$. We do not expect that the integer discretization influences the results because the discretization step, that in our case is $dn=1$, should be compared with the natural scale for $n$ that is $K$ (fixed point of the single species dynamics) that we chose to be large, $K=280$, exactly for this reason. The independence of the results on the discretization choice is shown in Appendix \ref{app:changing_dn_eps}.
Taking into account the discreteness of $n_i$ values, we modify the effective Hamiltonian of Eq.~\eqref{Ham-Altieri}: the term $\ln\theta(n_i -
\lambda)$ is dropped and the term $T\ln(n_i)$ is replaced by $T\ln(n_i+\epsilon)$, with $\epsilon$ small but finite ($\epsilon=0.0001$) in order to regularize the distribution at $n_i=0$. In Appendix \ref{app:changing_dn_eps} we show that the obtained species abundances are almost independent on the exact value of $\epsilon$ as long as it is small enough.
The elements of the symmetric matrix $\alpha_{ij}$ are taken from a Gaussian
distribution with mean and variance which, consistently with the literature on disordered systems~\cite{altieri2021properties,poley2024interaction}, are defined as:
\begin{equation}\label{MeanAndVariance}
    \text{mean}[\alpha_{ij}]=\frac{\mu}{k}=\hat{\mu},\,\,\,\ \text{var}[\alpha_{ij}]=\frac{\sigma^2}{k}=\hat{\sigma}^2.
\end{equation}
We thus rewrite the regularized version of Eq.~\eqref{Ham-Altieri} for discrete variables as:
\begin{equation}\label{Ham}
  H_\alpha({\bf n})=\sum_{i=1}^N h_i({n}_i)+\sum_{(ij)\in E} h^\alpha_{ij}({n}_i,{n}_j),
 \end{equation}
where the label $\alpha$ in $H_\alpha (\bf n)$ denotes a given instance of the quenched disordered couplings $\alpha_{ij}$. In particular we have
\begin{align}
  & h_i(n_i) = - r \left(n_i-\frac{n_i^2}{2 K}\right)+T \ln\, (n_i+\epsilon), \nonumber \\
  & h_{ij}^\alpha(n_i, n_j)= \frac{r}{K}\alpha_{ij}n_i n_j.
\end{align}
In the following we will use the so-called cavity method~\cite{mezard2009information} to sample the equilibrium Boltzmann distribution $P({\bf n}) \propto e^{-\beta H_\alpha({\bf n})}$ of the discrete variables $n_i$ and in particular to compute their marginals. 
The crucial observation is that a random regular graph is a \emph{locally tree-like} graph, for which the typical length of a loop is $\log(N)$ and thus it diverges in the large-$N$ limit (as also for a graph with a Poissonian degree distribution).
The main property of a random regular graph is therefore that in the thermodynamic limit, due to the absence of loops, local marginals become effectively factorized. It is in this perspective that, given two neighbouring nodes $i$ and $j$, one introduces the so called cavity marginal $\eta^\alpha_{i\rightarrow j}(n_i)$, where $\alpha$ denotes an instance of the quenched disorder. The cavity marginal is defined as the marginal probability distribution of the variable $n_i$ in a graph where the edge connecting the node $i$ with the node $j$ has been removed.
The cavity marginals satisfy the following self-consistent equations, that are called Belief-Propagation (BP) equations: 
\begin{align}\label{eq-cavitymarginal}
\eta_{i\rightarrow j}^\alpha(n_i)=\frac{e^{-\beta h_i(n_i)}}{z^\alpha_{i\rightarrow j}}\prod_{k\in \partial i\backslash j}\left[\sum_{\{n_k\}}\eta^\alpha_{k\rightarrow i}(n_k) e^{-\beta h^\alpha_{ik}(n_i, n_k)} \right], \nonumber \\ 
\end{align}
where $z^\alpha_{i\rightarrow j}$ is a normalization factor
\begin{align}\label{eq-norm}
  z^\alpha_{i\rightarrow j}=\sum_{\{n_i\}} e^{-\beta h_i(n_i)} \prod_{k\in \partial i\backslash j}\left[\sum_{\{n_k\}}\eta^\alpha_{k\rightarrow i}(n_k)
    e^{-\beta h^\alpha_{ik}(n_i, n_k)} \right] \nonumber \\ 
\end{align}
and where we indicate with $\partial i\backslash j$ the set of neighbours of $i$, excluding node $j$.
The solutions of Eqs.~\eqref{eq-cavitymarginal},~\eqref{eq-norm} can be obtained iteratively: details on the derivation of the equations and the implementation of the algorithmic procedure are given in Appendix~\ref{app:BP-algorithm}. 
In particular we have run the BP algorithm on a given graph, where each graph, being the connectivity fixed, is completely characterized by the choice of the quenched disordered couplings. At convergence we have a number $2\abs{E}$ of cavity marginals $\eta^\alpha_{i\rightarrow j}$, where $\abs{E}$ is the number of non-oriented edges. Once the cavity marginals have been computed, one can extract from them the marginal distribution $\eta^\alpha_i(n_i)$ for each
species $i$ as:
\begin{align}\label{eq:marginal-eta}
    \eta^\alpha_i(n_i) = \frac{1}{z^\alpha_i}e^{-\beta h_i(n_i)}\prod_{j\in
      \partial i}\left[\sum_{\{n_j\}}\eta^\alpha_{j\rightarrow i}(n_j)
    e^{-\beta h^\alpha_{ij}(n_i, n_j)} \right], \nonumber \\ 
\end{align}
where $z^\alpha_i$ is a normalization factor ensuring that $\sum_{n_i}\eta^\alpha_i(n_i)=1$. These are the exact marginal distributions associated to the equilibrium Gibbs-Boltzmann probability for a locally tree-like graph (as a random-regular graph or a Poissonian graph) as long as the whole Gibbs measure is composed of just one state.
Note that, because of the quenched disorder and the sparsity of the interaction network, the marginals for different species will be different. This is a crucial difference with respect to the fully-connected case where, for a given instance of the disorder, all marginals are identical by construction.
Taking into account that on sparse graphs the marginals may have fluctuations not only with respect to the graph realization but also from species to species, we consider two kinds of average for the
distribution $\eta^\alpha_i(n_i)$: the {\it sample} average,
\begin{align}\label{eq:sample-average}
\eta^\alpha(n) \equiv \frac{1}{N} \sum_{i=1}^N \eta^\alpha_i(n),
\end{align}
and the {\it disorder} average,
\begin{align}\label{eq:disorder-average}
  \eta(n) \equiv \overline{\eta^\alpha(n)},
\end{align} 
where the overline $\overline\cdot$ denotes the average over different realizations of the disordered interaction network (the graph) and the disordered interaction couplings $\{\alpha_{ij}\}$.
Throughout the whole paper we will always present marginal distributions averaged both over the sample and the disorder. 

We stress that on the fully-connected networks where the thermodynamics can be exactly solved (\cite{altieri2021properties})
the (global) local marginal $\eta(n_i)$ has a truncated Gaussian form, with the extinction peak at the origin already on a single realization of a graph. The main advantage of studying the generalized Lotka-Volterra model on a sparse topology is that the shape of local marginal $\eta_i(n_i)$ is not constrained to be a truncated Gaussian, even though the interactions are drawn from a Gaussian distribution. Indeed, in the next section we will show how in the presence of a sparse topology, at variance with fully connected networks, species abundance distributions develop interesting non-Gaussian and Gamma-like features already in the single equilibrium phase.

\section{Strong Non-Gaussianity in the single-equilibrium phase}
\label{sec:non-gaussianity}

The disordered system approach to large ecosystems has provided many
interesting and suggestive insights on their behaviour and their
response/stability with respect to external
perturbations~\cite{may1973qualitative,diederich1989replicators,altieri2021properties, opper1992phase}. Nevertheless, the robustness of the exact results obtained so far for the gLV problem depends on a very crude approximation in the model: the consideration of fully-connected interaction networks between species~\cite{bunin2017ecological,galla2018dynamically}, something which is clearly very unrealistic from the ecological systems perspective. A few points need to be carefully investigated: which aspects of the phenomenology obtained so far by exact results from disordered systems techniques are consequences only of the fully-connected networks?
Are there features intrinsically related to the sparsity of ecological networks, as for instance fat-tailed or otherwise non-Gaussian tailed abundance distributions, which cannot be
reproduced by the fully-connected model with normally distributed random interactions? 
In addition to addressing these questions, our work aims to determine whether the multiple attractor phases found in fully connected models remain robust in sparse networks, making them more likely to occur in realistic ecosystems.

The first important result we want to show is how sparsity is capable to induce remarkable deviations from Gaussianity in abundance distributions even in the presence of Gaussian random couplings. 

\begin{figure}[htbp]
    \includegraphics[width=\columnwidth]{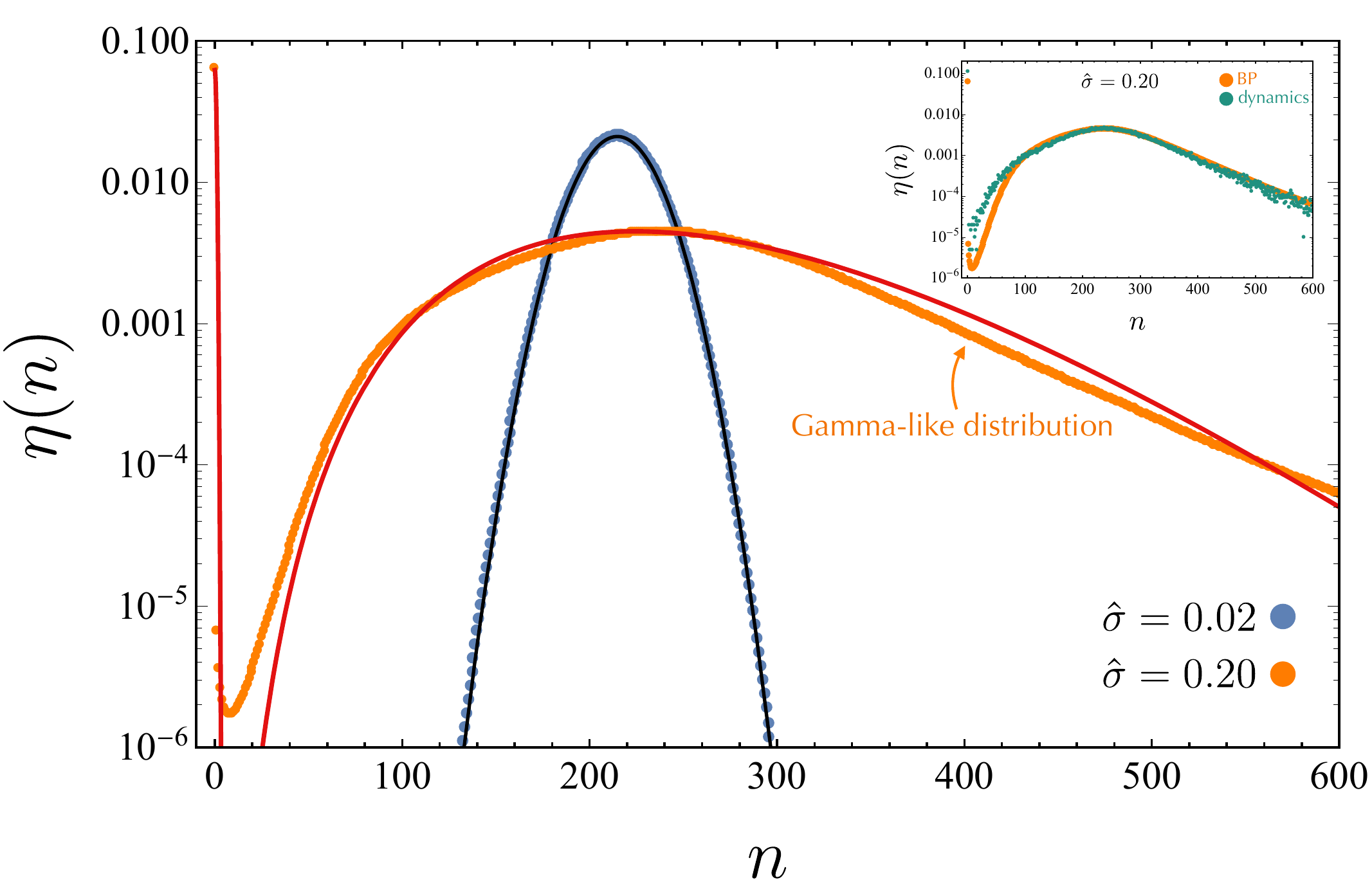}
    \caption{\textbf{Species abundance average distribution for two values of the coupling variance $\hat{\sigma}$ in the single equilibrium phase.}
    For small $\hat{\sigma} = 0.02$ (blue points) the distribution follows a Gaussian distribution (black line), while for large $\hat{\sigma} = 0.20$ (orange points) the species abundance follows a Gamma distribution (red line) as in \eqref{eq:Gamma}, with parameters $\alpha=7.5$ and $\beta=0.029$ plus a gaussian peaked at $n=0$. \textbf{Inset:} marginal at $\hat{\sigma} = 0.20$ computed using BP (orange points) and from the dynamics (green points). The two agree very well. In both plots the parameters are $T=1$, $\hat{\mu}=0.1$, $N=256$.}
    \label{fig:exp-tail}
\end{figure}

In Fig.~\ref{fig:exp-tail} we show the averaged species abundance marginal distribution $\eta(n)$, Eq.~\eqref{eq:disorder-average}, at $T=1$ and $\hat{\mu}=0.1$, for two different values of the standard deviation of the disordered couplings: $\hat{\sigma}=0.02$
and $\hat{\sigma}=0.20$. By studying the convergence of the Belief
Propagation algorithm, which will be discussed more thoroughly
in Sec.~\ref{sec:NOglass}, we know that both these two
values of the parameter $\hat{\sigma}$ correspond to a single equilibrium
phase. The remarkable finding is that, despite the lack of any multiple equilibria, upon increasing $\hat{\sigma}$ we find a crossover from a {\it low disorder} phase, where the
mean marginal $\eta(n)$ is well described by a Gaussian, to a {\it high disorder} phase
where $\eta(n)$ is highly non-Gaussian and is well described by a Gamma distribution. In particular, for $\hat\sigma=0.20$, the distribution $\eta(n)$ is well fitted by the red curve in Fig.~\ref{fig:exp-tail}, given by:
\begin{equation}\label{eq:Gamma}
\gamma(n;\alpha,\beta)=\frac{\beta^\alpha}{\Gamma(\alpha)}n^{\alpha-1}e^{-\beta n},
\end{equation}
where $\Gamma(\alpha)$ is the Gamma function and the fitted parameters are $\alpha=7.5$ and $\beta=0.029$. This is a very significant result as real ecosystems data exhibit species abundance distributions that follow a Gamma distribution \cite{grilli2020macroecological,volkov2007patterns}.

As we show in the top-right inset of Fig.~\ref{fig:exp-tail}, the shape of the fat-tailed distribution is accurately reproduced by running the Langevin dynamics. This consists in fixing a disordered graph, running the set of equations given by Eq.~\ref{LV-equations} and stopping the simulation at a sufficiently large time $t_\text{max}$ to collect all the $n_i(t_\text{max})$ for every $i$. By repeating this process for multiple disorder realizations and combining all $n_i(t_\text{max})$ from each realization, we construct the normalized histogram of species abundances, shown in green in the inset of Fig.~\ref{fig:exp-tail}. Further details on the implementation of the dynamics are provided in Sec.~\ref{subsec:langevin-dynamics}, while in Appendix \ref{app:BP-vs-DYN} we show that the marginals extracted from BP coincide with those extracted from the dynamics in all the region of parameters for which there is a unique fixed point.

Moreover, we show in Appendix \ref{app:Scaling} that in the region of the parameters where the marginals exhibit strong non-gaussianity, variances of the single species abundance distributions follow a power-law dependence on the means of the distributions, well known as Taylor's law, documented extensively in real ecosystems (see \cite{Giometto-taylor-law}). In this region of parameters, what predicted by our sparse LV model is thus in perfect agreement with what observed in real ecosystems.

\begin{figure}
    \includegraphics[width=\columnwidth]{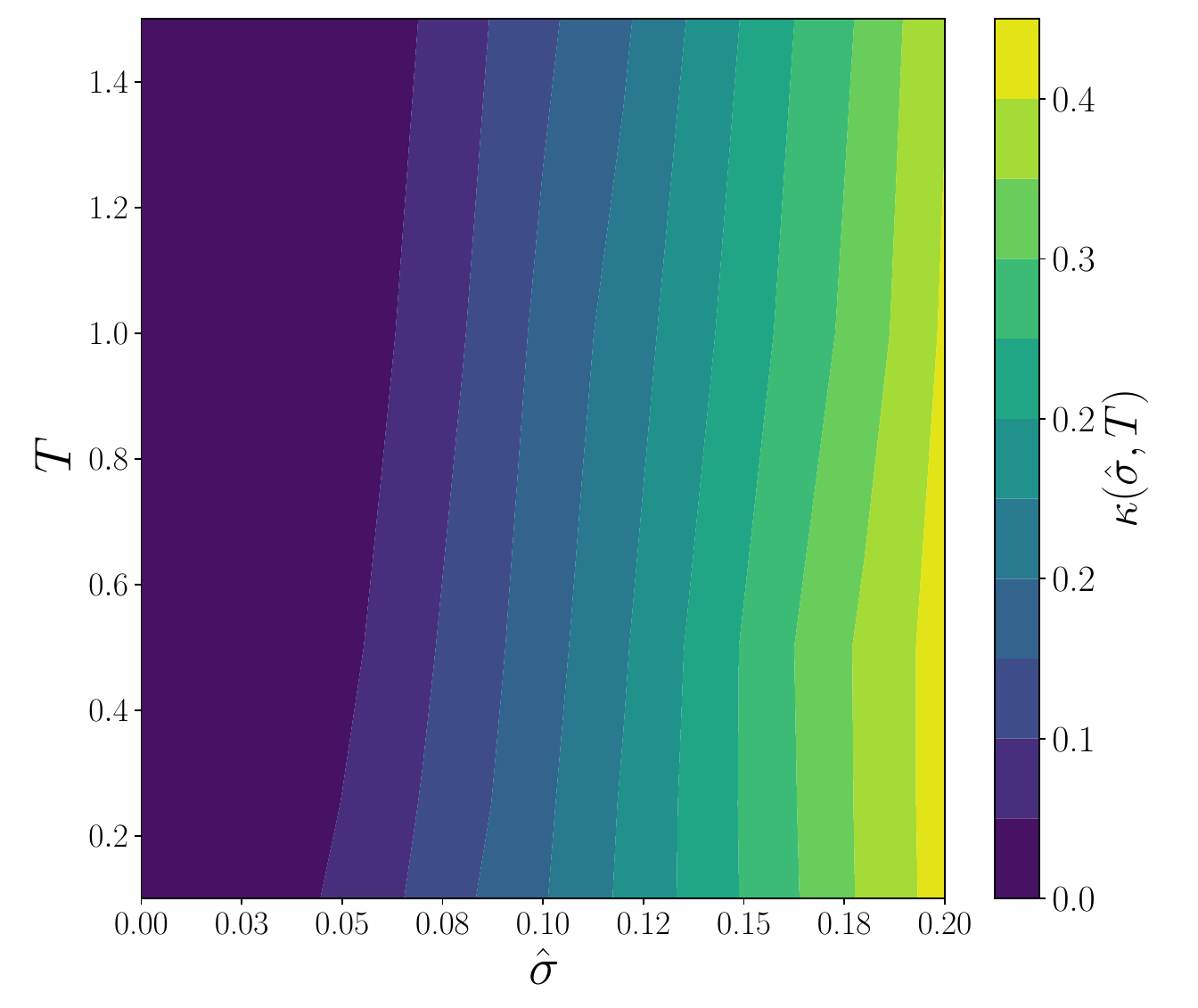}
    \caption{\textbf{Kurtosis $\kappa(\hat{\sigma},T)$ of the average species abundance in the unique fixed point phase.} For high variance of the couplings $\hat{\sigma}$ it is evident that $\kappa> 0$ implying that the marginals distributions are developing a non-Gaussian tail, while the kurtosis is almost independent on the temperature $T$. The parameters of the model are $\hat{\mu}=0.1$ and $N=256$.}
    \label{fig:2d-contour}
\end{figure}

The crossover from almost Gaussian to strongly non-Gaussian behaviour varying both the disorder strength $\hat{\sigma}$ and the temperature $T$ is represented in the phase diagram shown in Fig.~\ref{fig:2d-contour}, where color code represents the kurtosis $\kappa(\hat{\sigma},T)$ of the marginal distribution
\begin{align}
    \kappa(\hat{\sigma},T)\equiv\frac{\langle(n-\langle n\rangle)^4\rangle}{\langle(n-\langle n\rangle)^2\rangle^2} -3,
\end{align}
where the averages $\langle\cdot\rangle$ are taken with respect to $\eta(n)$. For high values of $\hat{\sigma}$ it can be easily recognized the strong deviation from Gaussianity due to the non-gaussian tail of $\eta(n)$ shown in Fig.~\ref{fig:exp-tail}. Looking at the marginal distribution form in eq.~\eqref{eq:marginal-eta}, it is evident that there is a factor $\frac{1}{n_i+\epsilon}$, associated to the demographic noise, preventing $\eta_i^\alpha$ from being a perfect Gaussian even in the absence of all the interactions; thus one could be tempted to claim that this could be the reason for the non-Gaussian beaviour of the marginals. In Appendix \ref{app:eta-tilde} we show that indeed the strong deviation from Gaussianity is mainly due to the sparse interactions and not to the demographic noise factor.

\begin{widetext}

\begin{figure}[htbp] 
    \centering
    \begin{minipage}{0.5\textwidth}
        \centering
        \includegraphics[width=\textwidth]{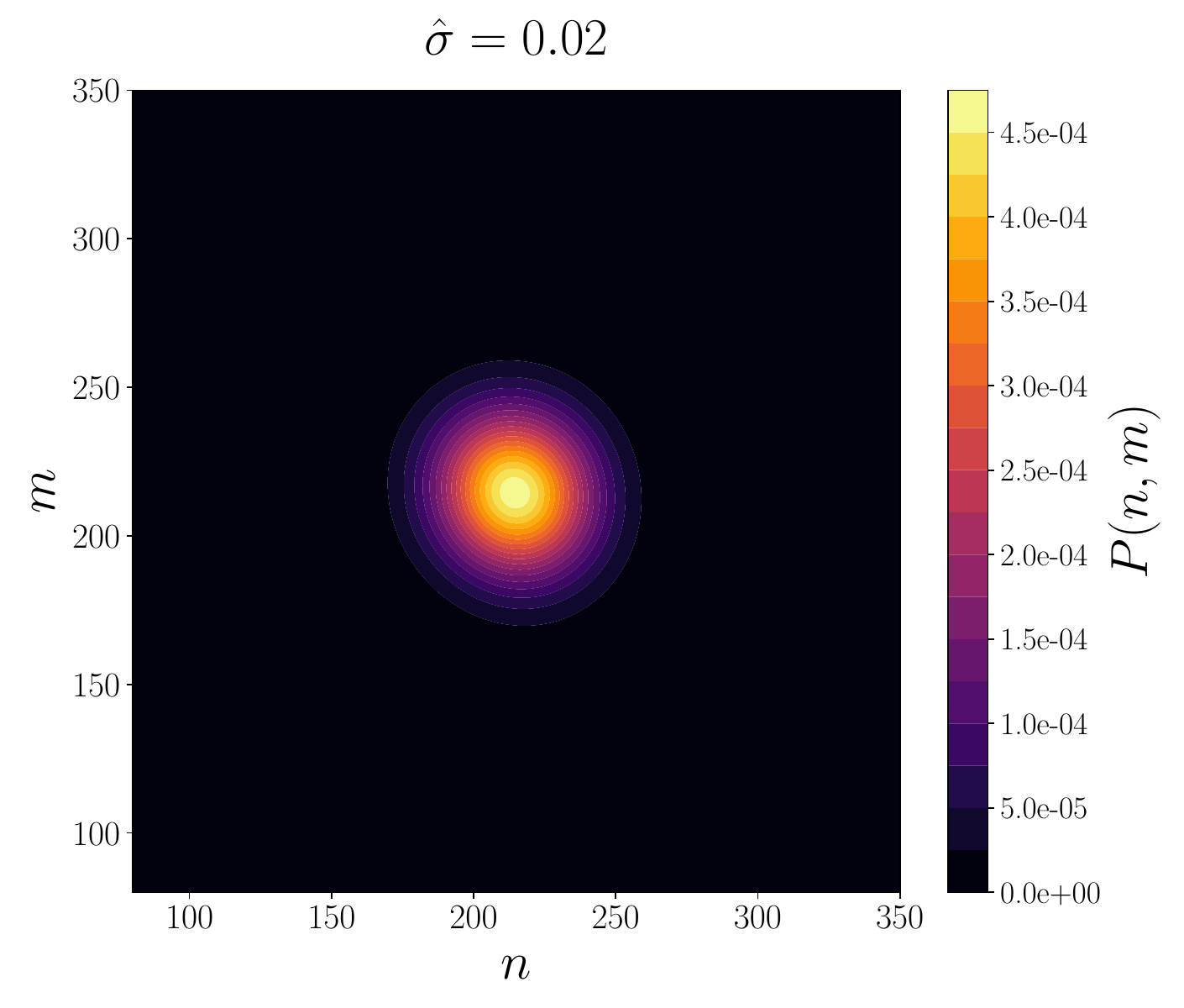}
    \end{minipage}
    \begin{minipage}{0.48\textwidth}
        \centering
        \includegraphics[width=\textwidth]{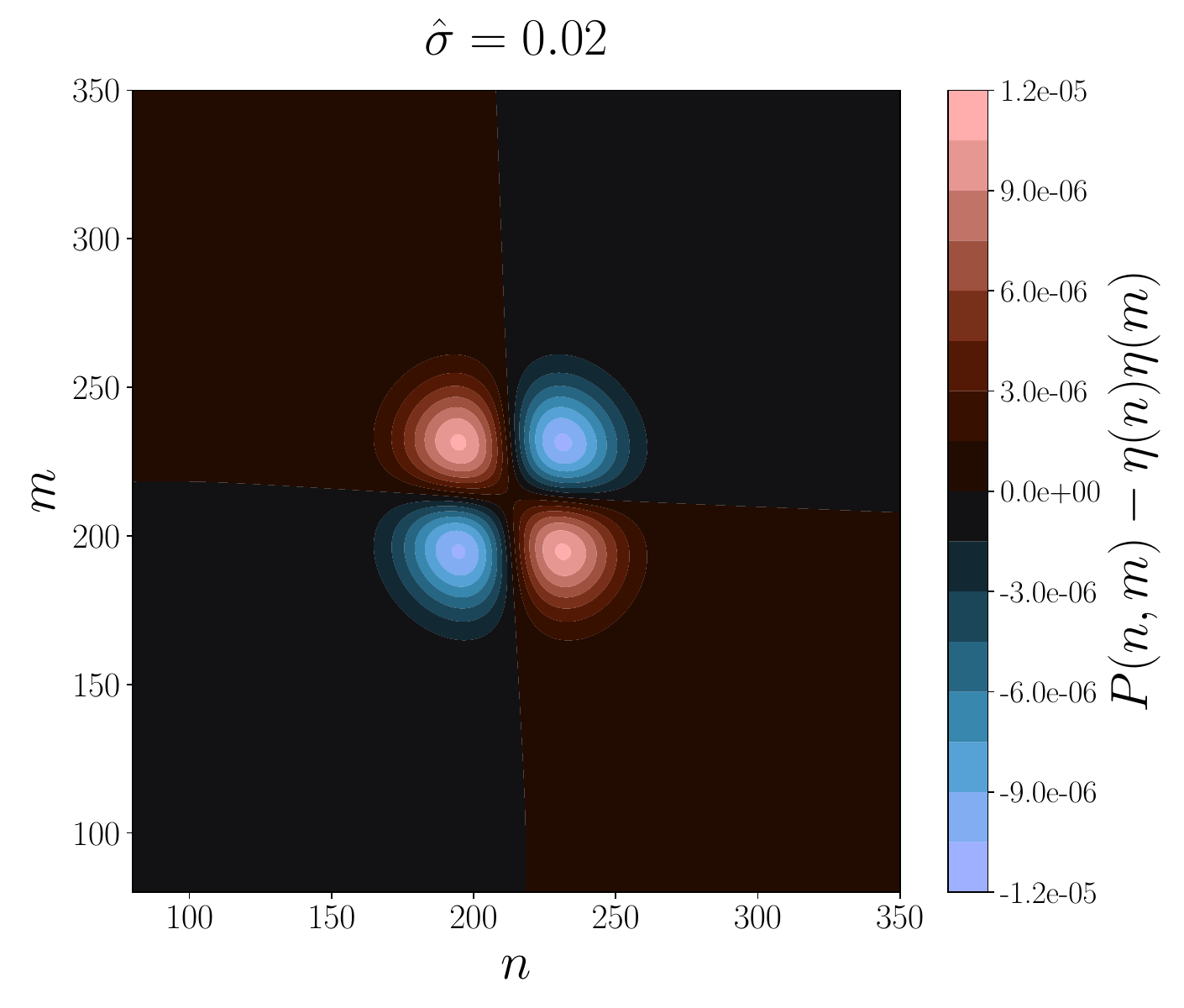}
    \end{minipage}
    \begin{minipage}{0.5\textwidth}
        \centering
        \includegraphics[width=\textwidth]{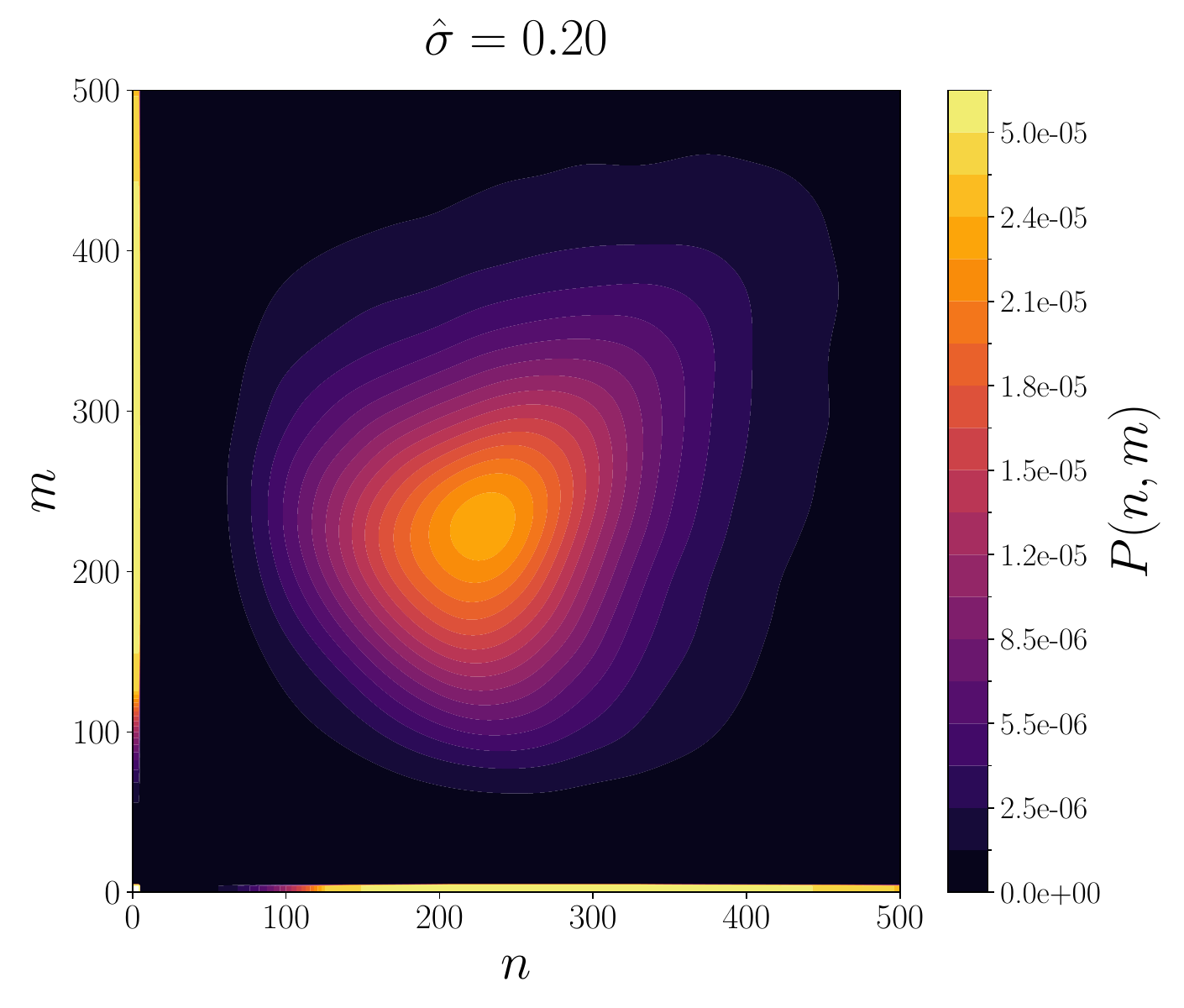}
    \end{minipage}
    \begin{minipage}{0.48\textwidth}
        \centering
        \includegraphics[width=\textwidth]{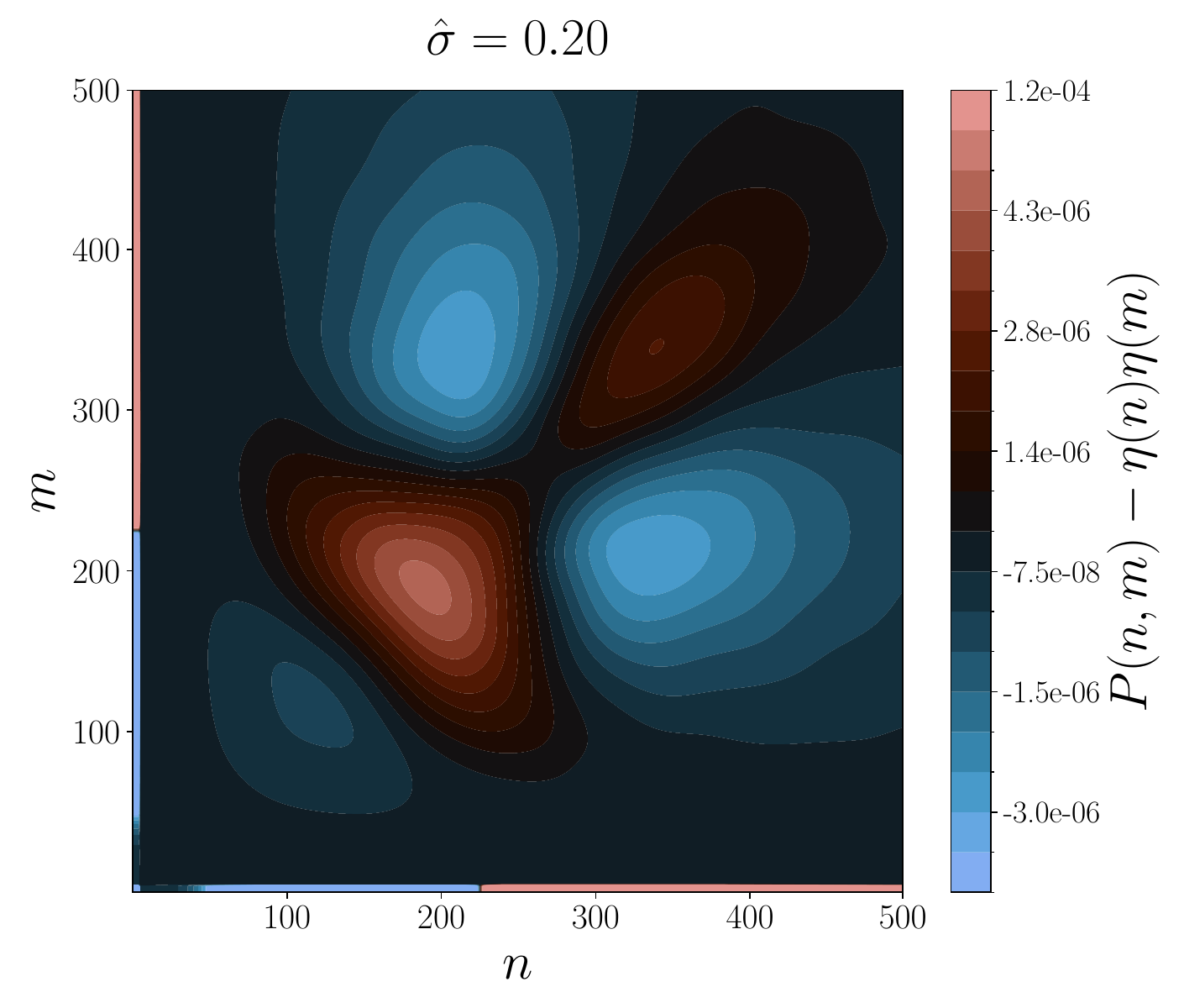}
    \end{minipage}
    \caption{\textbf{Left:} Contour plots for the joint distribution
      $P(n,m)$, see Eq.~\eqref{eq:disorder-average-joint}, for $\hat{\sigma}=0.02$, in the Gaussian regime (top), and $\hat{\sigma}=0.20$, in the non-Gaussian regime (bottom). \textbf{Right:} connected correlation function  $P(n,m)-\eta(n)\eta(m)$ for $\hat{\sigma}=0.02$ (top), and $\hat{\sigma}=0.20$ (bottom). Correlations are different from zero, at variance to what happens in the fully connected case. For all the panels $\hat{\mu}=0.1$, $T=1$ and $N=256$.}
    \label{fig:Pnm_contour-plots}
\end{figure}

\end{widetext}

The transition from almost Gaussian to strongly non-Gaussian
behaviour does not only affect the single
variable marginal distribution but has also an influence on
correlations. This fact is evident when studying the joint probability distribution $P^\alpha_{ij}(n_i,n_j)$ of a pair of nearest neighbour species on a given graph, which is  defined as:
\begin{align}
 P_{ij}^\alpha(n_i,n_j) = \frac{1}{\mZ^\alpha} \sum_{\lbrace n_1,\ldots,n_N \rbrace \setminus (n_i,n_j )} e^{-\beta H_{\alpha}({\bf n})},
\end{align}
where $\mZ^\alpha$ is a normalization factor. Contrary to what one finds in fully-connected networks, $P^\alpha_{ij}(n_i,n_j)$ does not factorize in a sparse graph. Indeed, on a dense network, where the connectivity grows with $N$, the joint
distribution $P^\alpha_{ij}(n_i,n_j)$ should simply obey a 
factorization property of the kind:
\begin{align}
P^\alpha_{ij}(n_i,n_j) \xrightarrow[N\to\infty]{} \eta^\alpha(n_i)\eta^\alpha(n_j).
\end{align}
On the Bethe lattice the {\it two-point joint probability} on a given graph
can be exactly computed as:
\begin{align}
  P^\alpha_{ij}(n_i,n_j) = \frac{1}{\Omega^\alpha_{ij}}\eta^\alpha_{i\rightarrow j}(n_i)\eta^\alpha_{j\rightarrow i}(n_j)
  e^{-\beta\frac{r}{K}\alpha_{ij} n_i n_j},
\end{align}
where $\Omega^\alpha_{ij}$ is the normalization factor.
Similarly to what we have done for the single species marginal, we
can define also for this two-species joint probability
the {\it sample} average as
\begin{align}
    P^\alpha(n,m)\equiv\frac{1}{\abs{E}}\sum_{(ij)\in E} P^\alpha_{ij}(n,m),
\end{align}
where both $n$ and $m$ represent species abundances and the summation runs over all edges in a given graph. The {\it
  disorder} average is defined as
\begin{align}\label{eq:disorder-average-joint}
  P(n,m)\equiv\overline{P^\alpha(n,m)}.
\end{align}
The contour plots showing the behaviour of $P(n,m)$ in the plane $(n,m)$ are reported on the two left panels of Fig.~\ref{fig:Pnm_contour-plots} for two different values of the standard deviation: $\hat{\sigma}=0.02$, in the
Gaussian regime, and $\hat{\sigma}=0.20$, in the strongly non-Gaussian regime. On the right panels of Fig.~\ref{fig:Pnm_contour-plots} we show the contour plot of $P(n,m)-\eta(n)\eta(m)$ for the same two different values of $\hat{\sigma}$. It is clear that $P(n,m) \neq \eta(n)\eta(m)$ and the presence of either correlations or anticorrelations can be fully appreciated. This is an important difference with respect to the fully-connected case where there are no correlations between different species. 
\begin{figure}[htbp]
    \centering
    \includegraphics[width=\columnwidth]{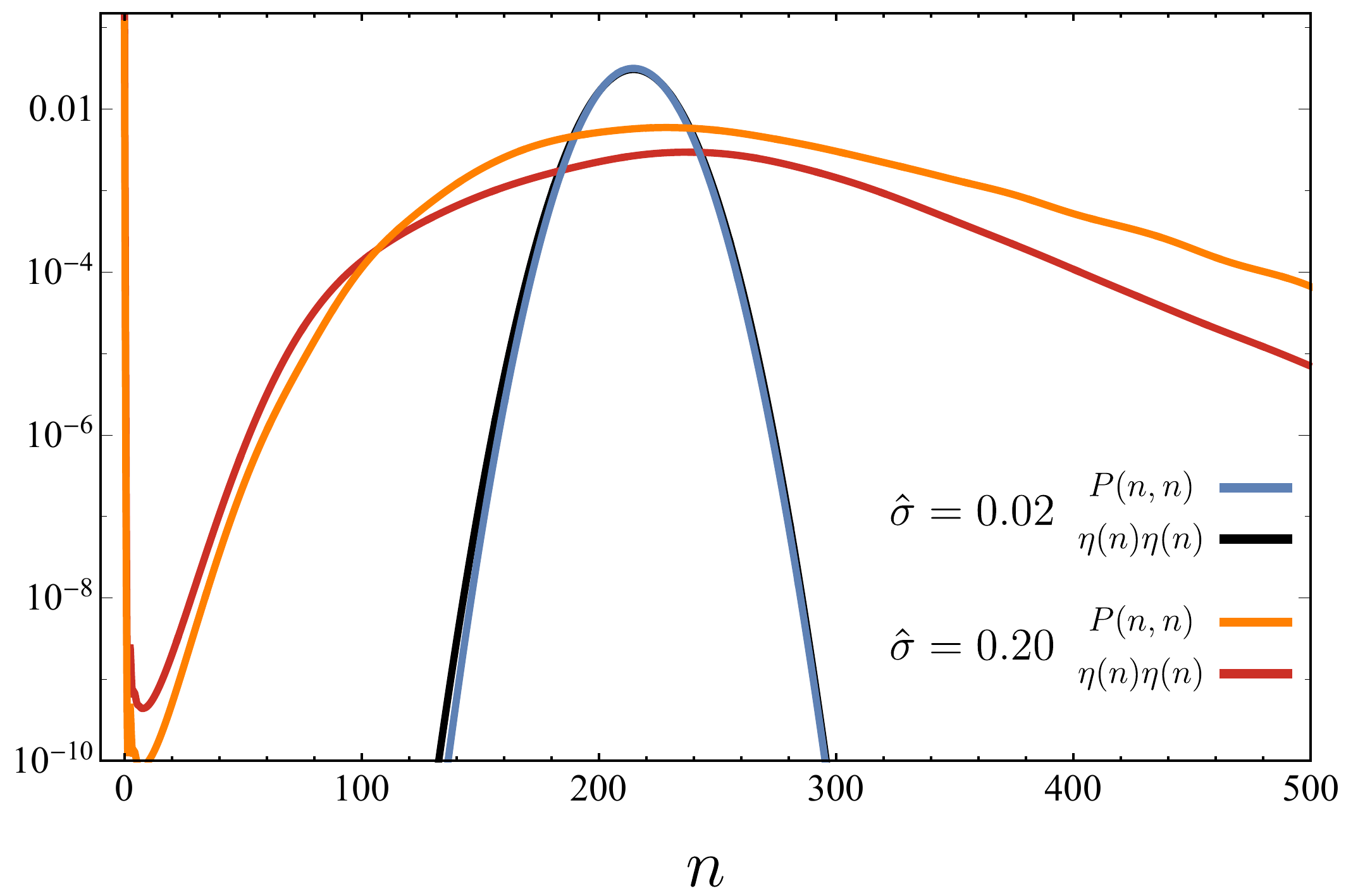}
    \caption{\textbf{Normalized sections of $P(n,n)$ and $\eta(n)\eta(n)$ for  $\hat{\sigma}=0.02$, Gaussian regime, and $\hat{\sigma}=0.20$, non-Gaussian regime.} Non trivial correlation effects are evident, differently from the fully connected case for which $P(n,n)=\eta(n)\eta(n)$. The parameters of the model are $\hat\mu=0.1$, $T=1$ and $N=256$.}
    \label{fig:Pnm_sections}
\end{figure}
In Fig.~\ref{fig:Pnm_sections} we compare $P(n,m)$ for $n=m$ with $\eta(n)\eta(n)$ in order to highlight again the relevance of correlations. In particular, looking at the case $\hat\sigma=0.20$ in Fig.~\ref{fig:Pnm_sections}, we see 
that $P(n,n)<\eta(n)\eta(n)$ at small values of $n$, which means that positive (competitive) interactions dominate the average correlation. On the contrary we find that for large values of the abundances one has $P(n,n)>\eta(n)\eta(n)$: in this regime negative (mutualistic) interactions are the dominant ones.  From the study of 
the connected probability $P(n,n)-\eta(n)\eta(n)$ it is therefore clear that strong non-Gaussianity affects (in a non trivial way) also the correlations when the disorder parameter $\hat{\sigma}$ is increased.
We have then studied the function $P(n,m)-\eta(n)\eta(m)$ in the specific cases of either exclusively positive ($\alpha_{ij}>0$) or exclusively negative ($\alpha_{ij}<0$) interactions. Namely we have specialized the average of $P^\alpha(n,m)$ with respect to the sign of the interaction:

\begin{align}
 P^{(+)}(n,m)=\overline{P^\alpha(n,m)}\big|_{\lbrace\alpha_{ij}>0\rbrace}, \nonumber \\
 P^{(-)}(n,m)=\overline{P^\alpha(n,m)}\big|_{\lbrace\alpha_{ij}<0\rbrace}.
\end{align}

\begin{figure}[htbp]
    \centering
    \includegraphics[width=\columnwidth]{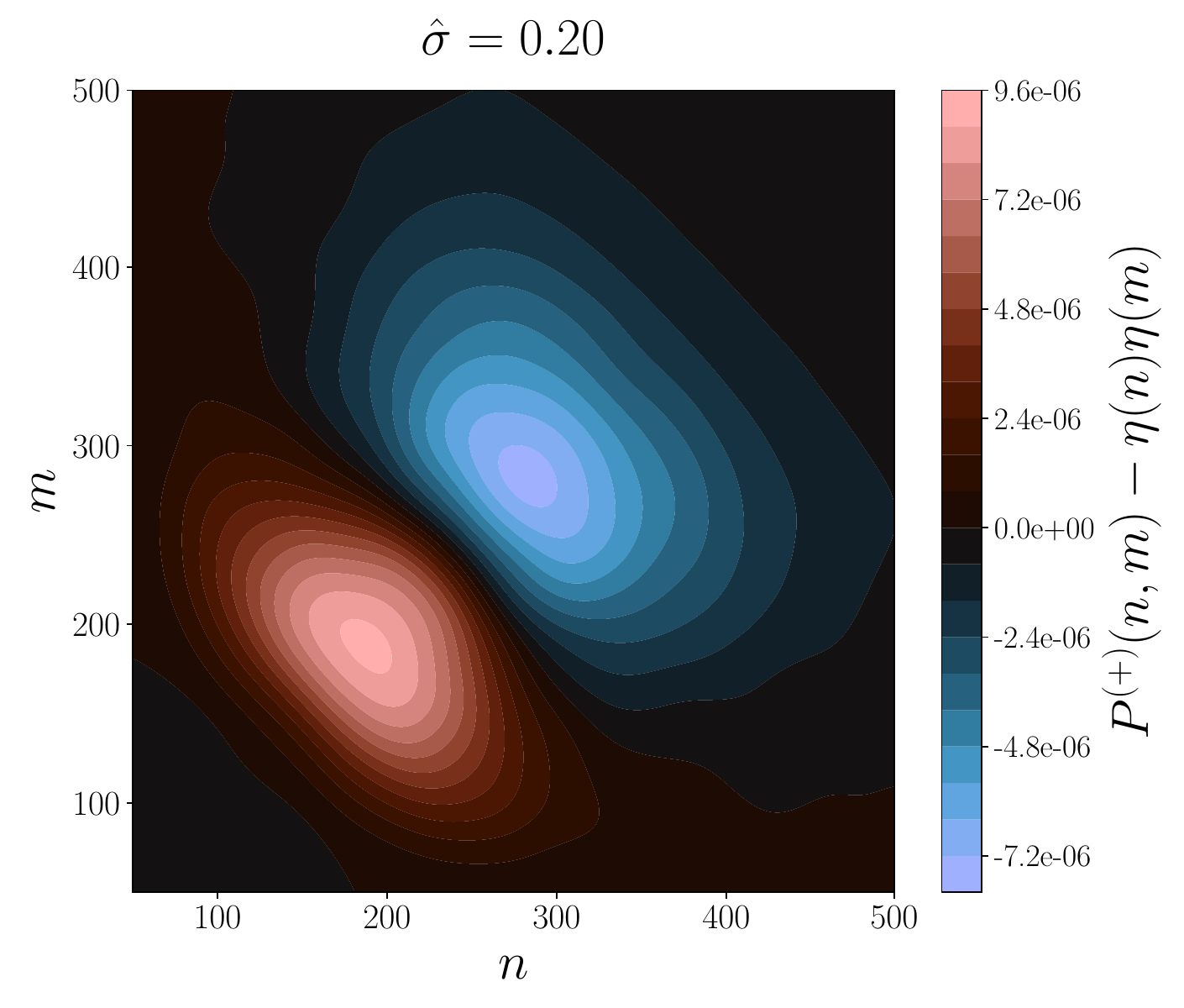}
    \includegraphics[width=\columnwidth]{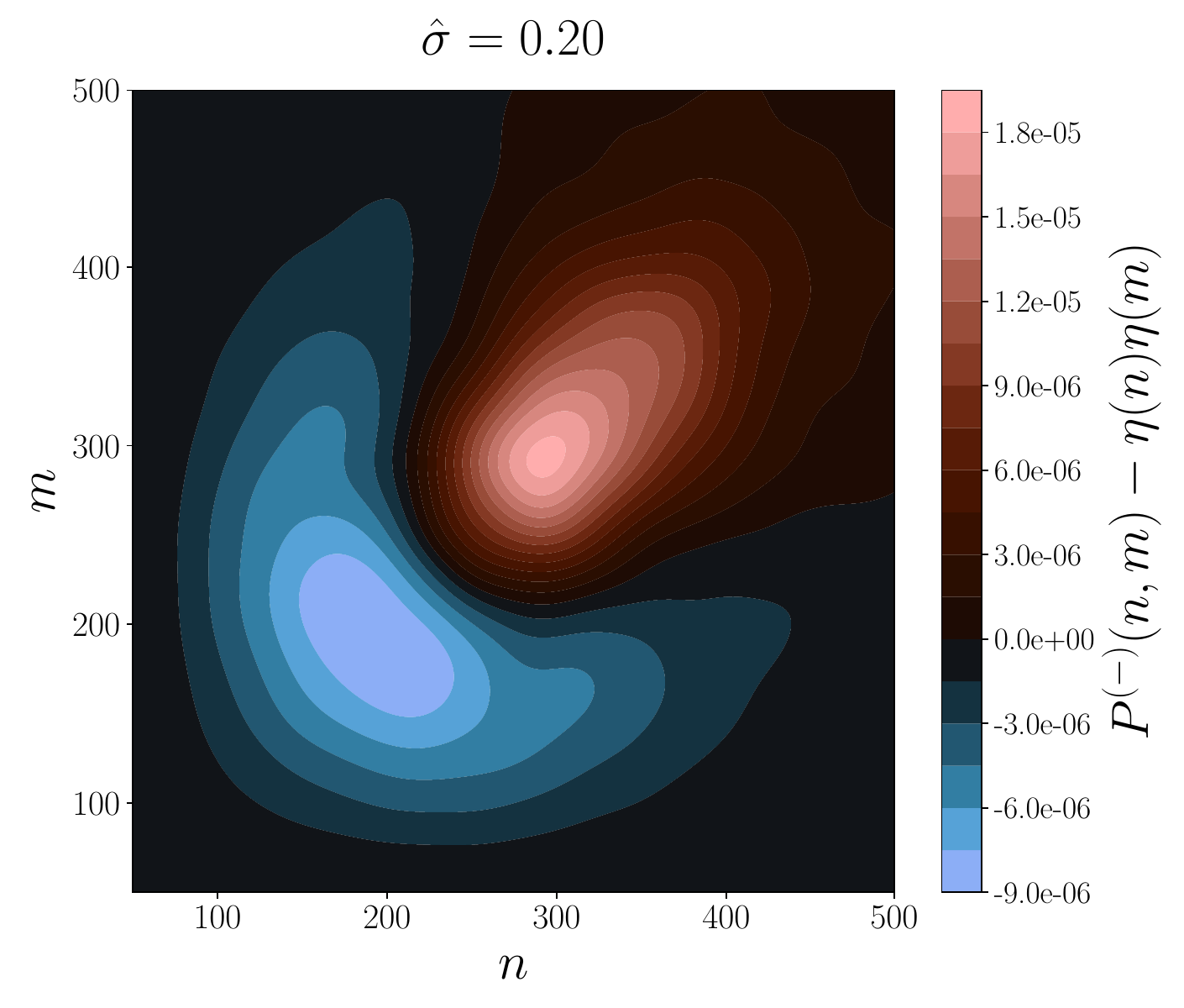}
    \caption{\textbf{Connected correlation functions for exclusively competitive $P^{(+)}(n,m)-\eta(n)\eta(m)$ (top) and exclusively mutualistic $P^{(-)}(n,m)-\eta(n)\eta(m)$ (bottom) interactions.} We are inside the strong-nonGaussian region $\hat\sigma=0.20$.}
    \label{fig:diff_pos_neg_s020}
\end{figure}

\begin{figure}[htbp]
    \centering
    \includegraphics[width=\columnwidth]{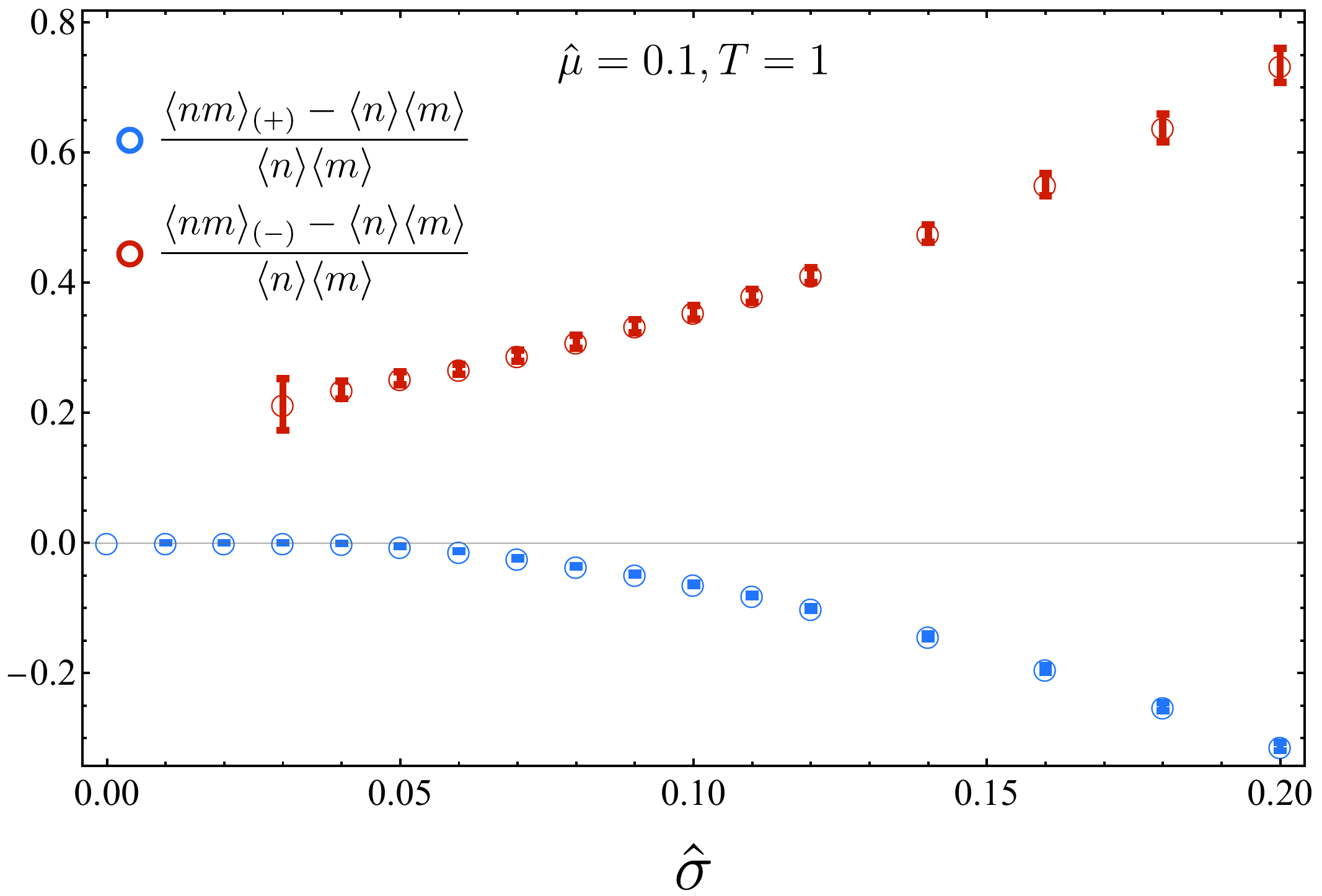}
    \caption{\textbf{Normalized covariances of the average abundances for neighboring species with positive (blue dots) and negative (red dots) interactions.} Data are displayed in function of $\hat\sigma$. Notice that for low values of $\hat\sigma$ no red dots are present because interactions are never mutualistic for low values of $\hat\sigma$. Correlations between neighboring species grow (in modulus) increasing the variance of the interactions $\hat\sigma$. Error bars are included in the plot.}
    \label{fig:corr-pos-neg-varying-sigma}
\end{figure}

The results for $P^{(+)}(n,m)$ and $P^{(-)}(n,m)$ in the case $\hat{\sigma}=0.20$ are shown in Fig.~\ref{fig:diff_pos_neg_s020}: it is clear that competitive (positive sign) interactions lead to anti-correlations, while mutualistic (negative sign) interactions induce positive correlations. From the joint probabilities $P^{(+)}(n,m)$ and $P^{(-)}(n,m)$, we can extract the mean values $\langle n m\rangle_{(+)}$ and $\langle n m\rangle_{(-)}$, which correspond to the average product $nm$ for neighbouring species with positive and negative interactions, respectively. Figure~\ref{fig:corr-pos-neg-varying-sigma} shows the corresponding normalized covariances defined as $\left(\langle nm\rangle_{(+)/(-)}-\langle n\rangle \langle m\rangle\right)/\langle n\rangle \langle m\rangle$, plotted as function of $\hat\sigma$. Competitive interactions lead to negative correlations while mutualistic interactions to positive correlations. The magnitude of these correlations increases with $\hat\sigma$.
\\

\subsection{Absence of a multiple attractor phase for a small competition strength $\hat\mu$}
\label{sec:NOglass}

We recall that in the case of a fully-connected network with symmetric interactions~\cite{altieri2021properties}, a multiple-equilibria phase is found at $T>0$ for any positive $\mu$, provided that $\sigma$ is large enough. Further increasing $\sigma$ leads then to the emergence of an unbounded growth phase, where species grow indefinitely. On the contrary, in the sparse network we find that for finite temperatures and $\hat{\mu}=0.1$, by increasing $\hat{\sigma}$ the system passes directly from the single equilibrium phase to the unbounded growth one, without any signature of an intermediate multiple attractor phase. The convergence of the belief propagation equations on a random graph directly implies the stability of the replica symmetric solution \cite{zdeborova2009statistical}; We have thus checked the convergence of the BP equations (or the entrance in an unbounded growth phase) in the whole range of  temperatures for $\hat{\mu}=0.1$ and this is sufficient to show the stability of the RS phase. This is a crucial difference between the BP method and the replica method usually used in fully-connected models: while in the latter, after having done a replica computation with the RS ansatz, one should always check that the RS solution is stable, for BP equations in the RS ansatz instead, the convergence of BP equations automatically implies that the RS phase found is locally stable. We have also checked this result by running the Langevin dynamics many times from different initial conditions, finding either a single fixed point, or an unbounded growth region without signatures of multiple fixed points, as shown in Appendix \ref{app:BP-vs-DYN}.
An investigation of the phase diagram for different (larger) values of $\hat{\mu}$ is presented in Sec.~\ref{sec:topological-glass}.

\begin{figure}[htbp]
    \centering
    \includegraphics[width=\columnwidth]{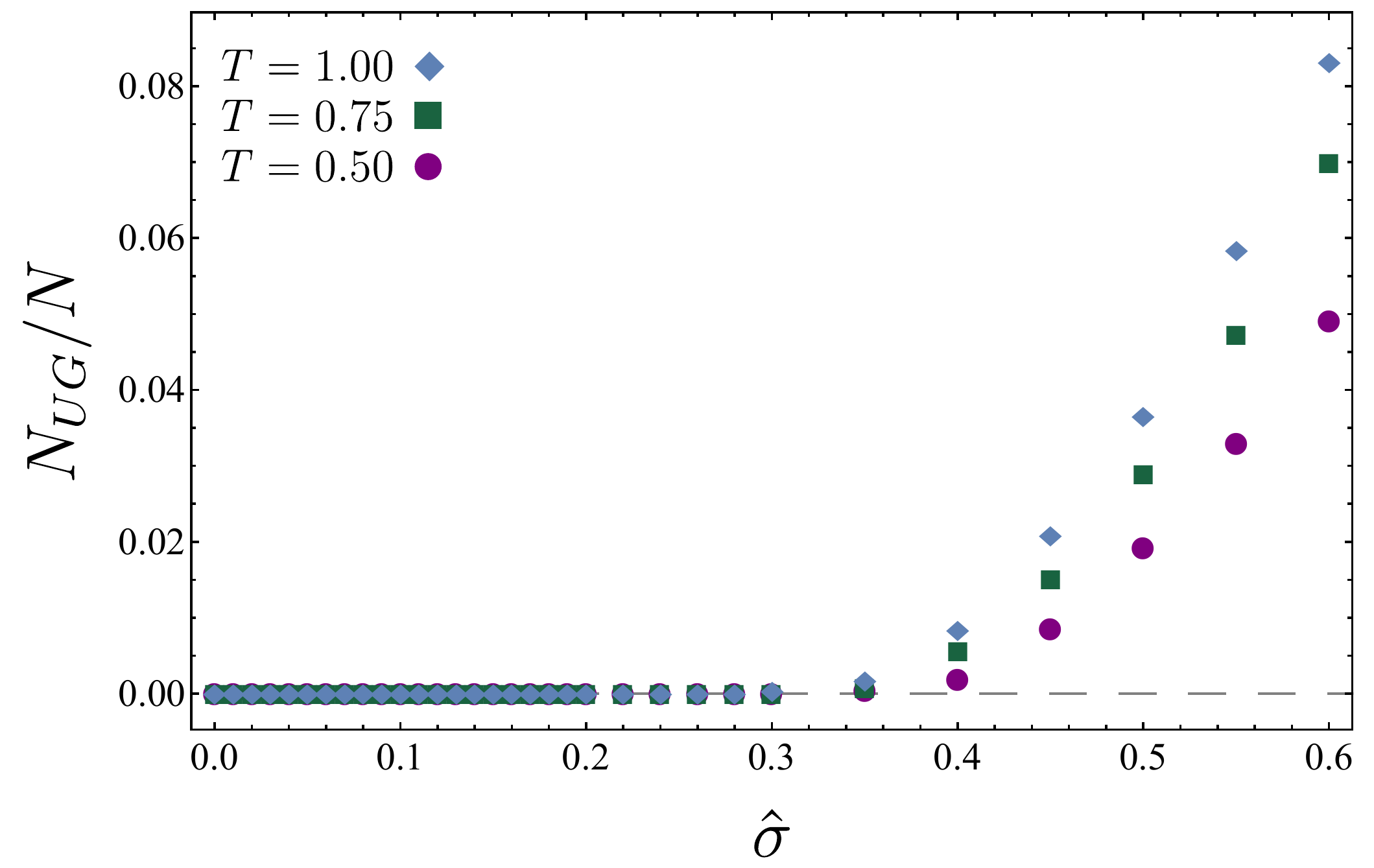}
    \caption{\textbf{Fraction of species that grow indefinitely.} Data are displayed for three different temperatures, for $\hat{\mu}=0.1$ and $N=128$. Around $\hat{\sigma}\sim0.35$ the fraction becomes finite (it deviates from the dashed grey line that represents $N_{UG}=0$) and the unbounded growth phase appears.}
    \label{fig:countUG-vs-sigma}
\end{figure}

At this point, it is necessary to clarify what is meant by \textit{unbounded growth phase} in the context of sparse networks. While in fully-connected networks the identification of the unbounded growth phase is straightforward, since all species behave identically at equilibrium, the same is not true for sparse networks: here species which undergo unbounded growth can coexist with species which do not. Therefore we define the unbounded growth phase of the gLV model on a sparse network as the phase where at least one species shows unbounded growth (details on the numerical identification of the unbounded growth are reported in the appendix \ref{app:UG}).
According to our data, for $\hat{\mu}=0.1$ and $T=0.5, 0.75, 1.0$, when $\hat{\sigma}$ is increased the system passes directly from single equilibrium to the unbounded growth phase. In Fig.~\ref{fig:countUG-vs-sigma} the behaviour of the fraction $N_{UG}/N$ as a function of $\hat{\sigma}$ is plotted, where $N_{UG}$ is the number of species with unbounded growth and $N$ is the total number of species. 

\begin{figure}[htbp]  
    \centering
    \includegraphics[width=0.8\columnwidth]{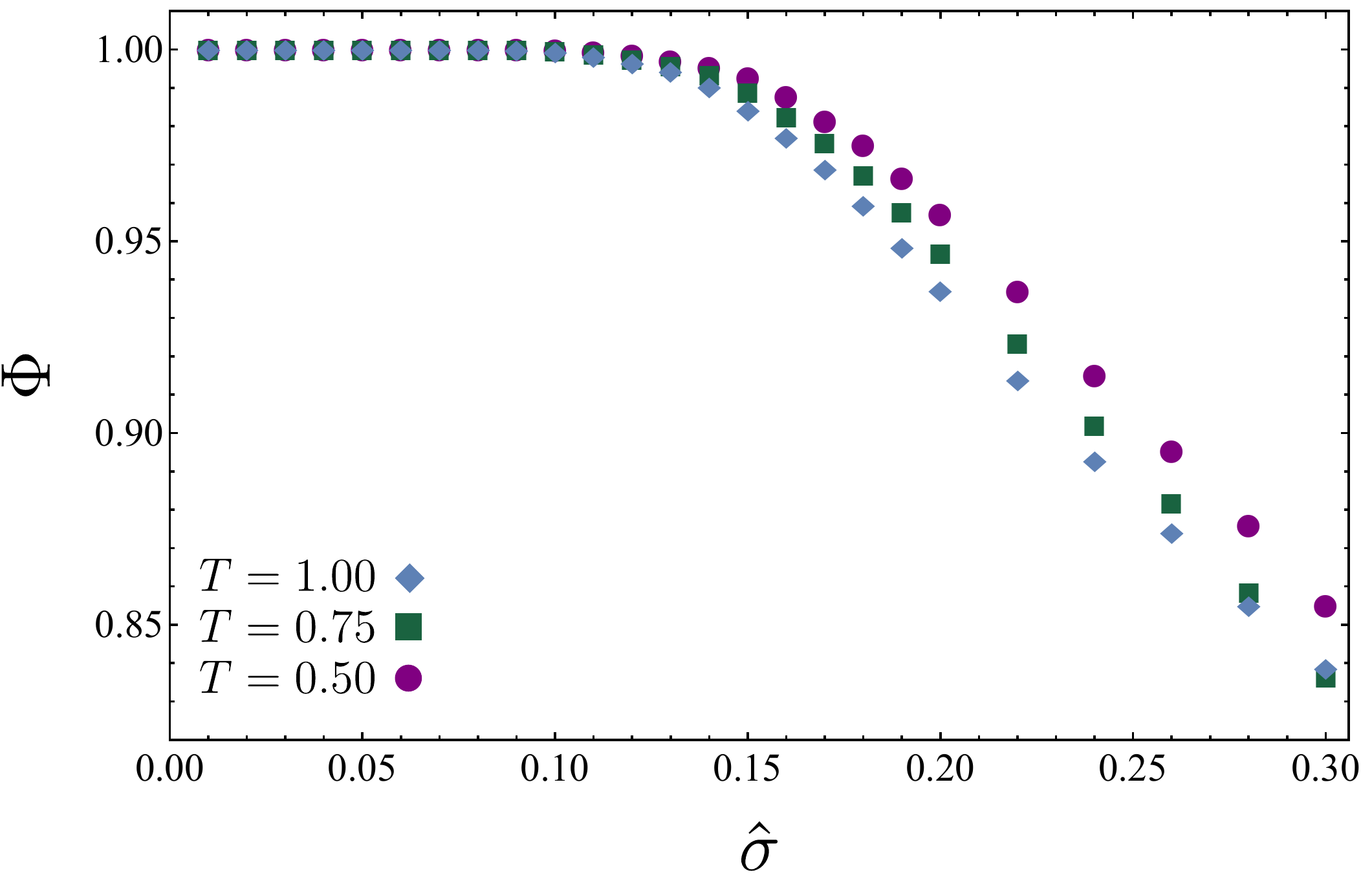}
    \includegraphics[width=0.8\columnwidth]{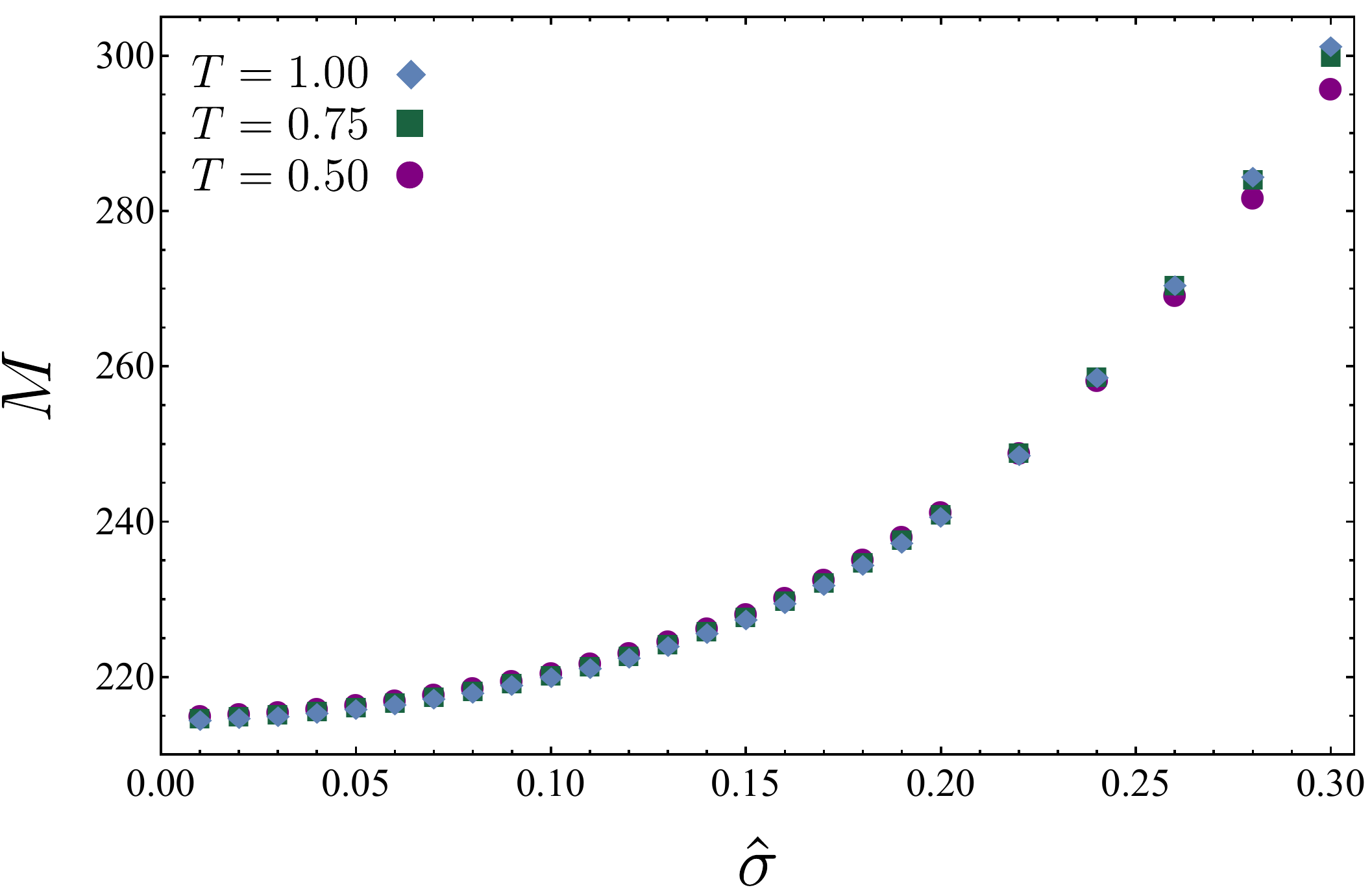}
    \caption{\textbf{Survival probability $\Phi$ (top) and the mean abundance $M$ (bottom) in the unique fixed point phase.} The observables are displayed for three different temperatures, varying $\hat{\sigma}$, up to the appearance of the unbounded growth phase. The parameters of the model are $\hat{\mu}=0.1$ and $N=128$.}
    \label{fig:phi_and_means-vs-sigma}
\end{figure}

In order to provide further evidence of the consistency between our results and others from the literature we have also computed, in the single equilibrium phase, the survival probability $\Phi$, i.e., the ratio between the number of surviving species and the total number of species, and the average population abundance $M=\sum_{i=1}^N M_i/N$, as a function of the disorder strength $\hat{\sigma}$. Data for $\Phi$ and  $M$ are shown respectively in the upper and the lower panel of Fig.~\ref{fig:phi_and_means-vs-sigma}. In both panels the behaviour of the corresponding observable is represented up to the value of $\hat{\sigma}$ at which unbounded growth is found (see Fig.~\ref{fig:countUG-vs-sigma}).

\section{Topological multiple equilibria phase at zero disorder}
\label{sec:topological-glass}

In this section we look at the zero-disorder phase diagram in the $(\hat{\mu},T)$ plane. Let us recall that in the fully connected case for $\sigma=0$ and $\mu$ positive the system is always in the single equilibrium phase. However the zero-temperature and zero-disorder thermodynamics of the generalized
Lotka-Volterra model is very different in sparse networks.
In~\cite{marcus2022local}, for a sparse interaction network, it is shown that at $T=0$
and $\sigma=0$ there is a transition from single to multiple
equilibria at the critical value:
\begin{equation}\label{eq:MU_critico}
    \hat{\mu}_c=\frac{1}{2 \sqrt{k-1}},
\end{equation}
where $k$ is the connectivity of the graph. This is a quite peculiar transition to a multiple-equilibria phase, since it is not due to disordered interactions (as we are considering the $\sigma=0$ case) but rather to high competition between species. For $\hat{\mu}>\hat{\mu}_c$, the increased competitive interactions cause the phase space to fragment into a large set of different equilibrium states, each characterized by the extinction of a different set of species and different pools of surviving ones. 
In ref. \cite{marcus2022local}, the transition point $\hat{\mu}_c$ is exactly located looking at the local stability of the
single equilibrium phase.

\subsection{Thermodynamics from Belief Propagation equations}

\begin{figure}[htbp] 
    \centering
    \includegraphics[width=\columnwidth]{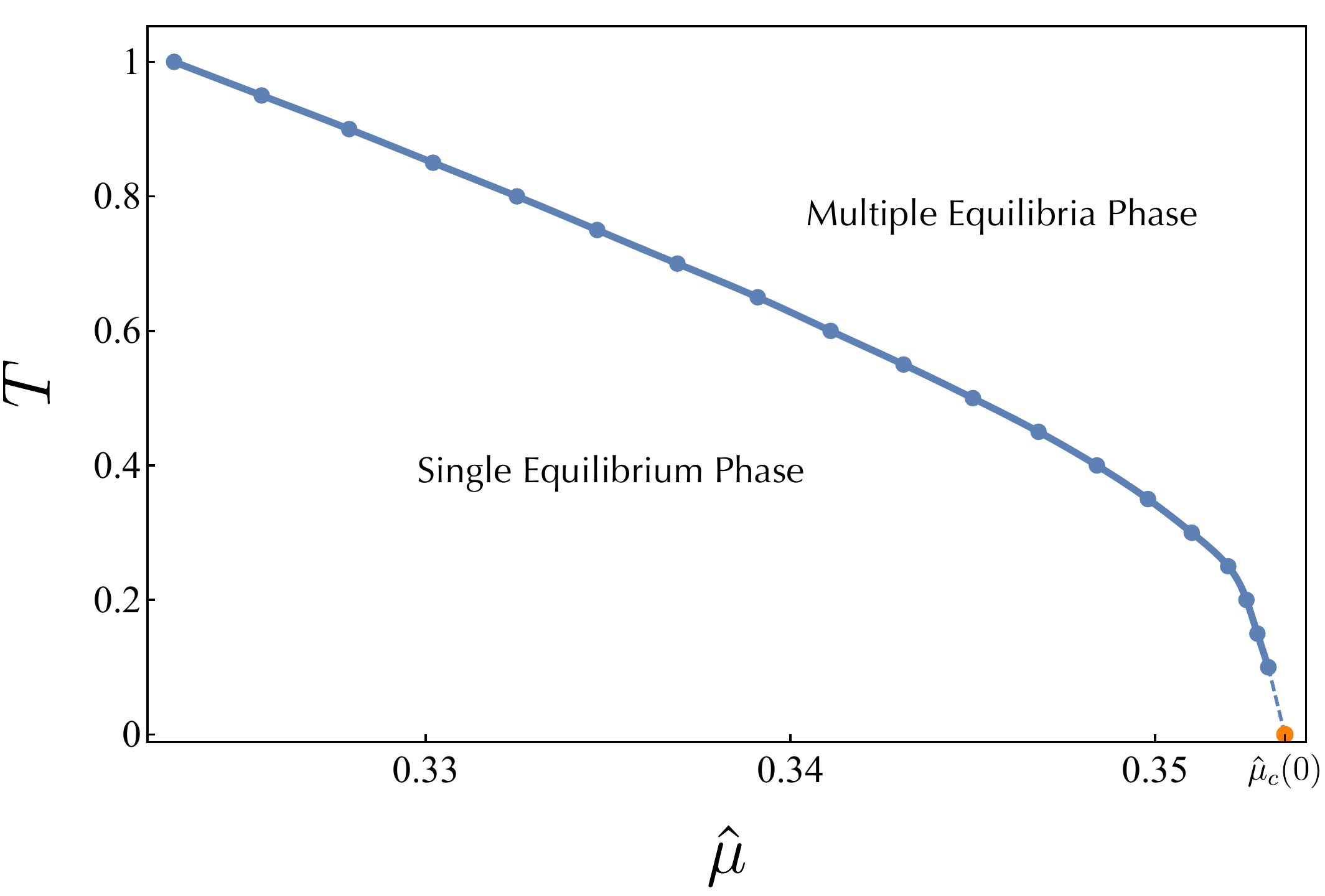}
    \caption{\textbf{Transition line from single to multiple equilibria in the $(\hat{\mu},T)$ space}. The transition line has been computed for zero disorder $\hat{\sigma}=0$ and $N=256$.}
    \label{fig:transMUcdashed}
\end{figure}

Studying the equilibrium thermodynamics by means of the single-equilibrium (or \emph{replica-symmetric}) Belief-Propagation equations for the cavity marginals, Eq.~\eqref{eq-cavitymarginal} (See Appendix~\ref{app:BP-algorithm}
for details on the algorithm), the criterion for detecting the emergence of a multiple-equilibria phase is provided by the lack of convergence of the above algorithm. This failure to converge means that the equations derived by
assuming the existence of a single equilibrium state are not appropriate to
describe the equilibrium thermodynamics of the system and more
complicated replica-symmetry-breaking ansatzs (\emph{multiple-equilibria adapted} assumptions on the equilibrium distribution) are needed to study the
cavity marginals~\cite{mezard2009information}. 
In Appendix \ref{app:Stability} we study the local stability of the unique fixed point found by BP, and we show that indeed when the BP algorithm stops to converge the unique fixed point becomes unstable.
On the basis of this criterion we have computed, for the zero disorder case, $\hat\sigma=0$, the
critical line in the plane $(\hat{\mu},T)$ which separates the single equilibrium phase from a multiple-equilibria one: data are shown in Fig.~\ref{fig:transMUcdashed} \footnote{It could be possible that a one-step replica symmetry breaking (1RSB) phase arises \emph{on top of} the replica symmetric one. In this case, even if the replica symmetric phase is still stable and BP equations converge, in the same region of parameters there could exist an additional stable 1RSB state. Even if we cannot exclude this situation, we see that, on the left of the critical line in Fig. \ref{fig:transMUcdashed} the Langevin dynamics converges to the unique solution found by BP, a strong evidence that no other states are thermodynamically relevant. The fact that the BP equations do not converge anymore, and the unique fixed point looses stability beyond the critical line are strong indications of the fact that the topologically multiple equilibria phase is of the full-replica-symmetry breaking type \cite{mezard2009information}, also called Gardner phase in ref. \cite{altieri2021properties}. A precise study of the multiple attractor phase will be performed in a subsequent paper.}. We highlight that a linear extrapolation of the transition line down to $T=0$ nicely matches the critical value $\hat{\mu}_c$ found in~\cite{marcus2022local}. The extrapolation is needed because eqs. \eqref{eq-cavitymarginal} are only well defined for  $T\neq0$. We stress that the multiple
equilibria phase found here upon increasing $\hat{\mu}$ at zero disorder is
quite different from the multiple equilibria phase found upon
increasing $\sigma$ at finite $T$ in the fully-connected case ~\cite{altieri2021properties}. In the sparse network the presence of many equilibria is determined by the following mechanism: when $\hat{\mu}>\hat{\mu}_c(T)$ the interactions between species become \textit{too competitive} and this causes the extinction of some species. For even higher values of $\hat{\mu}$ the graph breaks into disconnected groups of surviving species. Since all species are equivalent—having the same number of neighbours and identical interactions—the extinction of specific species is solely determined by the randomness associated to the initial values of the marginals and to thermal fluctuations. In particular, different instances of the thermal noise can lead to the extinction of different subsets of species in the sparse network, giving rise to a multiple attractor phase which we have termed {\it topological}, with no counterpart on dense networks. 
In Fig. \ref{fig:transMUcdashed} the critical line is plotted in the $(\hat{\mu}, T)$ plane: the critical value $\mu_c(T)$ decreases by increasing the temperature. This result can be easily understood as thermal fluctuations enhance the extinction of some species, shifting $\hat{\mu}_c$ further to the left as temperature increases.

\subsection{Thermodynamics from Langevin Dynamics}\label{subsec:langevin-dynamics}

In order to validate the equilibrium results obtained for the zero disordered case by means of the Belief Propagation algorithm, we simulated the Langevin dynamics at finite temperature for the same case without disorder. More precisely, we used a generalization of the Runge-Kutta method for stochastic dynamics to integrate the following set of Langevin equations:
\begin{align}
 \frac{d n_i(t)}{d t}=\frac{r}{K} n_i(t)\big[K-
      n_i(t)-\hat{\mu} \sum_{j\in\partial i} n_j(t) \big]+\xi_i(t),
      \label{eq:lang-numerics}
\end{align}
where the noise variance is  
\begin{align}
\langle \xi_i(t)\xi_j(t')\rangle = 2T\delta_{ij}n_i(t)\delta(t-t').
\end{align}
Following~\cite{altieri2021properties}, we have introduced a reflecting wall condition at $n_i=\lambda$ for each species.

\begin{widetext}

\begin{figure}[]
    \centering
    \includegraphics[width=\textwidth]{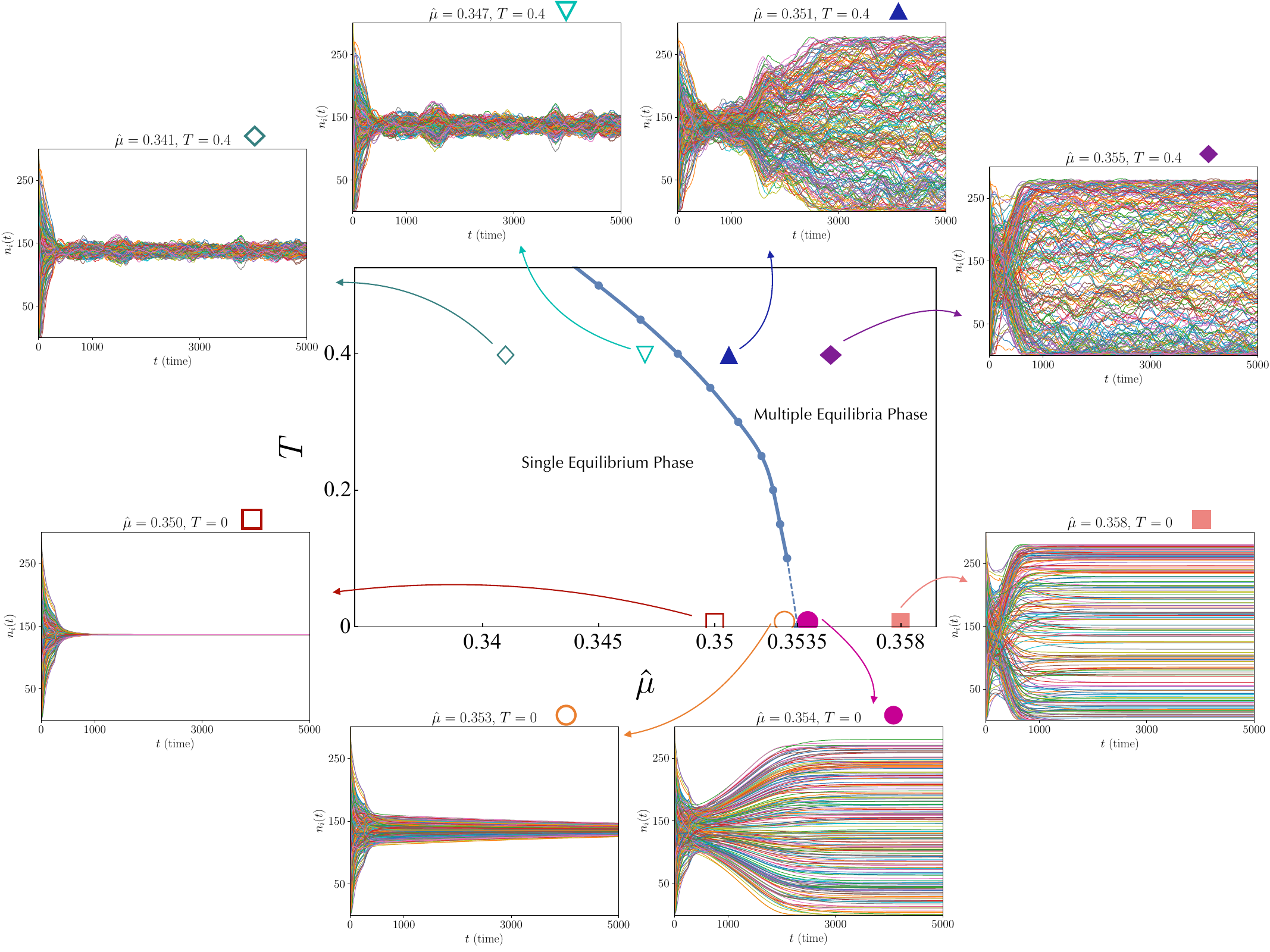}
    \caption{\textbf{Comparison between the dynamics and the critical line obtained by BP.} In the center, the same $(\hat{\mu},T)$ diagram of Fig. \ref{fig:transMUcdashed} is reported. We identify eight specific values of $(\hat{\mu},T)$ (identified by different markers), for which we show a single dynamics for the single-species abundance populations of a single ecosystem. When the multiple equilibria line is crossed, extinctions appear. For each species $i$, at each time $t$, $n_i(t)$ is the average over the previous $250$ time-steps. Here $N=256$.}
    \label{fig:dynamics_plots}
\end{figure}

\end{widetext}

The results of Langevin dynamics are in agreement with the results obtained from the Belief Propagation equations. In the phase which, according to BP, there is a single-equilibrium, the Langevin dynamics produces trajectories with small fluctuations and no extinctions. On the contrary, as soon as one tries to run the Langevin dynamics in the multiple-equilibria phase, an 
abrupt spreading of $n_i(t)$ trajectories is found, accompanied by the immediate appearing of extinctions. This peculiar behaviour is shown in the panels of Fig.~\ref{fig:dynamics_plots}: data are taken from the simulations of a random regular graph with $N=256$ species and the value of the parameters $\hat{\mu}$ and $T$ corresponding to each panel is indicated with a symbol (different in shape and color) in the ($\hat{\mu}$,$T$) phase diagram in the center.\\

Let us comment the simulations of dynamics in Eq.~\eqref{eq:lang-numerics} by starting from the $T=0$ case. By looking at the panels at the bottom of Fig.~\ref{fig:dynamics_plots} corresponding to $T=0$ and $\hat{\mu}=0.35,\,0.353$, it is clear that Langevin dynamics converges to the same abundance value for every species. On the contrary, the bottom-right panels of Fig.~\ref{fig:dynamics_plots}, corresponding to $T=0$ and $\hat{\mu}=0.354,\, 0.358$, clearly show the appearance of extinctions and the spreading of $n_i(t)$ trajectories, consistently with the nature of the topological multiple-equilibria phase discussed above. 
The same transition from single to multiple equilibria can be found by increasing $\hat\mu$ at fixed finite temperature, as can be seen by looking at the differences between the top-left panels ($\hat{\mu}=0.341, \, 0.347$, $T=0.4$) and the top-right ones ($\hat{\mu}=0.351, \, 0.357$, $T=0.4$) of Fig.~\ref{fig:dynamics_plots}. 
Please note that in all the right panels, with parameters inside the multiple-equilibria phase, both at zero and at finite temperature, the dynamics firstly seem to converge for small times towards values of the abundances similar to the ones in the unique fixed point, before the appearance of extinctions and the spreading of $n_i(t)$ trajectories at large times, clear indications of the multiple-attractor phase. This first approximate convergence at small times is due to a reminiscence of the unique fixed point that looses progressively stability as $\hat\mu$ grows, as shown in detail in Appendix \ref{app:Stability}. For this reason, the time spent in the vicinity of the unique fixed point decreases progressively moving to the right of the critical line.

All panels in Fig.~\ref{fig:dynamics_plots} represent trajectories of abundances $n_i(t)$ obtained at different values of $\hat{\mu}$ and temperature $T$ by exploiting the same initial conditions and stochastic noise realization. Clearly, different initial conditions and different realization of the noise would have led to the extinction of different species, as we show in Appendix \ref{app:multiple_attractor}. 
While the results of Langevin dynamics are perfectly compatible with the results obtained from the Belief Propagation equations, we want to stress that from the Langevin approach the transition line between single and multiple equilibria cannot be defined as sharply as from the lack of convergence of BP, which turns out to be a much more powerful tool to study this phase diagram.

We also want to point out that it is not always true that as long as extinctions appear, the single equilibria breaks into multiple equilibria states. In fact, we showed that at low $\hat{\mu}$ and high enough $\hat{\sigma}$, there are extinctions but the single equilibrium phase is the only stable phase, see Fig. \ref{fig:phi_and_means-vs-sigma}.
\subsection{A re-entrant transition}
\label{sec:3d-phase-diagram}

\begin{figure}[htbp] 
    \centering
    \includegraphics[width=\columnwidth]{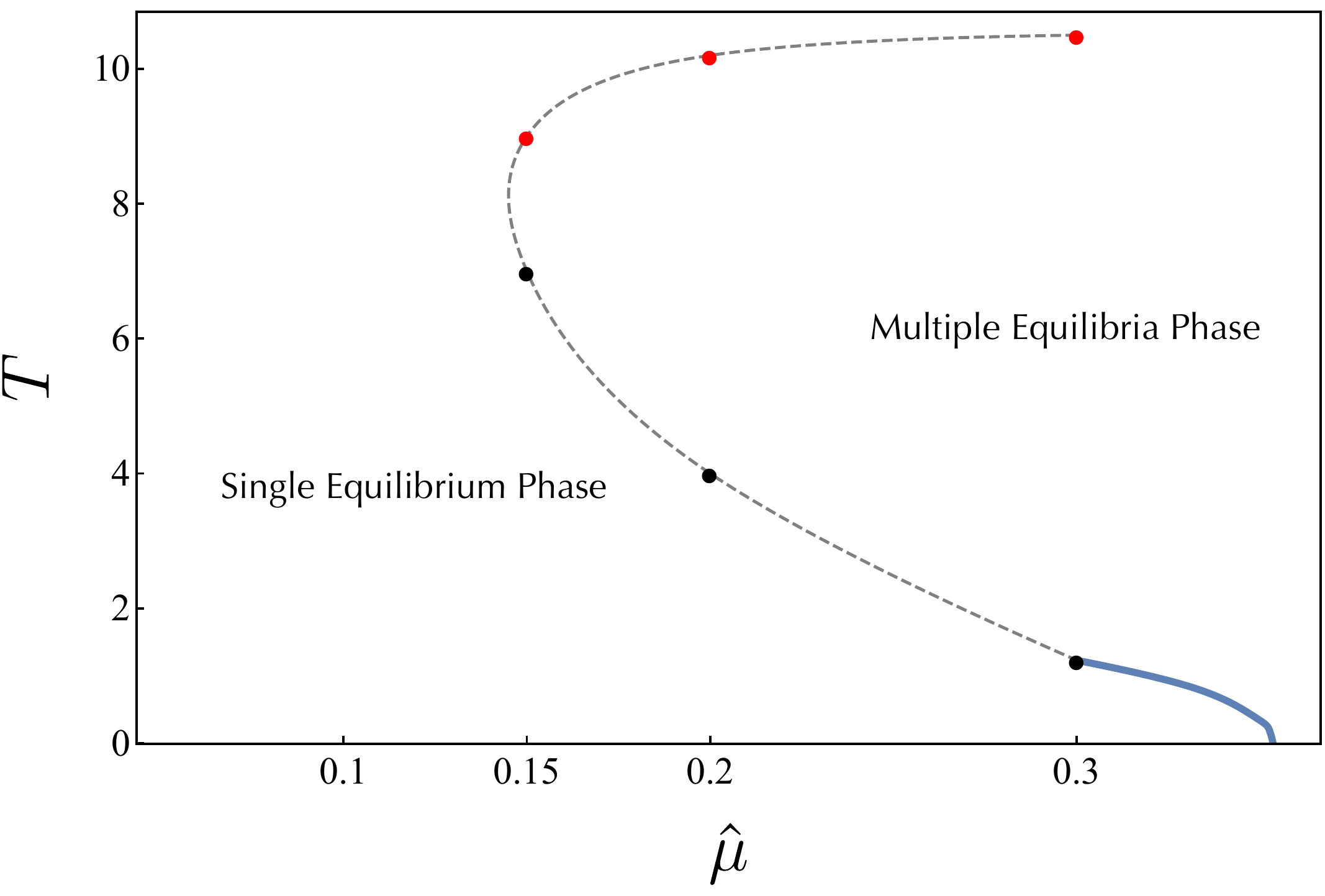}
    \caption{\textbf{Full transition line, at $\hat{\sigma}=0$, between the single equilibrium and the multiple equilibria phase.} The blue line reproduces the transition line from Fig.~\ref{fig:transMUcdashed}. The black dots are the transition points $T_c^{\text{down}}$ found by increasing the temperature from below, while the red dots correspond the transition points $T_c^{\text{up}}$ found by decreasing the temperature from above. Based on these points, we suggest that the transition line eventually bends to the right, forming a nose-like shape, which we represent as the grey dashed line in the figure.}
    \label{fig:reentrant-trans}
\end{figure}

We have shown that in the case of a sparse interaction network it is possible to find a \emph{topological} multiple-equilibria phase, characterized by the presence of extinctions. This phase, in contrast with the multiple-equilibria one typically found in fully-connected models, is due to highly competitive interactions rather than disordered ones. 
To fully characterize the transition line in the plane $(\hat{\mu},T)$ at $\sigma=0$, we performed a systematic study of the system as the temperature is varied for four fixed values of the (ordered) interaction strength $\hat{\mu}=0.1,0.15, 0.2, 0.3$. At $\hat{\mu}=0.1$ we are able to heat up the system without ever losing convergence of the algorithm, a signal that no multiple-equilibria phase is encountered (we already showed that no multiple equilibria is found for $\hat{\mu}=0.1$ and $\hat\sigma\neq 0$ in Sec. \ref{sec:NOglass}). Instead, at $\hat{\mu}=0.15,0.2,0.3$, we find a critical temperature that divides a single-equilibria phase from a multiple equilibria one increasing $T$ from low temperature, $T_c^{\text{down}}(\hat{\mu})$ (black dots in Fig.~\ref{fig:reentrant-trans}), and another critical temperature that divides a single-equilibria phase from a multiple equilibria one decreasing $T$ from high temperature, $T_c^{\text{up}}(\hat{\mu})$ (red dots in Fig.~\ref{fig:reentrant-trans}). These findings suggest the following scenario: there is a critical line enclosing the topological multiple-equilibria phase which starts at $\hat{\mu}_c$ at $T=0$, moves at smaller $\hat{\mu}$ for increasing $T$ and then bends to the right at a critical value of the interactions strength $\mu^*$ such that $0.1 < \mu^* < 0.15$ and goes to some horizontal asymptote at a temperature smaller than $T=15$, for which we checked that the system is in the replica symmetric phase for any $\hat{\mu}$. 
The re-entrant transition in the $(\hat{\mu},T)$ plane is pictured as a gray dashed line in Fig.~\ref{fig:reentrant-trans}. A more refined study of the phase diagram will be discussed elsewhere. \\ 

From a general perspective, the replica-symmetry-breaking critical line sketched in Fig.~\ref{fig:transMUcdashed} has a shape remarkably different from the standard replica-symmetry-breaking transition line found in fully-connected graphs \cite{altieri2021properties}. In the latter case, the transition from single to multiple equilibria is always triggered by the decrease of thermal fluctuations. In contrast, the sparse gLV model exhibits the opposite behavior. The reentrant transition of Fig.~\ref{fig:reentrant-trans} is very similar to the inverse freezing transition, present in the Random Blume-Capel model \cite{Crisanti2005Inverse, leuzzi2011random}. It is known that, in order to have an inverse-freezing transition, it is essential the presence of active and inactive nodes, that are modeled as bosonic spins that can take values $(+1, -1, 0)$ in the Blume-Capel model, while in the sparse gLV the active and inactive nodes correspond to surviving and extincted species. \\

\section{Conclusions}\label{sec:conclusions}

In this work we have studied the equilibrium thermodynamics of the generalized Lotka-Volterra (gLV) model on a sparse random graph with quenched disordered interactions, with particular attention to the behaviour of species abundance distributions. By exploiting the cavity method and the Belief Propagation equations, we have studied the properties of the gLV model on a random regular graph with small and fixed connectivity $k$, varying three parameters: the average interaction strength $\hat{\mu}$, the width $\hat{\sigma}$ of the randomly distributed disordered couplings and the extent of thermal fluctuations, parametrized by the temperature $T$. Our first important observation has been that, by increasing $\hat{\sigma}$ at finite and small $\hat{\mu}$ and finite temperature, the abundance marginal distributions become strongly non-Gaussian while staying in a single-equilibrium phase. In models with the same sort of random couplings but living on a dense network, there is by construction no room for non-Gaussian effects. We have thus argued that this result is purely due to the sparsity of the network, therefore providing a new perspective on the emergence of non-Gaussian and Gamma-like distributions in ecological systems. In particular, the finding of Gamma-like marginals is strongly consistent with the empirical observations from real ecosystems~\cite{grilli2020macroecological,McGill2007SAD}. We have observed non-Gaussian effects also in two-species joint probabilities, due to the emergence of non trivial correlations. In addition to the observation of these non-Gaussian effects in the single equilibrium phase, we have ascertained that by further increasing $\hat{\sigma}$ for the same choice of $\hat{\mu}$ and $T$ the multiple attractor phase encountered in fully-connected networks is never met: on the Bethe lattice with fixed and small connectivity, at small values of $\hat{\mu}$, upon increasing $\hat{\sigma}$ the system has a transition from the single-equilibrium phase directly to the unbounded growth one.\\ 

In the second part of our study we concentrated on the phenomenology of the gLV model on a sparse topology and with zero disorder, i.e., $\hat{\sigma}=0$. By varying both the temperature $T$ and the average strength $\hat{\mu}$ of the ordered interactions we have been thus able to extend the zero temperature results of~\cite{marcus2022local}, showing that the formation of the topological multiple equilibria phase pointed out in~\cite{marcus2022local} at $T=0$ for high enough interaction strength, $\hat{\mu}>\hat{\mu}_c$, is robust against the introduction of temperature.
This phase emerges due to the occurrence of extinctions, resulting from high competition within the system (large $\hat{\mu}$). The higher is $\hat{\mu}$, the larger is the amount of extinctions, which eventually determines the formation of disconnected groups of surviving species and the \emph{freezing} of the system into a configuration where a certain number of species have gone extinct, forming non-communicating islands of surviving ones. The pattern of surviving/disappearing species strongly depends on the initial conditions and the sequence of thermal fluctuations, so that there is a huge multiplicity of possible target stationary states: that is why it makes sense to talk about a multiple attractor phase, but of very different nature with respect to the one induced by the disordered couplings in fully connected networks~\cite{altieri2021properties}. 

 In conclusion, the main outcomes of our investigation have been the highlight of realistic non-Gaussian effects in the single-equilibrium phase and the uncovering of a non-trivial multiple attractor phase at finite temperature and in the absence of disorder, therefore questioning two of the main ingredients of the models so-far used to mimic the behaviour of large ecosystems: dense networks and large fluctuations in the disordered couplings. From a model-building perspective, we provided the evidence that the study of sparse networks offers a new avenue for understanding the behaviour of large ecosystems within a more realistic framework.

\section*{Acknowledgments}

We thank David Machado Perez and Pietro Valigi for useful discussions.
We acknowledge support from the project MIUR-PRIN2022, {\it ``Emergent Dynamical Patterns of Disordered Systems with Applications to Natural Communities''}, code 2022WPHMXK, funded by European Union — Next Generation EU, Mission 4 Component 1 CUP: C53D23001560006, D53D23002900006 and B53D23005310001.

\newpage

\appendix

\begin{widetext}
  
\section{Cavity Equations and Belief Propagation Algorithm}
\label{app:BP-algorithm}

Here we briefly show how the cavity equations on a random regular graph (graph in which all the nodes have the same fixed connectivity) are derived. We follow the approach of \cite{zamponi-cavity}.
Let's start from a general model with two-body interactions and variables $n_i$ associated to each node $i$, with $i\in\{1,...,N\}$. If the Hamiltonian is
\begin{equation}\label{eq-app:general-Ham}
  H({\bf n})=\sum_{i=1}^N h_i({n}_i)+\sum_{(ij)\in E} h_{ij}({n}_i,{n}_j),
 \end{equation}
where $h_i(n_i)$ is a local field term and $h_{ij}(n_i,n_j)$ is an interaction term between node $i$ and $j$, we can write the partition function of the system as:
\begin{equation}
    Z=\sum_{\{{n}_i\}}\prod_i \psi_i(n_i) \prod_{\langle ij\rangle} \psi_{ij}(n_i,n_j),
\end{equation}
where
\begin{equation}\label{}
    \psi_i(n_i)=\exp\left(-\beta h_i({n}_i)\right)\hspace{1.8cm} \psi_{ij}(n_i,n_j)= \exp\left(-\beta h_{ij}({n}_i,{n}_j)\right).
\end{equation}

Defining the message $\eta_{j\rightarrow i}(n_j)$ as the normalized \emph{cavity marginal} of the variable $n_j$ on a modified graph where the edge between nodes $j$ and $i$ has been cut, one can demonstrate that the following self-consistent equations hold \cite{mezard2009information}:
\begin{align}\label{app-eq:Etaij}
\eta_{i\rightarrow j}(n_i)&=\frac{\psi_i(n_i)}{z_{i\rightarrow j}}\prod_{k\in \partial i\backslash j}\left(\sum_{\{n_k\}}\eta_{k\rightarrow i}(n_k) \psi_{ik}(n_i,n_k) \right),\nonumber\\
\eta_{i}(n_i)&=\frac{\psi_i(n_i)}{z_i}\prod_{j\in \partial i}\left(\sum_{\{n_j\}}\eta_{j\rightarrow i}(n_j) \psi_{ij}(n_i,n_j) \right),
\end{align}
where $z_{i\rightarrow j}$ and $z_i$ are normalization factors that ensure $\sum_{n_i}\eta_{i\rightarrow j}(n_i)=\sum_{n_i}\eta_{i}(n_i)=1$. One can easily show that Eqs.~\eqref{app-eq:Etaij} are exact on a tree exploiting its structure. Indeed, on a tree, for each node $i$, the messages $\eta_{k\rightarrow i}(n_k)$ coming from the neighbours $k$ of $i$ are all independent from each other once the edge $(i,k)$ has been cut, and so their contributions can be simply multiplied.
This is not necessarily true in the case of RRG's. In particular in such graphs loops are present and so the messages $\eta_{k\rightarrow i}(n_k)$ of the neighbours $k$ of node $i$ could be in principle not independent from each other when we cut the edges $(i,k)$. However, in RRG's the typical length of loops is $O(\log N)$ and thus RRG are locally tree like (Fig.\ref{fig:bethe-lattice}) and in the thermodynamic limit Eqs~\eqref{app-eq:Etaij} are exact also for RRG's, as long as a the whole Gibbs measure is composed of just one state \cite{mezard2009information}. In particular for $N$ large enough we can use Eqs.~\eqref{app-eq:Etaij} also for our gLV-system on a sparse random regular graph.
\begin{figure}  
    \centering
    \includegraphics[width=6cm]{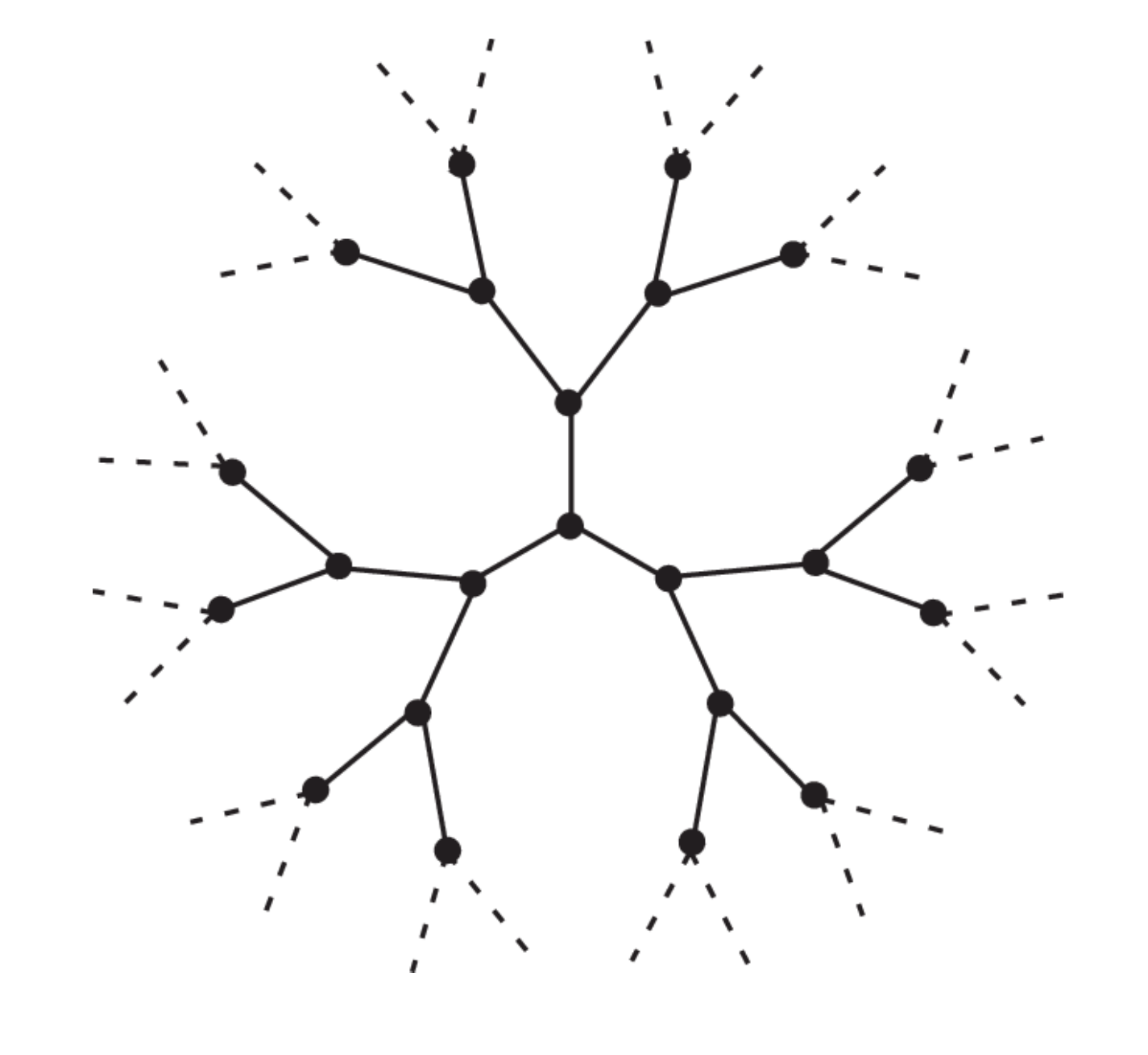}
    \caption{Local tree-like structure of a Random Regular Graph with 2-body interactions and connectivity equal to 3.}
    \label{fig:bethe-lattice}
\end{figure}
As we explained in Sec.~\ref{sec:bethe-lattice}, the Hamiltonian that describes the equilibria of our system has exactly the form in eq. \eqref{eq-app:general-Ham} where
\begin{align}
  & h_i(n_i) = - r \left(n_i-\frac{n_i^2}{2 K}\right)+T \ln\, (n_i+\epsilon), \nonumber \\
  & h_{ij}(n_i, n_j)= \frac{r}{K}\alpha_{ij}n_i n_j.
\end{align}
Following Eqs.~\eqref{app-eq:Etaij}, the cavity marginals are then:
\begin{equation}\label{app:eq-cavitymarginal}
\eta_{i\rightarrow j}(n_i)=\frac{1}{z_{i\rightarrow j}}\exp{\left(\beta r \left(n_i-\frac{n_i^2}{2 K}\right)- \ln\, (n_i+\epsilon)\right)}\prod_{k\in \partial i\backslash j}\left(\sum_{\{n_k\}}\eta_{k\rightarrow i}(n_k) \exp{-\beta \alpha_{ik} \frac{r}{K} n_i n_k} \right),
\end{equation}
and the marginal distributions can be obtained by the cavity marginals as:
\begin{equation}\label{eq-app:marginal}
    \eta_i(n_i)=\frac{1}{z_i}\exp{\left(\beta r \left(n_i-\frac{n_i^2}{2 K}\right)- \ln\, (n_i+\epsilon)\right)}\prod_{j\in \partial i}\left(\sum_{\{n_j\}}\eta_{j\rightarrow i}(n_j) \exp{-\beta \alpha_{ij} \frac{r}{K} n_i n_j} \right).
\end{equation}
The marginal distribution $\eta_i(n_i)$ thus represents the probability for a species $i$ to have a number $n_i$ of individuals, once we have marginalized over all the other species.
Once we have the set of equations for the cavity marginals given by \eqref{app:eq-cavitymarginal}, we can solve them iteratively. In principle, the cavity marginals could be thought as continuous probability distributions. However, to practically solve the self-consistent equations, it is much more useful to discretize the values that can be taken by the variable $n_i$. In particular, we will assume that $n_i \in [0,1,...,n_\text{max}-1]$, and we will take $n_\text{max}$ large enough so that  $\forall i \quad \eta_i(n_\text{max}-1)=0$ inside the numerical precision of a computer. 
For each oriented edge $i\rightarrow j$, the cavity marginal $\eta_{i\rightarrow j}(n_i)$ has to be initialized to a normalized distribution. 
For example we can take, for every oriented couple $(i,j)$, $\eta_{i\rightarrow j}^{(0)}(n_i)=\frac{1}{n_\text{max}}$, $\forall \, n_i \in [0,1,...,n_\text{max}-1]$, i.e. an uniform initial condition.
But in general, we checked that every result of the paper is independent on the chosen initial conditions. 
Starting from $\eta_{i\rightarrow j}^{(0)}$, we can compute at the following time step the normalization factor $z_{i\rightarrow j}^{(1)}$ and the cavity marginals $\eta_{i\rightarrow j}^{(1)}$. In general, from step $t$ to step $t+1$ the iteration process goes like this:
\begin{align}
z_{i\rightarrow j}^{(t+1)}&=\sum_{\{n_i\}} \exp{\left(\beta r \left(n_i-\frac{n_i^2}{2 K}\right)- \ln\, (n_i+\epsilon)\right)}\prod_{k\in \partial i\backslash j}\left(\sum_{\{n_k\}}\eta_{k\rightarrow i}^{(t)}(n_k) \exp{-\beta \alpha_{ik}\frac{r}{K} n_i n_k} \right),\\
    \eta_{i\rightarrow j}^{(t+1)}(n_i)&=\frac{1}{z_{i\rightarrow j}^{(t+1)}}\exp{\left(\beta r \left(n_i-\frac{n_i^2}{2 K}\right)- \ln\, (n_i+\epsilon)\right)}\prod_{k\in \partial i\backslash j}\left(\sum_{\{n_k\}}\eta_{k\rightarrow i}^{(t)}(n_k) \exp{-\beta \alpha_{ik} \frac{r}{K} n_i n_k} \right).
\end{align}
At each step $t$, from the values of the cavity marginals $\eta_{i\rightarrow
  j}^{(t)}(n_i)$, we can compute the marginal distributions $\eta_i^{(t)}(n_i)$ as shown in equation \eqref{eq-app:marginal}. In particular we checked the algorithmic convergence directly on the final marginal distributions: at each iteration step $t$ we compute mean, variance and kurtosis of each marginal distribution $\eta_i^{(t)}(n_i)$, $\forall i$. We then compute a relative difference with respect to the previous iteration step. In order to have convergence we require that, for each marginal distribution $\eta_i^{(t)}(n_i)$, each of the three moments (mean, variance and kurtosis) does not differ more than the $10^{-6}\,\%$ from the previous iteration step ones, and we require this condition to be true for ten successive iteration steps. When these conditions are satisfied we say that we have reached convergence. Thus, for a given graph, we obtain all the single marginals $\eta_i(n_i)$ from which we can then compute the sample and disorder averages as described in Eqs.~\eqref{eq:sample-average},\eqref{eq:disorder-average}. In Fig~\ref{fig:single-marginals} we show a set of $20$ single marginals $\eta_i(n_i)$ for $\hat{\sigma}=0.02$ and $\hat{\sigma}=0.20$, corresponding to the cases that we discussed in Sec~\ref{sec:non-gaussianity}. In the Gaussian regime ($\hat{\sigma}=0.02$), the single marginals are themselves all Gaussian and are concentrated within a small range of population abundances. In the non-Gaussian regime ($\hat{\sigma}=0.20$) instead, the single marginals spread over a much larger range, extinctions appear, and single marginals themselves are not all Gaussian anymore.

\begin{figure}[htbp] 
    \centering
    \begin{minipage}{0.5\textwidth}
        \centering
        \includegraphics[width=\textwidth]{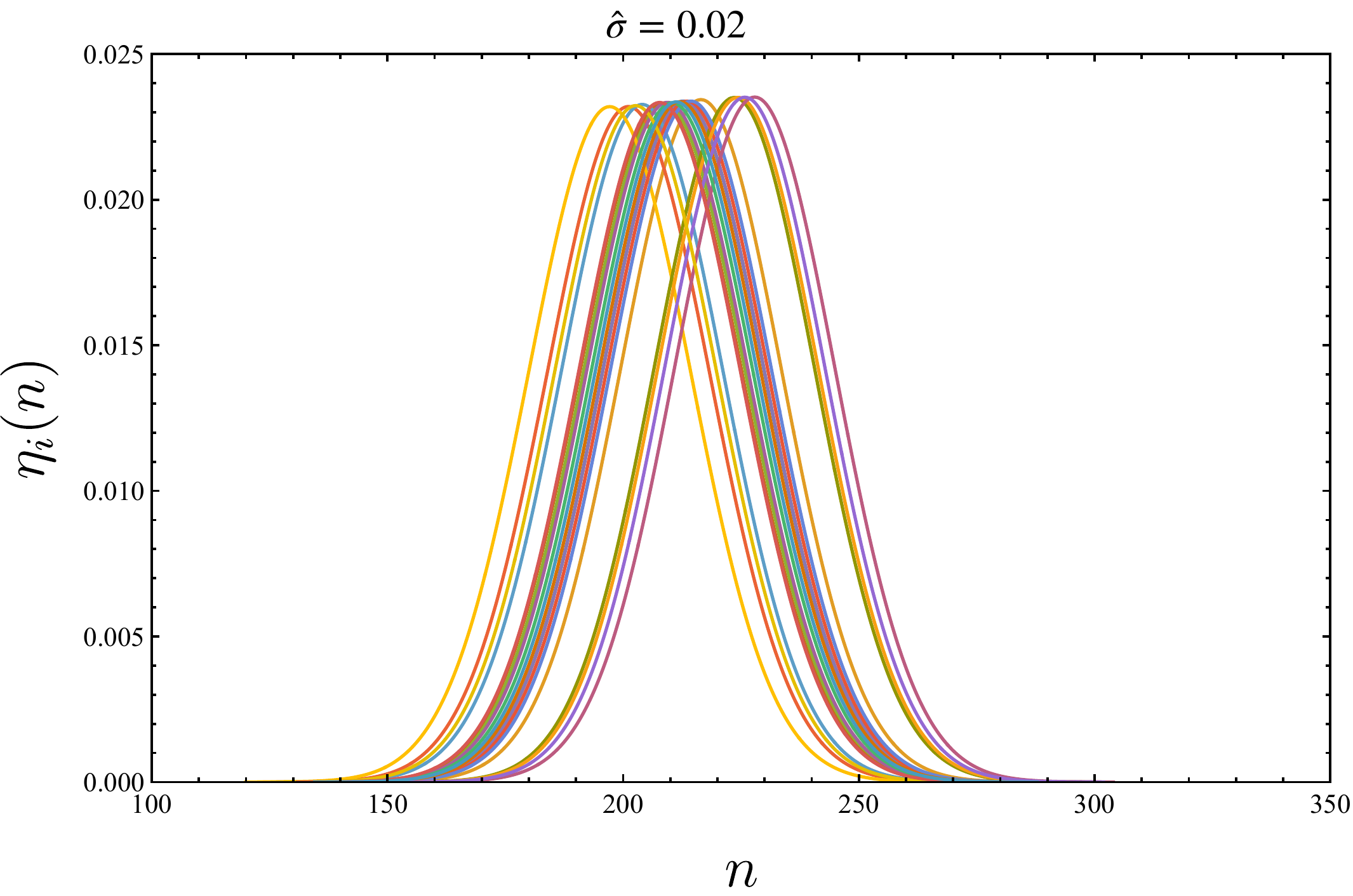}
    \end{minipage}
    \begin{minipage}{0.48\textwidth}
        \centering
        \includegraphics[width=\textwidth]{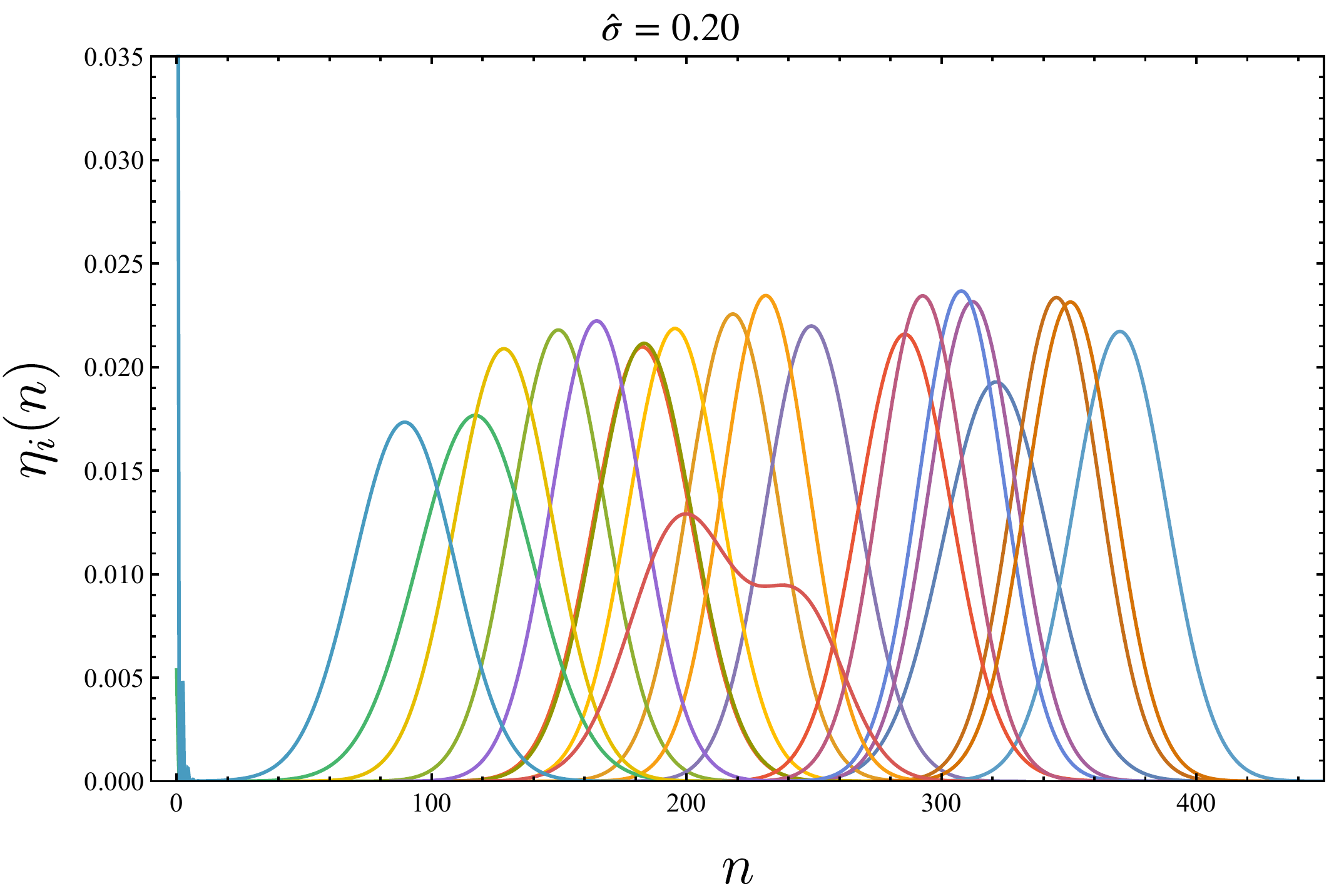}
    \end{minipage}
    \caption{\textbf{Single species marginal distributions $\eta_i(n)$ in the Gaussian regime (left) and in the non-Gaussian regime (right).} In the Gaussian regime ($\hat{\sigma}=0.02$) the single marginals are Gaussian and concentrated within a small range of $n$. In the non-Gaussian regime ($\hat{\sigma}=0.20$) the distributions spread over a larger range of $n$, extinctions appear, and not all single marginals are Gaussian anymore. The other parameters are $\hat{\mu}=0.1$, $T=1$ and $N=256$. }
    \label{fig:single-marginals}
\end{figure}

\section{Independence of the marginals from the regularization parameter $\epsilon$ and the discretization step} \label{app:changing_dn_eps}

As explained in Appendix \ref{app:BP-algorithm}, in our approach, in order to numerically solve the BP equations we chose to discretize the abundance distributions into nonnegative integers. We made this choice because in this way one can interpret  $n_i$ as the number of individuals for the species $i$ (that is, by definition, an integer). However, this is not an inevitable choice: indeed one can choose to discretize $n_i$ into small steps. We do not expect that the integer discretization influences the results because indeed the discretization step, that in our case is $dn=1$ should be compared with the natural scale for $n$ that is $K$ (fixed point of the single species dynamics) that we chose to be large exactly for this reason ($K=280$). To verify this, in this Appendix we show that the results do not change changing $dn$. In particular, in Fig. \ref{Fig:dn05} we show the average marginal distribution with the same two set of parameters used in Fig. \ref{fig:exp-tail} of the main text, but with a discretization step $dn=0.5$, that gives non integer numbers for the abundances. The distributions are completely equivalent to the ones with $dn=1$, for the exception of a very small variation at small values of $n$, that leads to a marginal distribution with $dn=0.5$ that is more comparable to the results of the dynamics than the one with $dn=1$ at small $n$.

\begin{figure}[htbp]
\includegraphics[width=0.7\columnwidth]{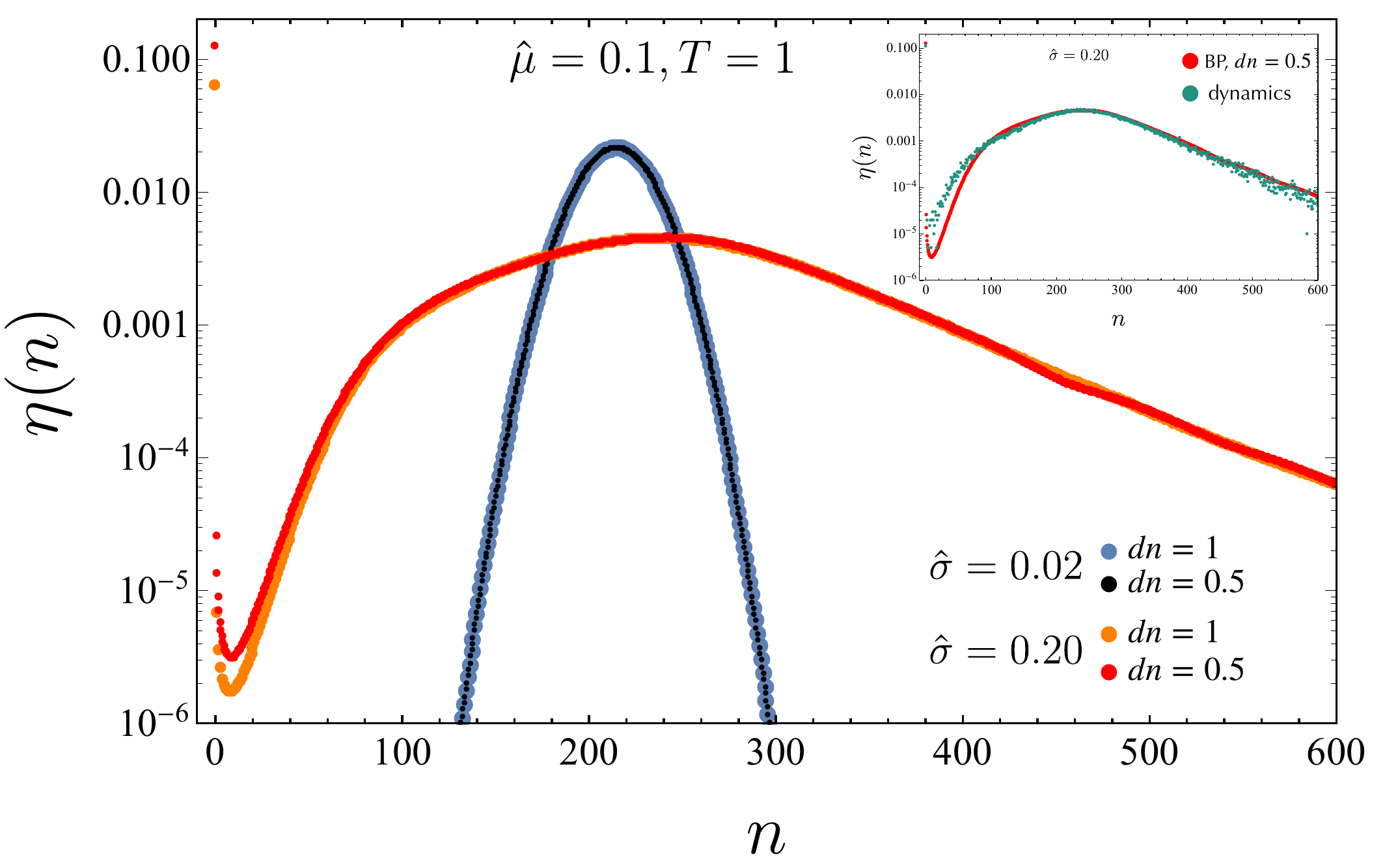}
\caption{\textbf{Comparison between species abundances computed with discretization steps $dn=1$ and $dn=0.5$.}
    For small $\hat{\sigma} = 0.02$ (blue and black points) the distributions with $dn=1$ and $dn=0.5$ are perfectly equivalent, for large $\hat{\sigma} = 0.20$ (orange and red points) there are small differences between the two distributions at small values of $n$. \textbf{Inset:} marginal at $\hat{\sigma} = 0.20$ computed using BP with $dn=0.5$ (red points) and from the dynamics (green points). The two agree very well and the BP marginal computed with $dn=0.5$ is more similar to the dynamics than the one computed with $dn=1$. In both plots the parameters are the same as in Fig. 1 of the main text: $T=1$, $\hat{\mu}=0.1$, $N=256$.}
\label{Fig:dn05}
\end{figure}

\begin{figure}[htbp]
\includegraphics[width=0.46\columnwidth]{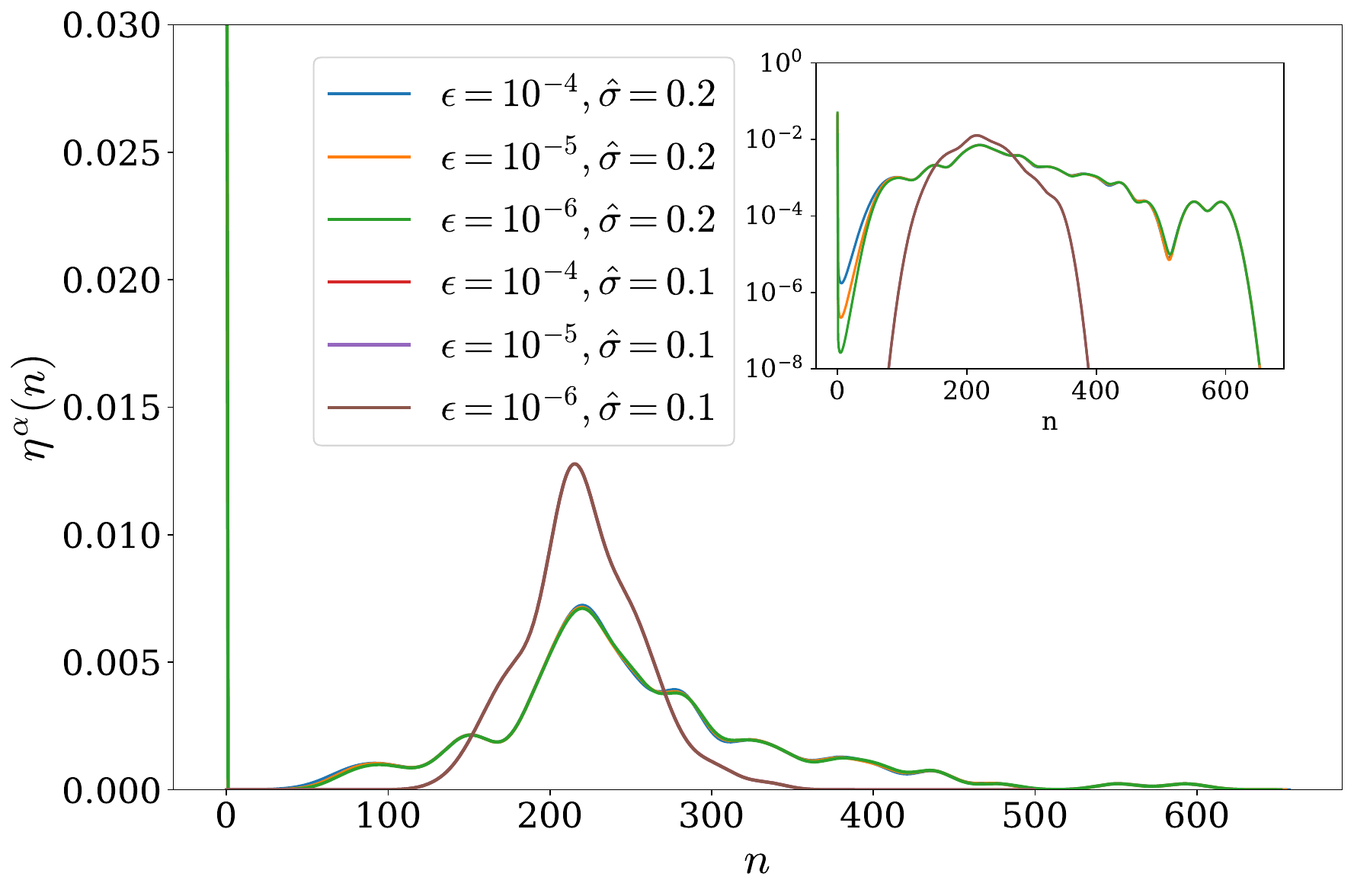}
\hspace{0.8cm}
\includegraphics[width=0.475\columnwidth]{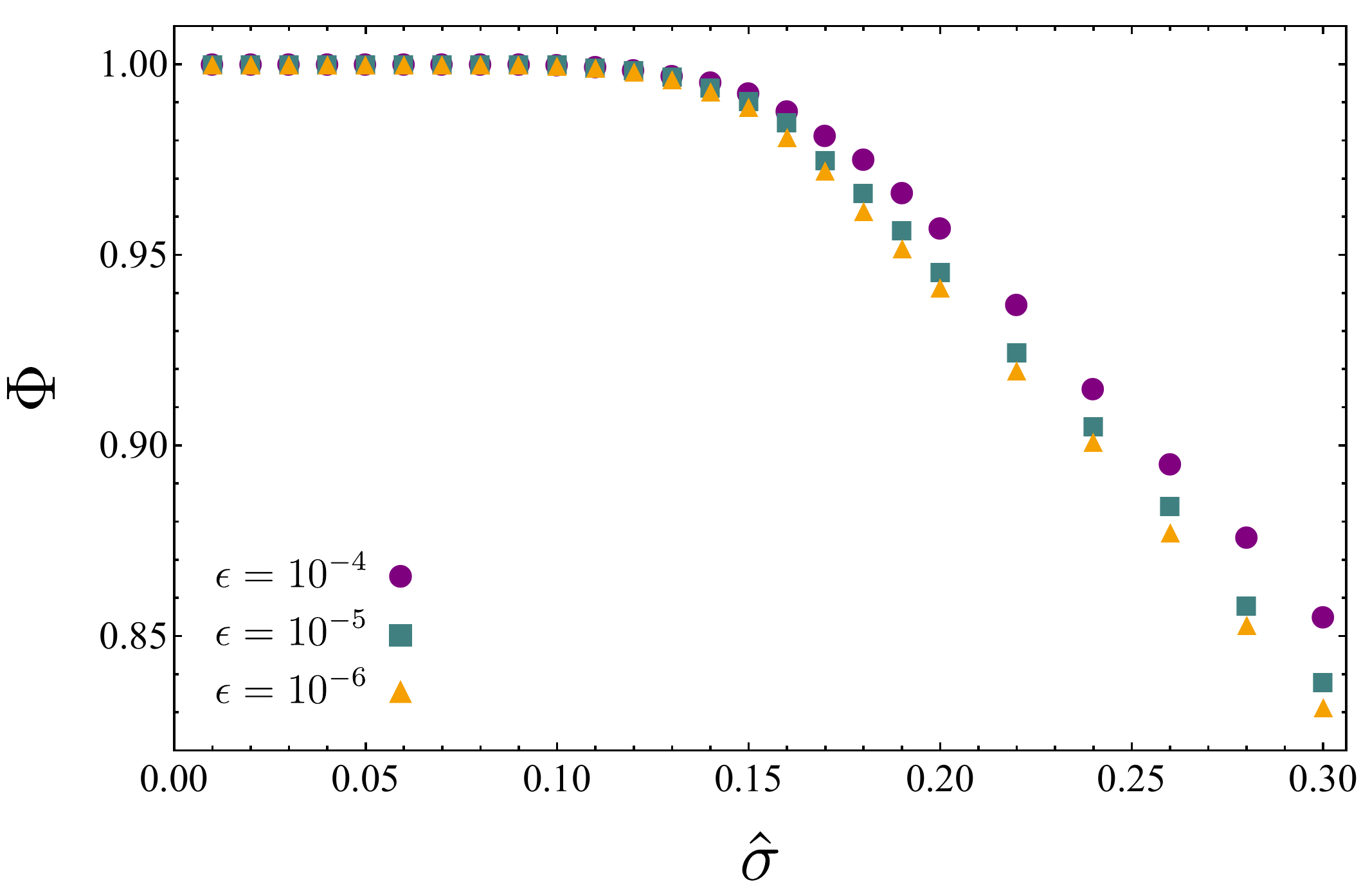}
\caption{\textbf{Left: Comparison between species abundance distributions computed on a single graph with different regularization parameters $\epsilon$.}
    For small $\hat{\sigma} = 0.02$ the distributions with $\epsilon=10^{-4}, 10^{-5}, 10^{-6}$ are perfectly equivalent, for large $\hat{\sigma} = 0.20$ there are small differences between the three distributions at small values of $n$. \textbf{Inset:} Same as in the main Figure, but in logarithmic scale to highlight the small differences at small $n$. 
    \textbf{Right: Surviving probability $\Phi$ as a function of $\hat\sigma$ for different values of $\epsilon$.}
    In both plots the parameters are: $T=0.5$, $\hat{\mu}=0.1$, $N=128$.} 
\label{Fig:marg_diff_eps}
\end{figure}

As explained in the main text, because of the discretized version of BP, we replaced the reflecting wall used for example in ref. \cite{altieri2021properties}, introducing a regularization parameter $\epsilon$ in the term $T\ln(n_i)$ inside the Hamiltonian of eq. \eqref{Ham-Altieri}, otherwise ill-defined at $n_i=0$. We show now that the obtained results for the marginal distributions are almost independent on the choice of $\epsilon$. In the left panel of Fig. \ref{Fig:marg_diff_eps} we show the average species abundance for two different values of the variance of the couplings $\hat\sigma$ for three different values of the parameter $\epsilon$. For small $\hat{\sigma}$ the distributions with $\epsilon=10^{-4}, 10^{-5}, 10^{-6}$ are perfectly equivalent, while for large $\hat{\sigma} = 0.20$ there are very small differences between the three distributions at small values of $n$ (only visible in log scale).

In the right panel of Fig. \ref{Fig:marg_diff_eps} we show how the Surviving probability $\Phi$, defined as $\Phi=1-\eta(0)$, changes for different values of $\epsilon$. The point at which $\Phi$ becomes different from 1 is independent of $\epsilon$, while for larger $\hat\sigma$, where extinctions are possible, $\Phi$ varies of a few percent passing from $\epsilon=10^{-4}$ to $\epsilon=10^{-6}$. However, we verified that the transition from the single fixed point phase to the unbounded growth phase (at small values of $\hat\mu$, such the one in Fig. \ref{Fig:marg_diff_eps}) is independent on $\epsilon$. Moreover, in Appendix \ref{app:Stability}, we will show that also the transition from the single fixed point phase to the multiple attractor phase at high values of $\hat\mu$ is independent on the value of $\epsilon$.

\section{Detecting the Unbounded Growth phase} \label{app:UG}

As said in the previous section, in order to have a numerical implementation of the BP equations, we both discretize $n$ and introduce a maximum value $n_{max}$. In this section we show how the final marginals are independent on $n_{max}$ (for $n_{max}$ sufficiently high) in the single fixed point phase, and we show how to detect the Unbounded Growth phase.

\begin{figure}[htbp] 
    \centering
    \begin{minipage}{0.5\textwidth}
        \centering
        \includegraphics[width=\textwidth]{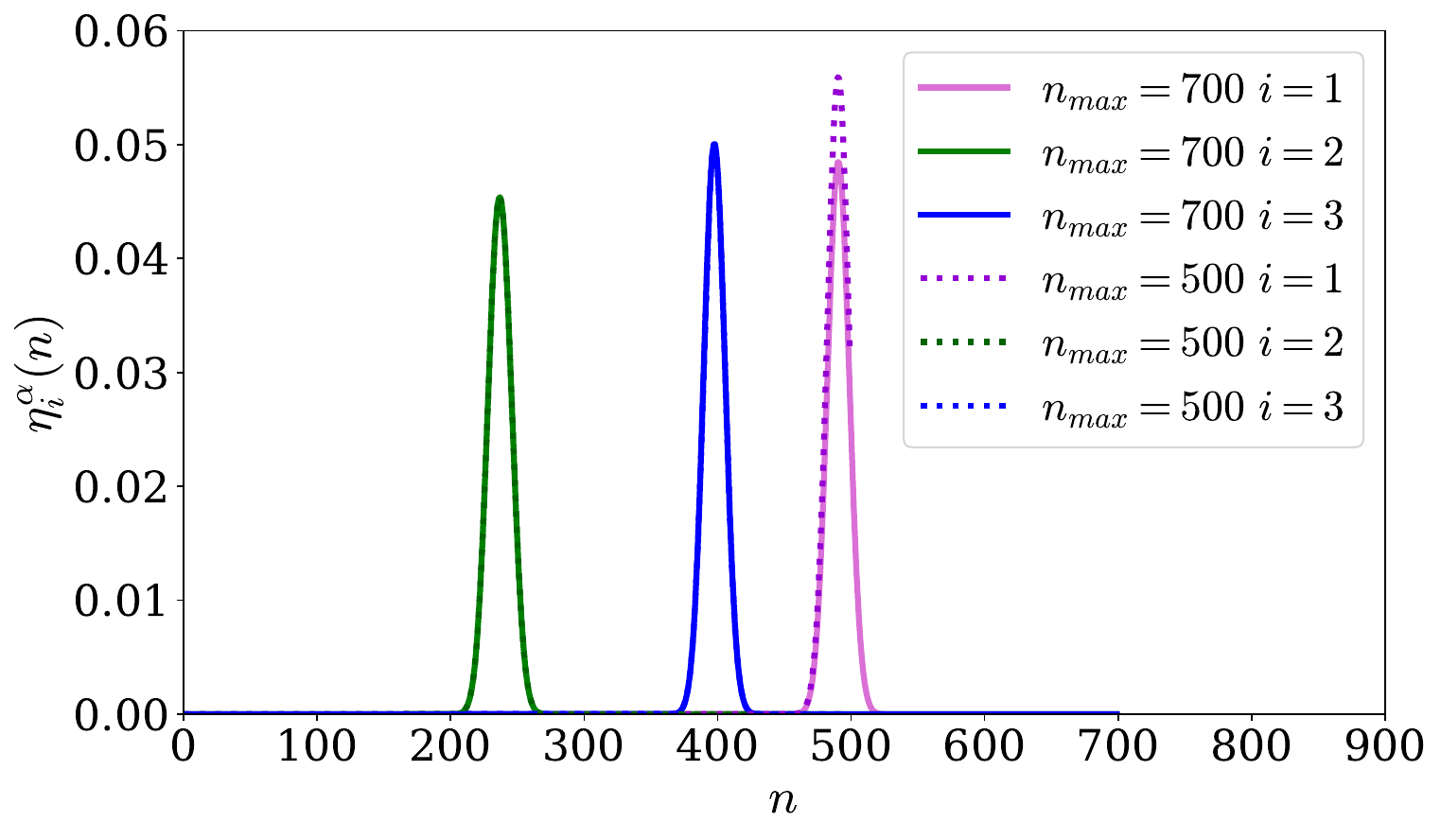}
    \end{minipage}
    \begin{minipage}{0.48\textwidth}
        \centering
        \includegraphics[width=\textwidth]{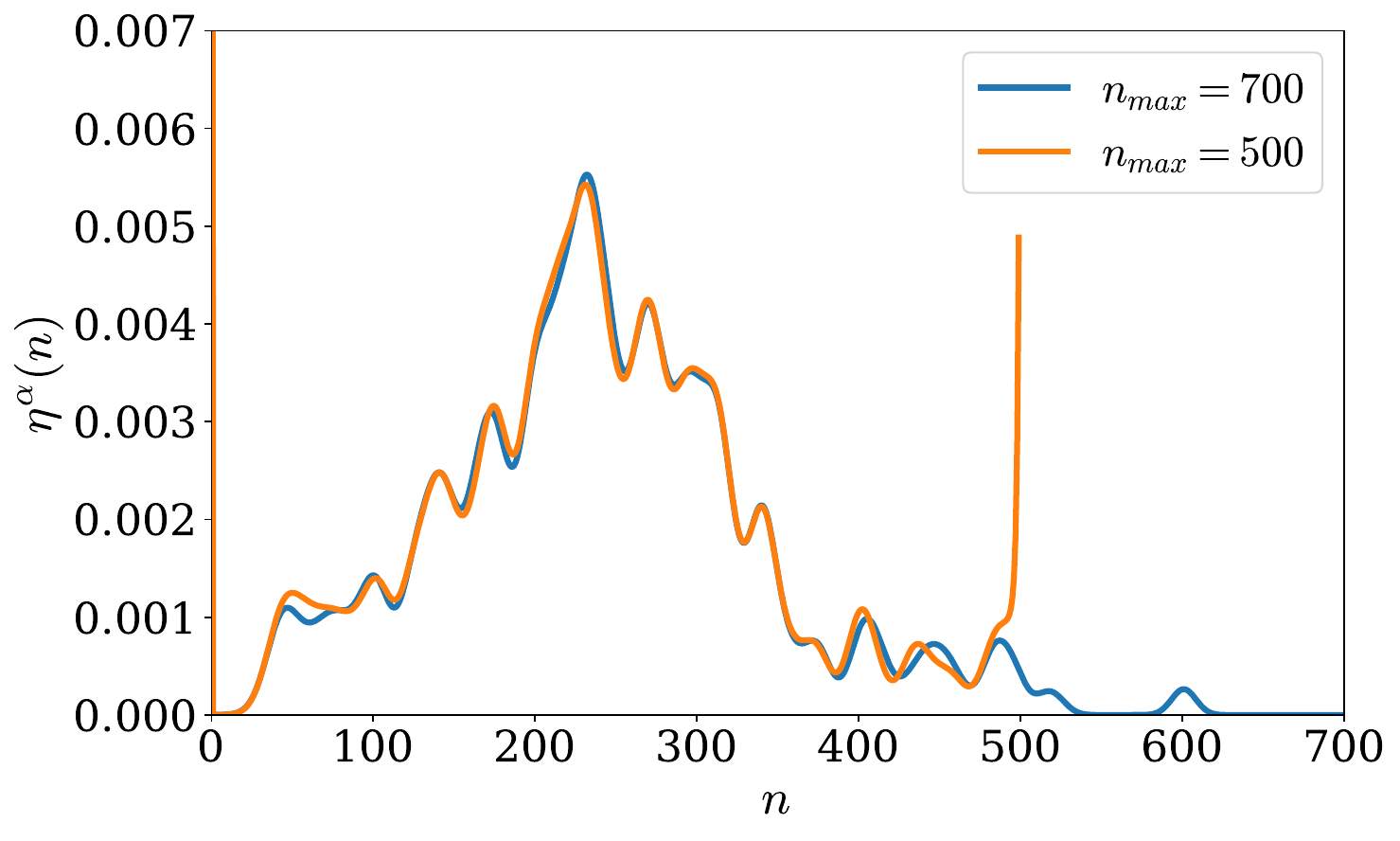}
    \end{minipage}
    \caption{\textbf{Left: Marginals $\mathbf{\eta_i^{\alpha}(n)}$ for three species changing $n_{max}$ in the unique fixed point phase.} In the unique fixed point phase, if $n_{max}$ is too small, the marginals for some species will be wrong: We detect the error looking if there are some species for which $\eta_i(n_{max}-1)\neq 0$, as happens for species 1 in the figure. Enlarging $n_{max}$ will then correct the marginal. \\
    \textbf{Right: Sample average $\mathbf{\eta^{\alpha}(n)}$ changing $n_{max}$ in the unique fixed point phase.} In the unique fixed point phase, if $n_{max}$ is too small, the mean abundance will show a peak in correspondence to the value $n=n_{max}-1$. Enlarging $n_{max}$ will correct the error. The mean of the distribution always remains at values near to $K=280$. Both figures refer to the same graph and interaction matrix with parameters $\hat{\mu}=0.1$, $T=0.2$, $\hat{\sigma}=0.2$, $N=200$. 
    }
    \label{fig:ChangingNmax}
\end{figure}

In Fig. \ref{fig:ChangingNmax} we show the mean abundance on a given network and the marginal probabilities for three different species in the unique fixed point phase. If $n_{max}$ is too small, the marginals for some species will be wrong: We detect the error looking if there are some species for which $\eta_i(n_{max}-1)\neq 0$. However, enlarging $n_{max}$ will then correct the marginal. Once we identify a value of $n_{max}$ such that  $\eta_i(n_\text{max})=0$ inside computer precision, enlarging $n_{max}$ further will have no effect on the marginals: this is a clear indication of the fact that the solution found for the BP equations is not influenced by our numerical implementation, as long as $n_{max}$ is large enough. A good practical choice for $n_{max}$ could be $n_{max}\simeq 2K$. In fact, without interactions, the marginal for each species will be centered around $n=K$: the interactions could modify slightly the mean of the marginals, but not so much: one can verify from Fig. \ref{fig:ChangingNmax} right that the mean of the distribution remains at values near to $K=280$ in this phase.

The behaviour in the unbounded growth phase is instead different. The numerical implementation of the BP equations with finite $n_{max}$ is converging even in the unbounded growth phase. However, this should not bother the reader too much: once implemented with a finite $n_{max}$, the BP equations are searching a solution for the marginal distributions generated not by the original Hamiltonian in eq. \eqref{Ham}, but by a modified Hamiltonian in which we are manually injecting a reflecting wall at $n=n_{max}$. While in the unique fixed point phase the final marginals are independent on the value of $n_{max}$ as long as it is large enough, as explained in the precedent paragraph, in the Unbounded Growth phase some marginals will always depend on $n_{max}$, irrespectively on how large $n_{max}$ is, and in particular they will always have $\eta_i(n_{max}-1)\simeq 1$, as shown in the left panel of Fig. \ref{fig:UG}: this is a clear evidence that the abundance distribution for $n_{max}\to\infty$ would run off to infinity.

\begin{figure}[htbp] 
    \centering
    \begin{minipage}{0.5\textwidth}
        \centering
        \includegraphics[width=\textwidth]{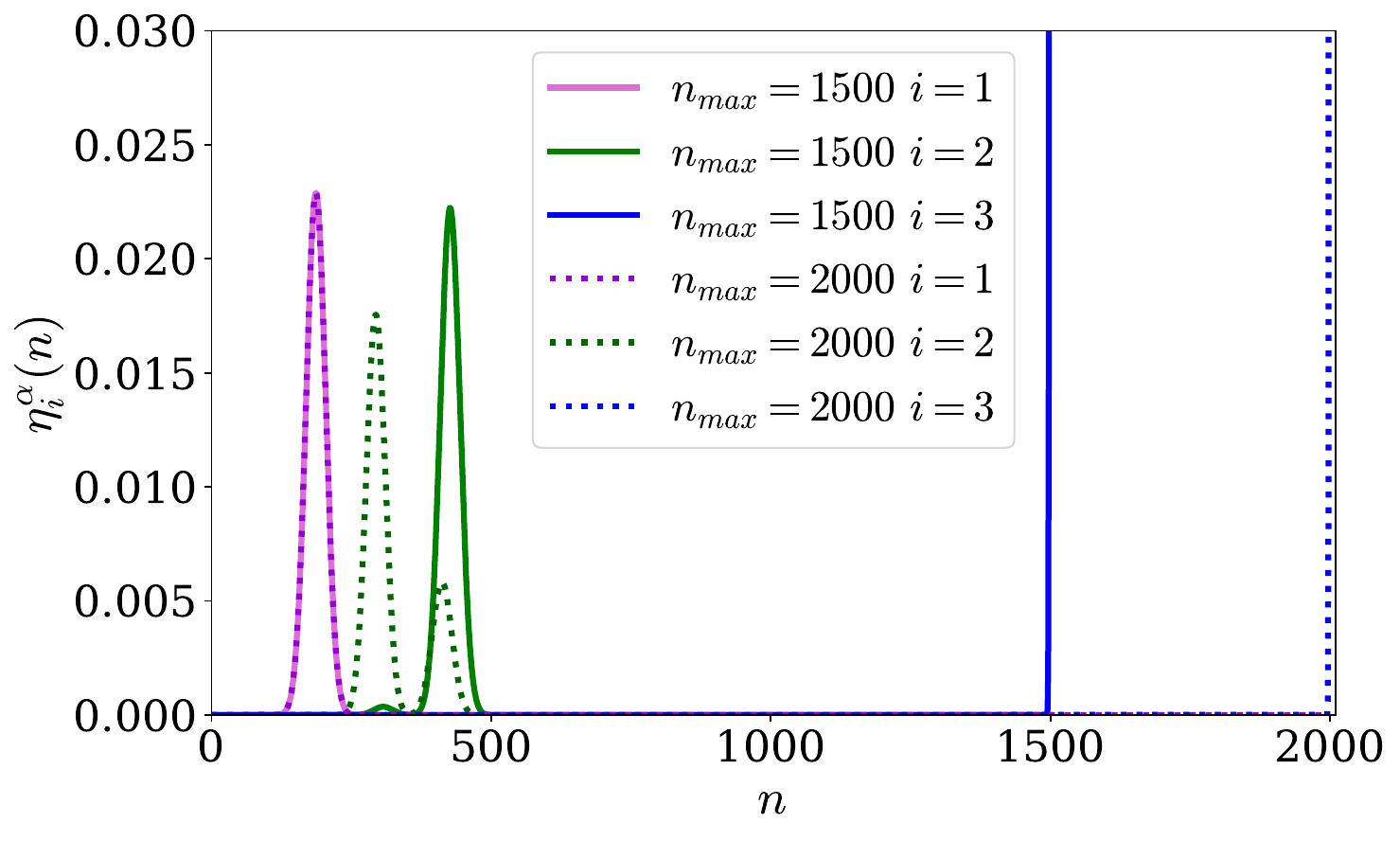}
    \end{minipage}
    \begin{minipage}{0.48\textwidth}
        \centering
        \includegraphics[width=\textwidth]{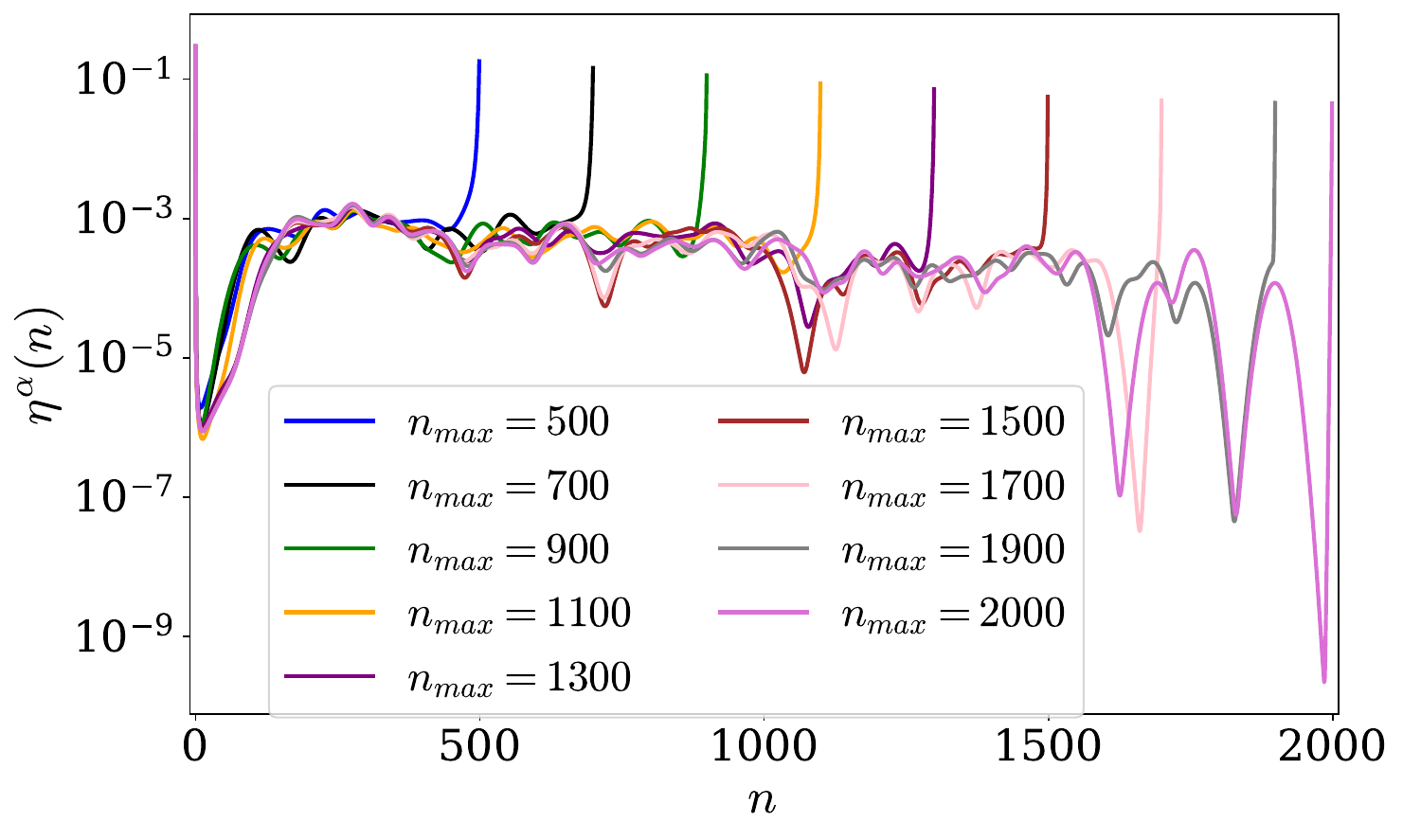}
    \end{minipage}
    \caption{\textbf{Left: Marginals $\mathbf{\eta_i^{\alpha}(n)}$ for three species changing $n_{max}$ in the Unbounded Growth phase.} While for some species changing $n_{max}$ does not change their marginals, like species 1 in the figure, some other species have $\eta_i(n_{max}-1)\simeq 1$ irrespectively on the value of $n_{max}$, like species 3.\\
    \textbf{Right: Sample average $\mathbf{\eta^{\alpha}(n)}$ changing $n_{max}$ in the Unbounded Growth phase.} In the Unbounded Growth phase, the mean abundance will continue to change changing in $n_{max}$. In particular, $\eta^{\alpha}(n_{max}-1)$ will have a finite weight that, for $n_{max}$ large enough, does not depend on $n_{max}$. Both figures refer to the same graph and interaction matrix with parameters $\hat{\mu}=0.1$, $T=1$, $\hat{\sigma}=0.5$, $N=200$. }
    \label{fig:UG}
\end{figure}

In the right panel of Fig. \ref{fig:UG}, we show the mean abundance in the Unbounded Growth phase changing $n_{max}$: it is evident that $\eta^{\alpha}(n_{max}-1)$ has a finite weight that, for $n_{max}$ large enough, does not decrease anymore increasing $n_{max}$, differently with what happened in the Unique Fixed point phase. This can be chosen as a practical rule for detecting the Unbounded Growth phase: if $\eta^{\alpha}(n_{max}-1)$ is different from zero and its value does not decrease increasing $n_{max}$ the system is in the Unbounded Growth phase. We stress that, at variance with what happens in fully connected models, there are only some species whose marginals run off to infinity while, the marginals for other species remain finite and independent on $n_{max}$: this is due to the sparsity of the model that could create subsystems whose species grow indefinitely while not influencing the others that stay finite. The convergence of BP equations even in the Unbounded Growth phase allows us to count the fraction of diverging species, that is shown in Fig. \ref{fig:countUG-vs-sigma} of the main text.

\section{Comparison between Belief Propagation and Langevin dynamics simulations} \label{app:BP-vs-DYN}

This Appendix shows that the exact equilibrium marginal distributions computed by BP are fully consistent with the marginal distributions obtained from SDE simulations of the Langevin dynamics. In Fig.~\ref{fig:app-bp-vs-dyn} we show this agreement for 4 different values of $\hat\sigma$. For each value of $\hat\sigma$, at $\hat\mu=0.1$ and $T=1$, we display both the species abundance trajectories $n_i(t)$ (left panels) and the corresponding equilibrium marginal distributions (right panels). In particular, the marginal distributions are obtained in two ways: directly from BP (shown in solid lines) and by sampling the long-time dynamical evolution of the abundances (shown in dots). As shown in the right panels, the two methods perfectly coincide. Small differences only appear for marginal distributions with a finite probability of extinctions (the peak at $n=0$); such differences come from the different regularizations used at $n=0$: in the SDE simulations through the reflecting wall condition $n^{\text{min}}=\lambda\ll 1$, and in BP through the parameter $\epsilon$. These differences on the regularization parameters have no impact on the phenomenology, as also discussed in Appendix~\ref{app:changing_dn_eps} and \ref{app:Stability}.

The simulations in Fig.~\ref{fig:app-bp-vs-dyn} are performed on a single graph realization, which is kept identical for both BP and the dynamics. We focus on five species of the ecosystem, highlighting their temporal evolution in color, and plotting with the same color their equilibrium abundance distributions, both from BP and from the dynamics. For $\hat\sigma=0$, where no disorder is present, the temporal evolutions differ because of the different demographic noise sequence, but their equilibrium distributions are equivalent for all species. For $\hat\sigma>0$ the five marginal distributions become distinct across species, but are still in perfect agreement between BP and the dynamical sampling.

\begin{figure}[htbp]
\includegraphics[width=0.7\columnwidth]{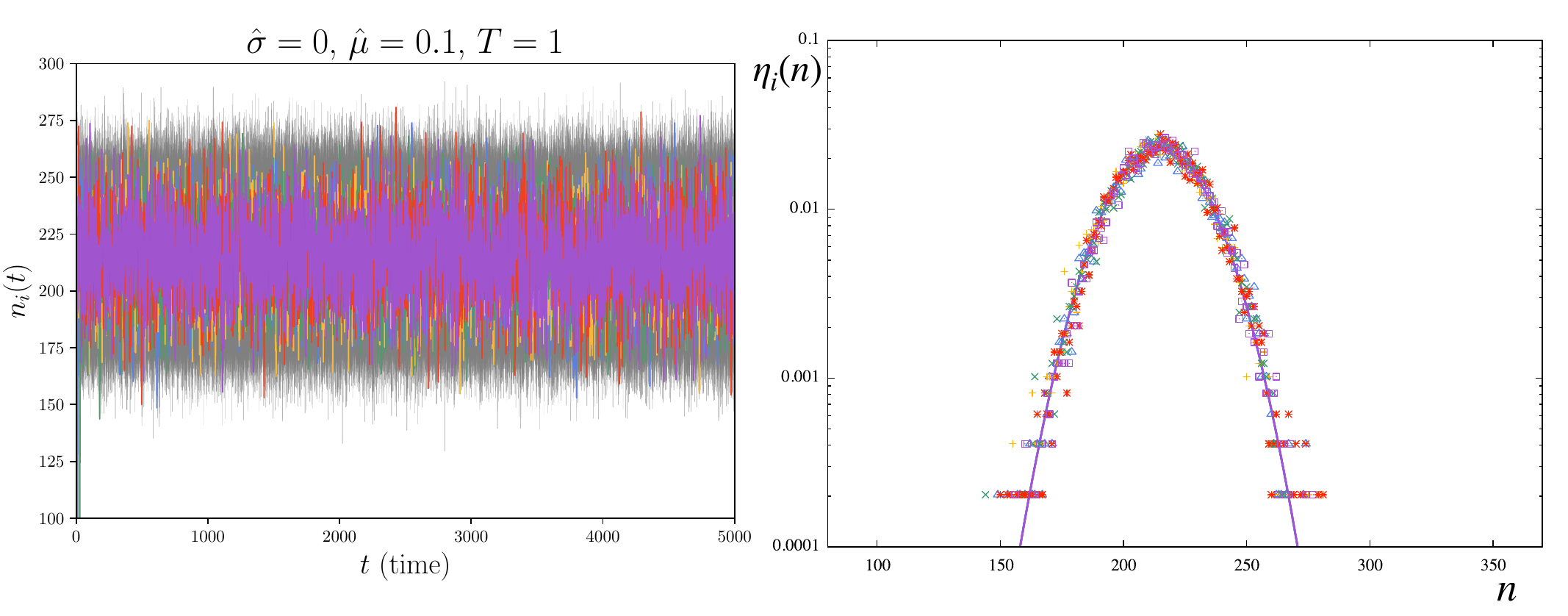}
\includegraphics[width=0.7\columnwidth]{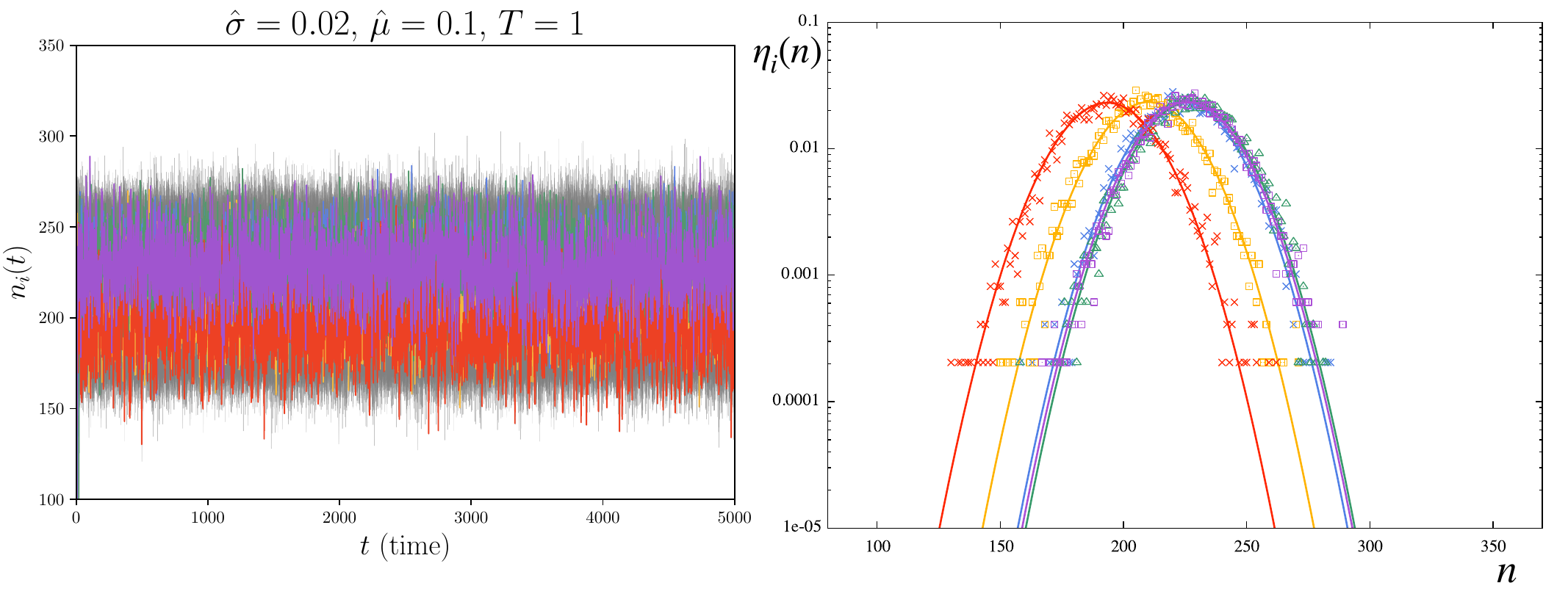}
\includegraphics[width=0.7\columnwidth]{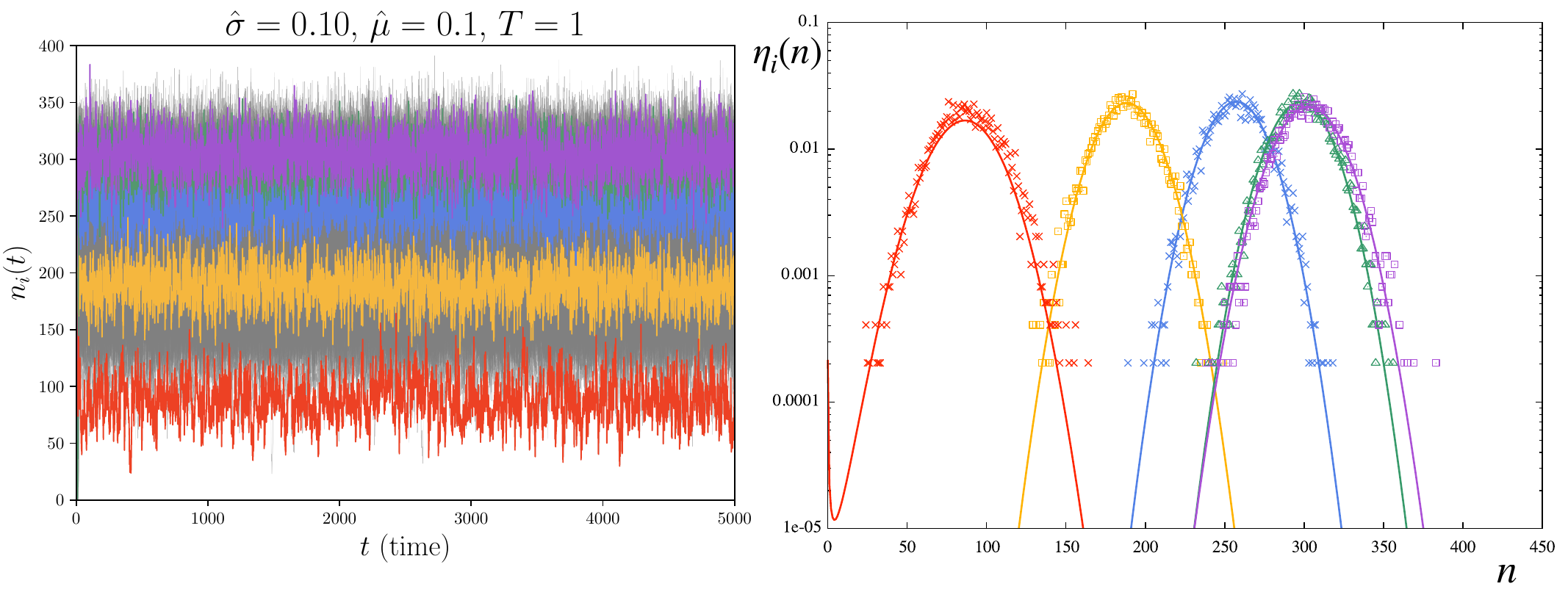}
\includegraphics[width=0.7\columnwidth]{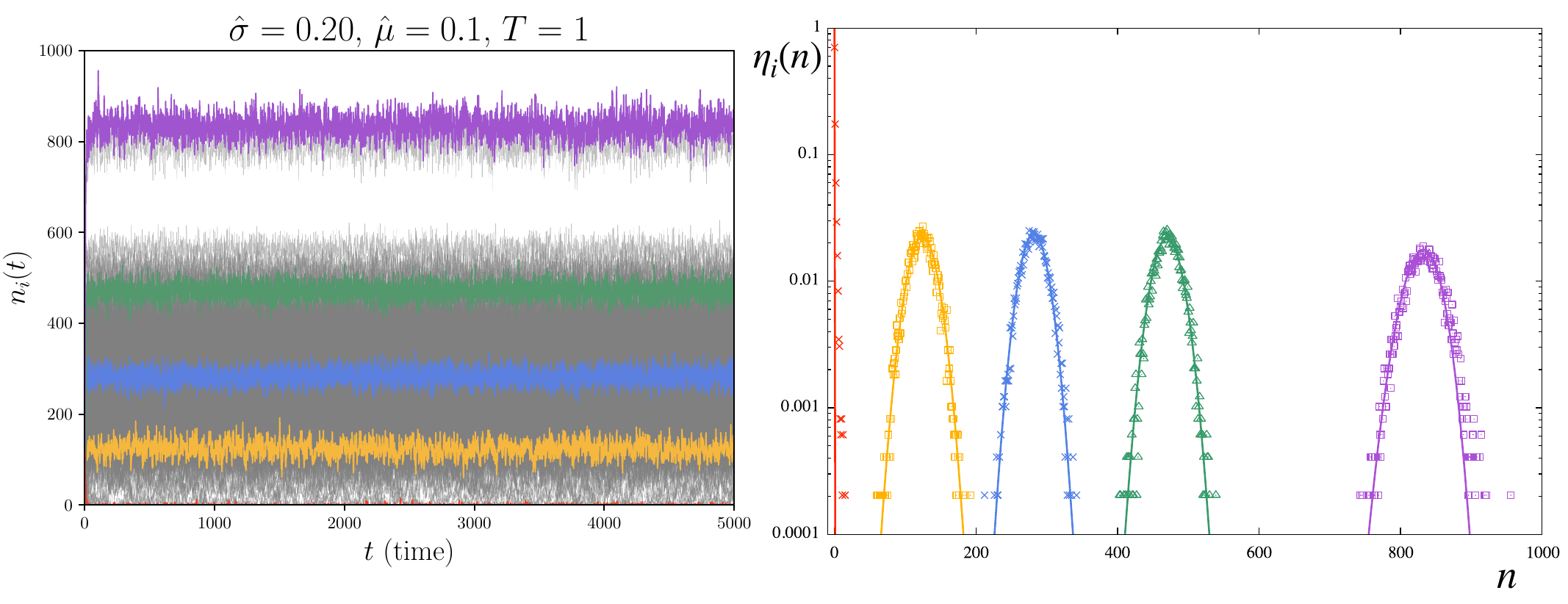}
\caption{
\textbf{Left: Trajectories of the species abundances $n_i(t)$ in time, simulated through the Langevin dynamics.} The same five species are selected for all the different sets of parameters and their temporal evolution is higlighted in color. \textbf{Right: Comparison between the marginal distributions computed with BP (in solid lines) and those sampled from the dynamical trajectories (as points).} The species distributions from BP and from the dynamics, plotted in the same color of the species temporal evolution, perfectly coincide. }
\label{fig:app-bp-vs-dyn}
\end{figure}

We then show an example of another result, obtained in the main text through BP, which is perfectly consistent with the simulation of the SDE dynamics. Such result is the appearance of the unbounded growth phase presented in Fig.~\ref{fig:UG}. In particular, focusing on an interaction graph with N=128 species and parameters $\hat\mu=0.1, T=1$, we have run the dynamics for three different values of $\hat\sigma= 0.30,0.35,0.40$. As shown in Fig.~\ref{fig:UG-dynamics-s030-s035-040}, at $\hat\sigma=0.30$ species simply fluctuate in time always keeping finite abundances. For $\hat\sigma=0.35$ instead, some species start to grow indefinitely already at small times, and in particular the number of species which undergo unbounded growth increases increasing $\hat\sigma$, as shown in the panel of Fig.~\ref{fig:UG-dynamics-s030-s035-040} with $\hat\sigma=0.40$. This analysis confirms what shown in Fig.~\ref{fig:UG}, namely that, for $\hat\mu=0.1$ and $T=1$, around $\hat\sigma\sim 0.35$ the unbounded growth phase appears.

\begin{figure}[htbp]
\includegraphics[width=0.45\textwidth]{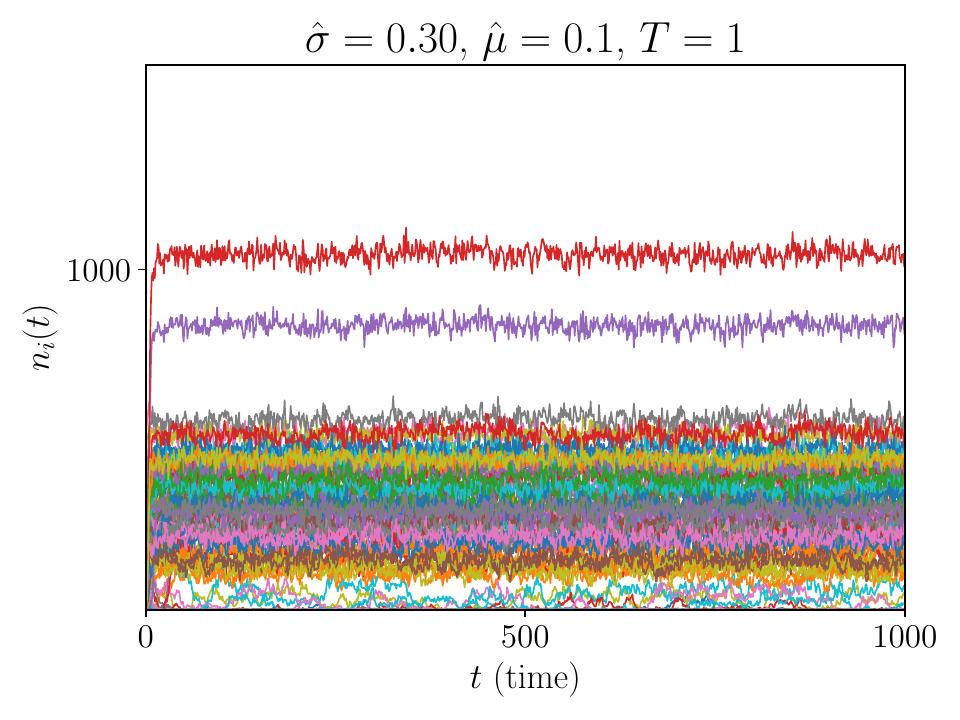}\\
\includegraphics[width=0.45\textwidth]{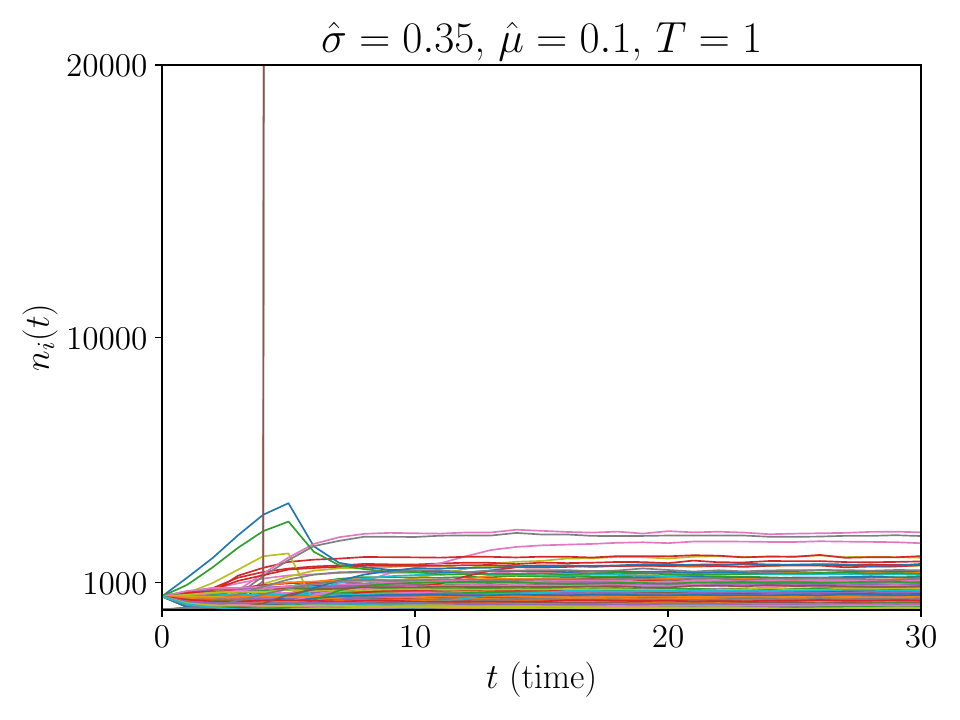}
\includegraphics[width=0.45\textwidth]{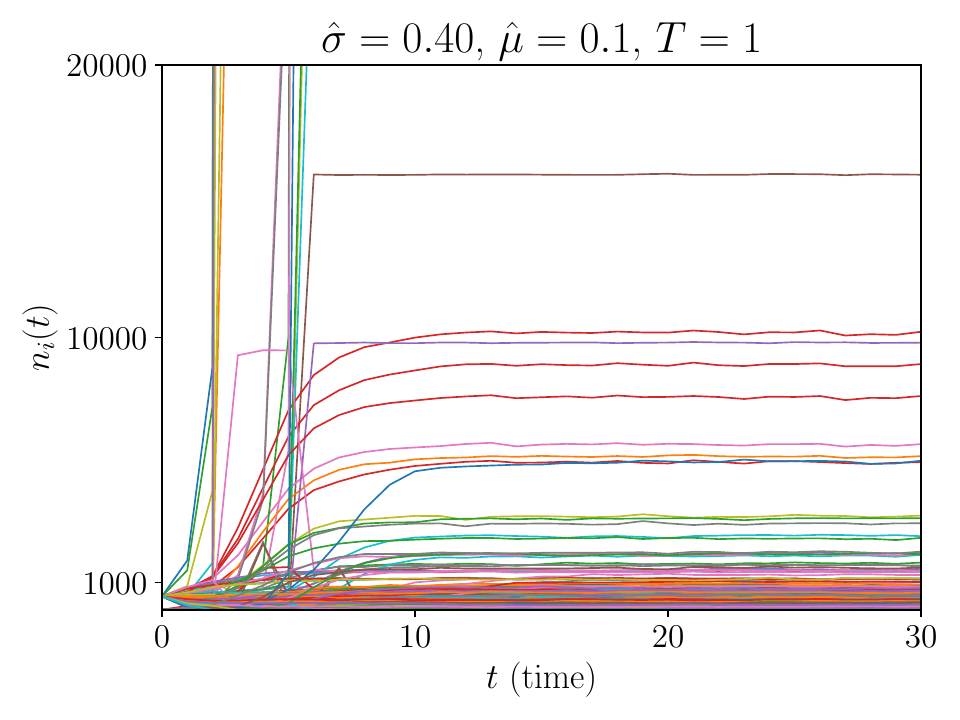}
\caption{\textbf{Species abundance trajectories near the unbounded growth transition.} For $\hat\mu=0.1$ and $T=1$, for $\hat\sigma=0.30$ no unbounded growth is present. For $\hat\sigma=0.35, 0.40$ instead some species in the ecosystem grow indefinitely, displaying an unbounded growth. In these simulations the interaction graph is generated for an ecosystem of $N=128$ species.}
\label{fig:UG-dynamics-s030-s035-040}
\end{figure}

\section{Universality and Scaling Law for the Sparse Ecosystems Dynamics}
\label{app:Scaling}

Here we focus on an ecologically relevant aspect of our system. In real ecosystems, extensive observations have documented the presence of Taylor's law (see \cite{Giometto-taylor-law}), which describes the relationship between the variance and mean of species abundances. Specifically, if one measures the variance $V_i$ and the mean $M_i$ for the abundance trajectory $n_i(t)$ of each species in the ecosystem, Taylor's law states that the variances follow a power-law dependence on the means:
\begin{equation}\label{eq:power-law}
    V=a M^b,
\end{equation}
where, in ecological systems, the exponent $b$ has widely been reported to be $b\simeq 2$~\cite{Anderson_1982,Giometto-taylor-law}.

To investigate this in our model, we set $\hat{\mu}=0.1$ and $\hat\sigma=0.25$, ensuring that the system is in the non-Gaussian (ecologically realistic) regime, see Sec.~\ref{sec:non-gaussianity}. We then ran Langevin dynamics (see Fig.~\ref{fig:dynamics-for-scaling}) for a system described by Eq.~\eqref{LV-equations}, computing the mean $M_i$ and variance $V_i$ for each species $i$:
\begin{equation}
    M_i\equiv\frac{1}{t_{\text{max}}-t_0}\sum_{t=t_0}^{t_{\text{max}}} n_i(t),\quad V_i\equiv\frac{1}{t_{\text{max}}-t_0}\sum_{t=t_0}^{t_{\text{max}}}\left(n_i(t)-M_i\right)^2,
\end{equation}
where $t_0$ is the initial transient time beyond which the distributions no longer depend on initial conditions and $t_\text{max}$ is the final simulation time. 
We performed these simulations for $N=400$ species and various temperatures $T$.
\begin{figure}[htbp]
\includegraphics[width=0.5\columnwidth]{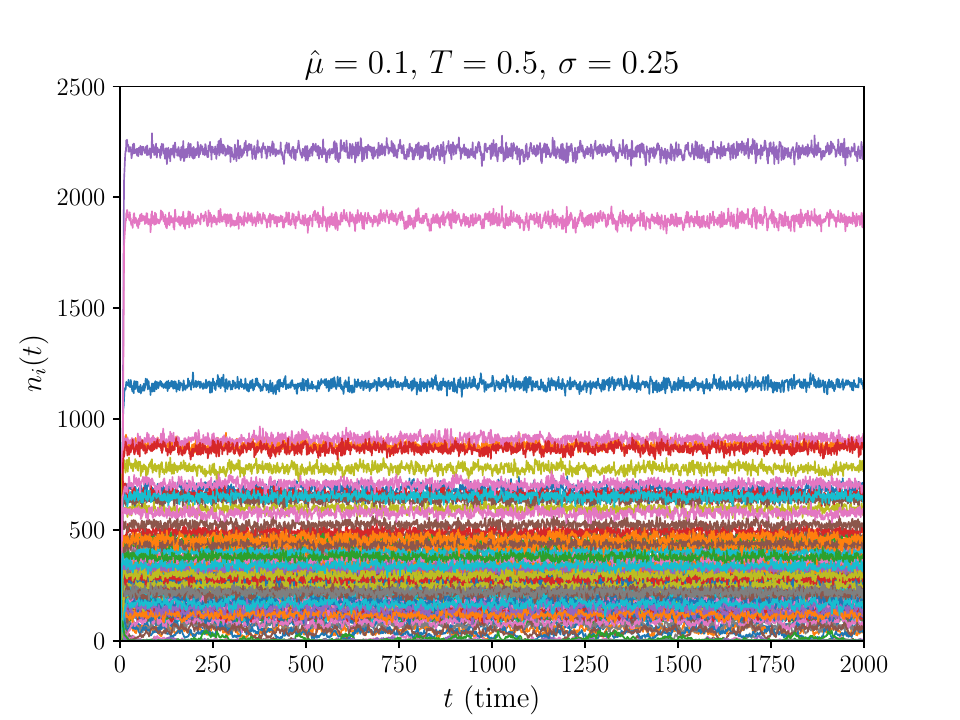}
\caption{\textbf{Species abundance dynamics} for $\hat{\mu}=0.1, \hat{\sigma}=0.25$ and $T=0.5$. The number of species here is $N=400$.}
\label{fig:dynamics-for-scaling}
\end{figure}
In the left panel of Fig.~\ref{fig:scaling}, we plot all pairs $(M_i, V_i)$, with different colors representing different temperatures.

Regardless of the temperature, the data always exhibit a \textit{power law} behaviour for small $M_i$, followed by a \textit{saturation} region for large $M_i$.
As shown in the left plot of Fig.~\ref{fig:scaling}, the power law remains the same for all $T$, with the exponent $b$ of Eq.~\eqref{eq:power-law} being $b\simeq 2$. In particular the black solid line represents the fit of the data for small $M_i$ and is given by $V=aM^b$, with $a=3.4$ and $b=2.05$. This is a remarkable result, given the widespread occurrence of this power laws in nature.
Moreover, the same power law is found for different values of disorder $\hat{\sigma}$, provided that they are sufficiently large ($\hat{\sigma}\gtrsim 0.15$), corresponding to the non-Gaussian regime. 

\begin{figure}[htbp] 
    \includegraphics[width=0.48\columnwidth]{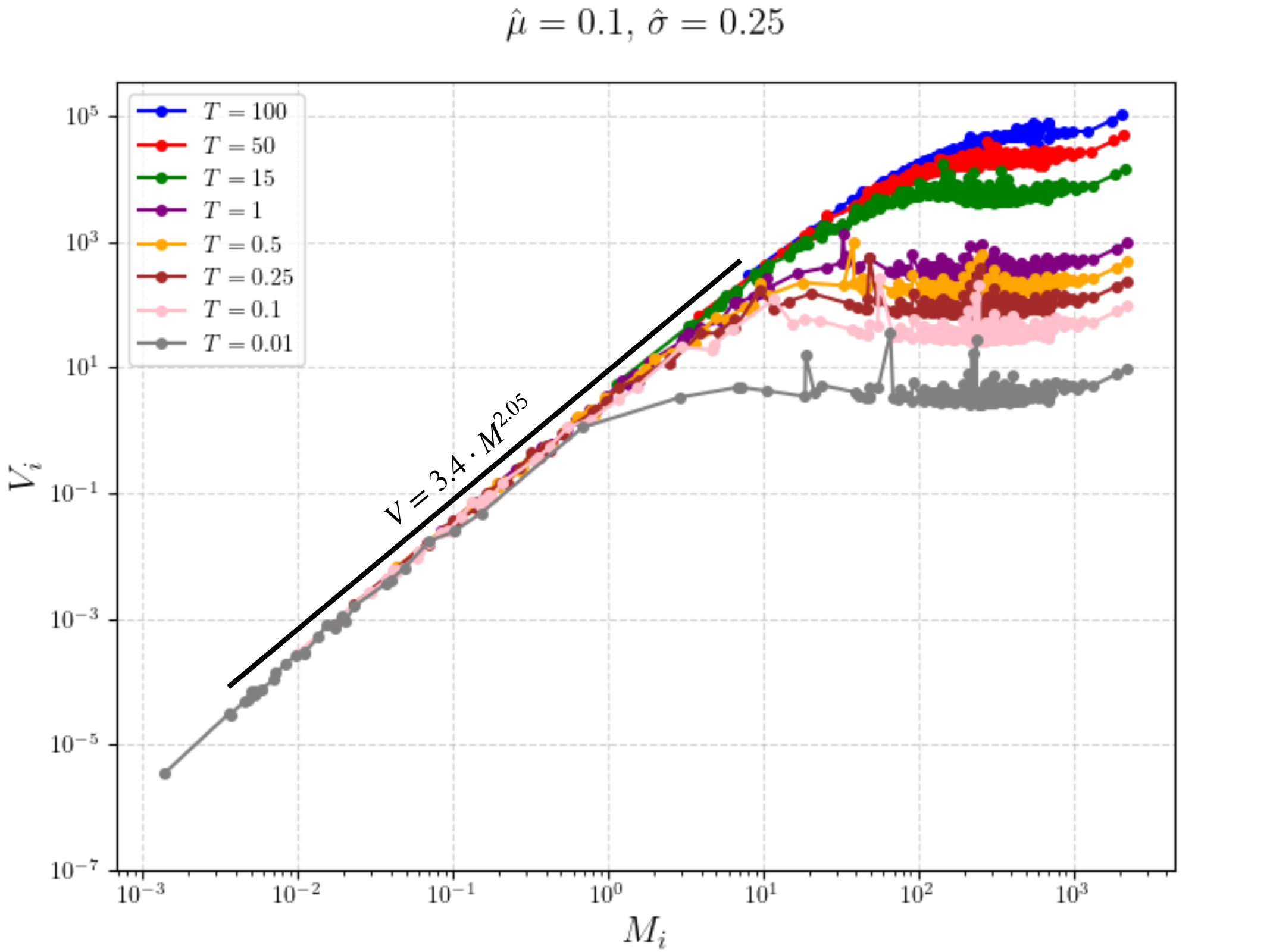}
    \includegraphics[width=0.505\columnwidth]{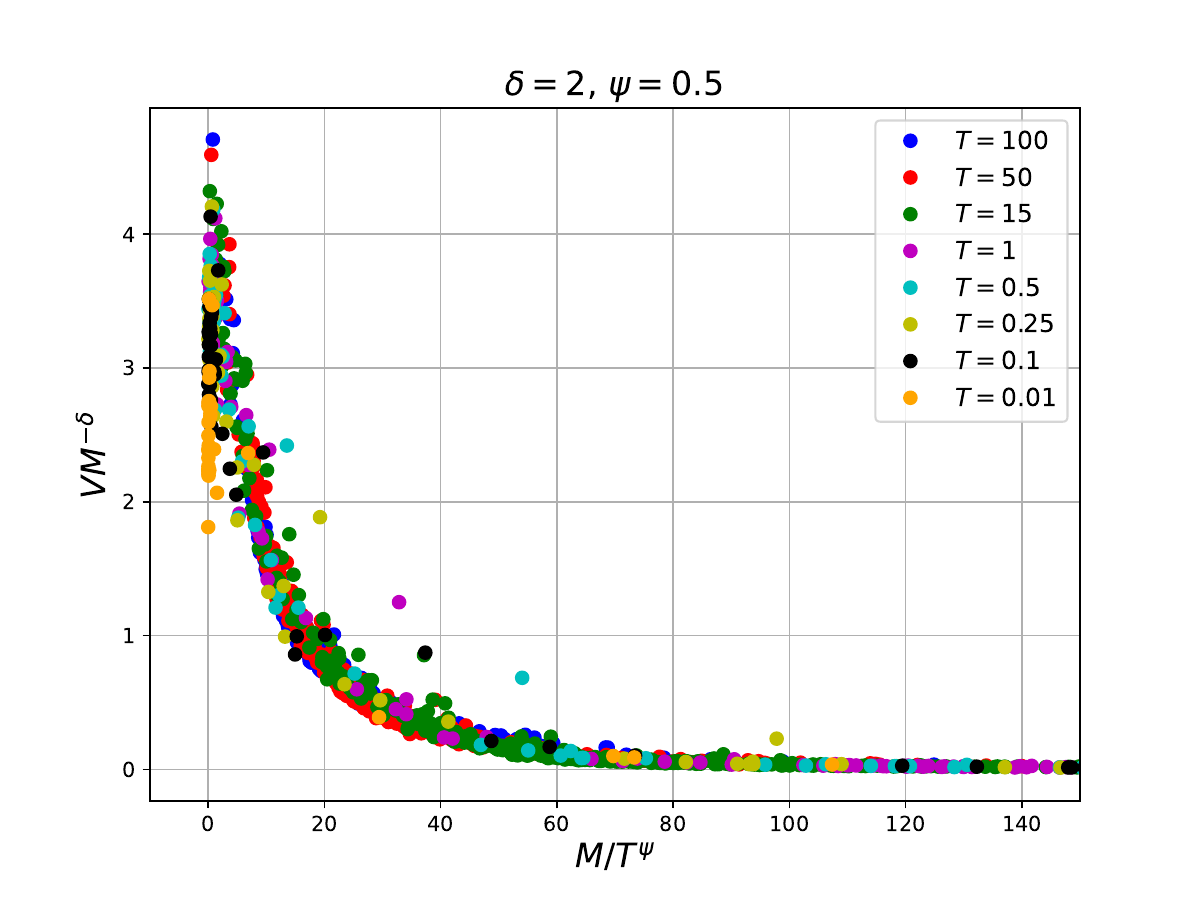}
    \caption{\textbf{Left: pairs of mean and variance $(M_i,V_i)$ of single species distributions for different temperatures.} For all the temperatures there is a power-law region for low values of $M_i$ and a saturation region for large values of $M_i$. \textbf{Right: pairs of mean and variance combined and rescaled with appropriate exponents.} Data for different temperatures, once plotted with the new scaling observables, collapse into a unique curve, that corresponds to the scaling function in eq. \eqref{eq:scaling-ansatz}. For these plots $\hat{\mu}=0.1, \hat{\sigma}=0.25$ and $N=400$. }
    \label{fig:scaling}
\end{figure}

In the saturation regime instead, the variance is more or less constant ($V(M,T)=V_S(T)$), no matter which is the value of $M$. In particular, the higher is the temperature, the higher is the saturation value $V_S(T)$.

Given these observations on the power-law and saturation regimes, we propose a scaling ansatz for the variance $V(M,T)$:
\begin{equation}\label{eq:scaling-ansatz}
    V(M,T)=M^\delta F\left(\frac{M}{T^\psi}\right),
\end{equation}
where $V$ is the variance, $M$ is the mean, $T$ is the temperature and $\delta$ and $\psi$ are critical exponents that have to be computed. Looking at our data we state that the function $F$ is such that:
\begin{itemize}
    \item $F(x)\sim$ constant for $x\rightarrow 0$,
    \item $F(x)\sim x^{-\delta}$ for $x\rightarrow \infty$.
\end{itemize}

The exponent $\delta$ is found by fitting the low-$M$ regime data. In particular, as we already said, we get:
\begin{equation*}
    \delta\simeq 2.
\end{equation*}

Regarding the exponent $\psi$, we can compute it by noticing that, for each temperature $T$, the variance saturation value is:
\begin{equation*}
    V_S(T)=\lim_{M\rightarrow\infty} V(M,T).
\end{equation*}
Recalling the behaviour of $F(x)$ for $x\rightarrow\infty$, from Eq. \eqref{eq:scaling-ansatz} we get:
\begin{equation*}
    V_S(T)\sim M^\delta \frac{M^{-\delta}}{T^{-\psi\delta}}=T^{\psi\delta}.
\end{equation*}
So, once we have the value of $\delta$, we can compute $\psi$ by fitting all the values $V_S(T)$ with the function $V_S(T)=T^{\psi\delta}$. In this way we get:
\begin{equation*}
    \psi\simeq 0.5.
\end{equation*}

To conclude, on the right panel of Fig. \ref{fig:scaling} we plotted, for all the data of the left panel, $V M^{-\delta}$ vs $M/T^{\psi}$, with $\delta$ and $\psi$ set to the values that we found. In this way we can really see that the scaling of Eq.~\eqref{eq:scaling-ansatz} works and all the data from different ecosystems at different temperatures collapse onto the same curve. These results highlight the emergence of a universality mechanism.

\section{The key role of sparsity of interactions for the strong non-Gaussian effects}\label{app:eta-tilde}

In Sec.~\ref{sec:non-gaussianity} we have analyzed how the shape of the marginal distributions strongly changes upon increasing the interaction disorder $\hat\sigma$. Here we want to further stress that the presence of strong non-Gaussian effects depends mainly on the sparsity of the interactions. In Sec.~\ref{sec:bethe-lattice} we introduced the marginal distributions $\eta^\alpha_i(n_i)$ as:
\begin{align}
    \eta^\alpha_i(n_i) = \frac{1}{z^\alpha_i}e^{-\beta h_i(n_i)}\prod_{j\in
      \partial i}\left[\sum_{\{n_j\}}\eta^\alpha_{j\rightarrow i}(n_j)
    e^{-\beta h^\alpha_{ij}(n_i, n_j)} \right], 
\end{align}
which, given that $h_i(n_i)=-r\left(n_i-n_i^2/2\right)+T\ln(n_i+\epsilon)$, can be rewritten as
\begin{align}
    \eta^\alpha_i(n_i) = \frac{1}{z^\alpha_i}\frac{1}{n_i+\epsilon}e^{\beta r\left(n_i-n_i^2/2\right)}\prod_{j\in
      \partial i}\left[\sum_{\{n_j\}}\eta^\alpha_{j\rightarrow i}(n_j)
    e^{-\beta h^\alpha_{ij}(n_i, n_j)} \right]. 
\end{align}

In particular, in Sec.~\ref{sec:non-gaussianity} we analyzed the average marginal distributions, averaged over the species and the disorder:
\begin{align}\label{eq:sample-average}
\eta(n)\equiv \overline{\frac{1}{N} \sum_{i=1}^N \eta^\alpha_i(n)}.
\end{align}

We can actually define a different marginal distribution $\tilde\eta_i(n_i)$, rescaled by $n_i+\epsilon$, where the contribution of the demographic noise is factored out:
\begin{align}\label{eq:app-def-eta-tilde}
    \tilde\eta^\alpha_i(n_i)=\frac{1}{\tilde z^\alpha_i} (n_i+\epsilon) \eta^\alpha_i(n_i).
\end{align}
From Eq.~\eqref{eq:app-def-eta-tilde} it follows that the corresponding averaged distribution is:
\begin{align}\label{eq:app-def-averaged-eta-tilde}
    \tilde\eta(n)=\frac{1}{\tilde z} (n+\epsilon) \eta(n).
\end{align}

Studying the shape of $\tilde\eta(n)$ instead of $\eta(n)$ allows us to focus only on the effect of interactions sparsity, leaving out the effects due to the demographic noise. Please note that now the rescaled marginal $\tilde\eta(n)$ has an exact Gaussian form in absence of the interactions. Fig.~\ref{fig:etatilde-vs-eta} shows both $\eta(n)$ and $\tilde\eta(n)$ for two different values of $\hat\sigma$, namely $\hat\sigma=0.02$ and $\hat\sigma=0.20$ (the same of Fig.~\ref{fig:exp-tail}). Notice that at $\hat\sigma=0.20$ the distribution $\tilde\eta(n)$ does not display any extinction peak in $n=0$, contrary to $\eta(n)$. The extinction peak is in fact mainly due to the demographic noise factor.
It is anyway clear that strong non-Gaussian effects are present for the distributions $\tilde\eta(n)$, and in particular $\tilde\eta(n)$ displays an even fatter tail than $\eta(n)$. This analysis confirms that, for large $\hat\sigma$, the non-Gaussianity of the marginal distributions is driven almost exclusively by network sparsity rather than by demographic noise.
\begin{figure}[htbp]
\includegraphics[width=0.7\columnwidth]{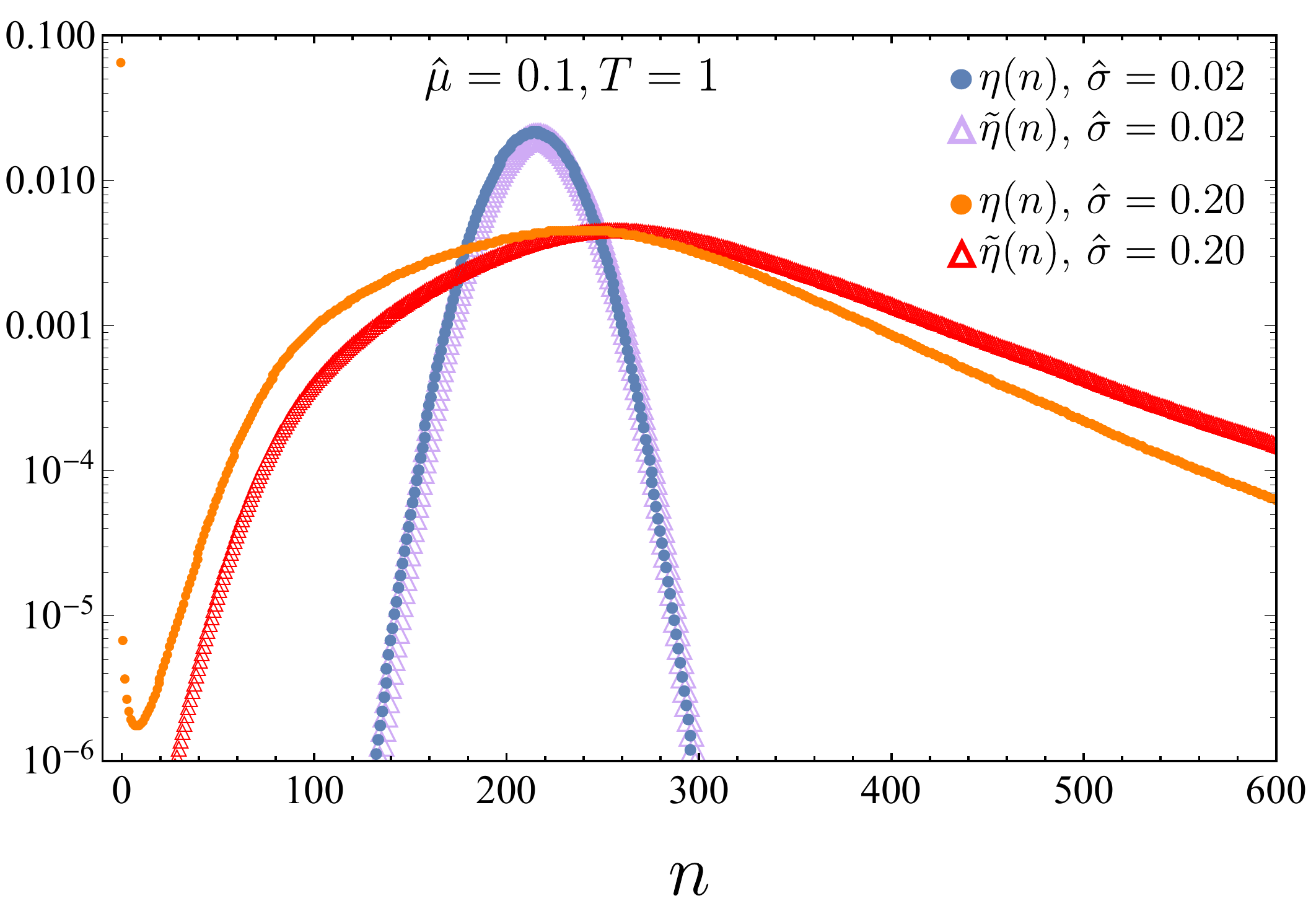}
\caption{\textbf{Comparison between marginal distributions $\eta(n)$ and rescaled marginal distributions $\tilde\eta(n)$ for two different values of $\hat\sigma$.}  Strong non-Gaussian effects are present for both $\eta(n)$ and $\tilde\eta(n)$ for large values of $\hat{\sigma} = 0.20$ (orange and red), while they are absent for small $\hat{\sigma}= 0.02$ (blue and purple). The parameters of the plots are $T=1$, $\hat{\mu}=0.1$ and $N=256$.}
\label{fig:etatilde-vs-eta}
\end{figure}
We can also replicate, for the rescaled  distributions $\tilde\eta(n)$, the study of the crossover from Gaussian to non-Gaussian behaviour varying $\hat{\sigma}$ and $T$ shown in Fig.~\ref{fig:2d-contour} for the marginals $\eta(n)$. In Fig.~\ref{fig:contour-plot-etatilde} we quantify, for two different values of $\hat\mu$, the deviation from Gaussianity of the marginal distributions $\tilde\eta(n)$ as the difference between their kurtosis, $\kappa_{\tilde\eta}(\hat\sigma,T)$, and the kurtosis $k_G(\hat\sigma,T)$ of a reference discrete Gaussian distribution $G(n_\text{peak}(\tilde\eta),\sigma^2_{\tilde\eta})$ with the same variance and centered at the peak of $\tilde\eta$. The two kurtosis are computed as
\begin{align}
  \kappa_{x}(\hat{\sigma},T)\equiv \frac{\langle (n - \langle n \rangle_{x})^4 \rangle_{x}}{\langle (n - \langle n \rangle_{x})^2 \rangle_{x}^{2}},
  \qquad x = \tilde{\eta},\, G,
\end{align}
where the averages $\langle\cdot\rangle_x$ are taken with respect to $\tilde\eta(n)$ and to $G(n_\text{peak}(\tilde\eta),\sigma^2_{\tilde\eta})$ respectively.

\begin{figure}[htbp]
\includegraphics[width=0.49\columnwidth]{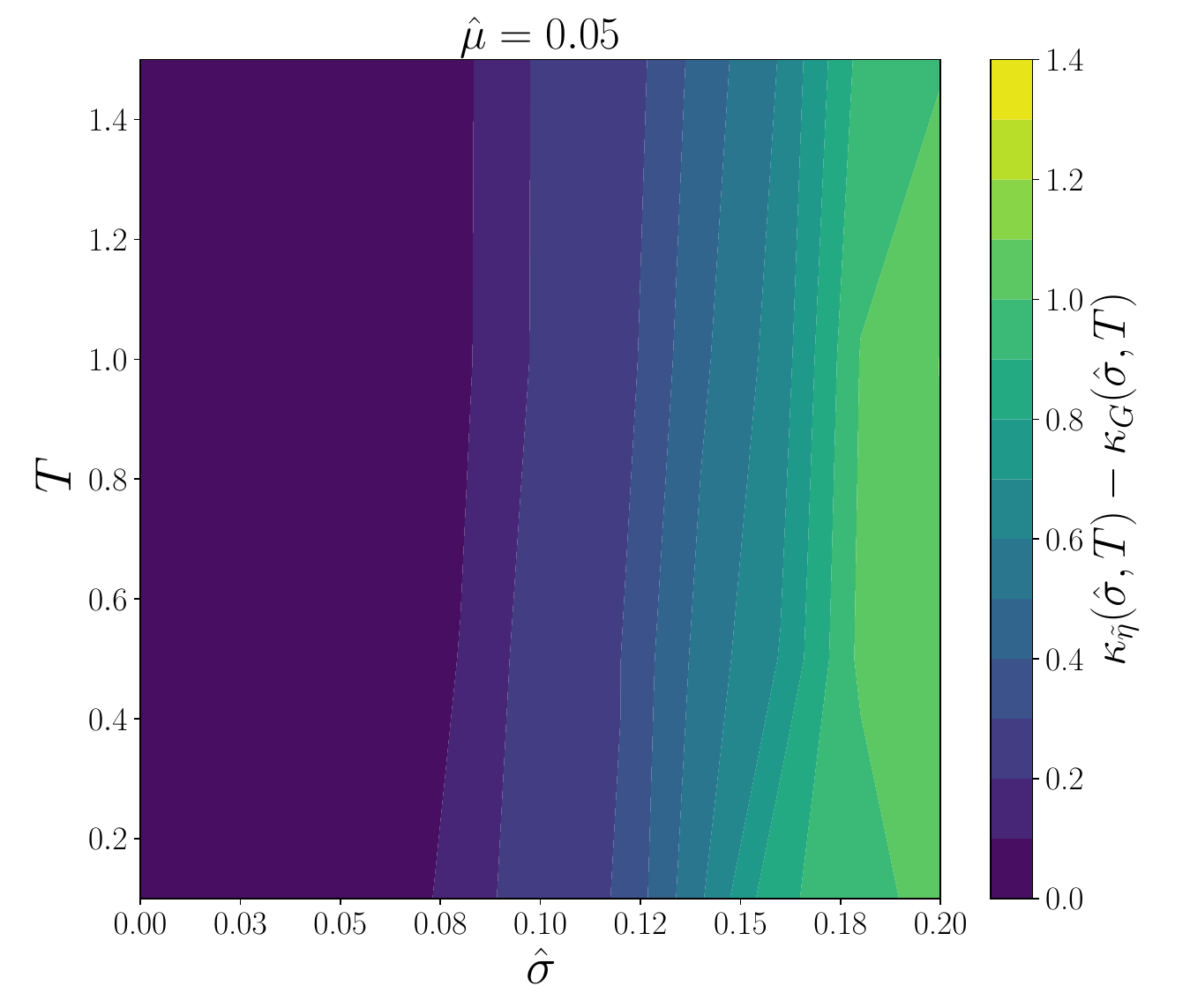}
\includegraphics[width=0.49\columnwidth]{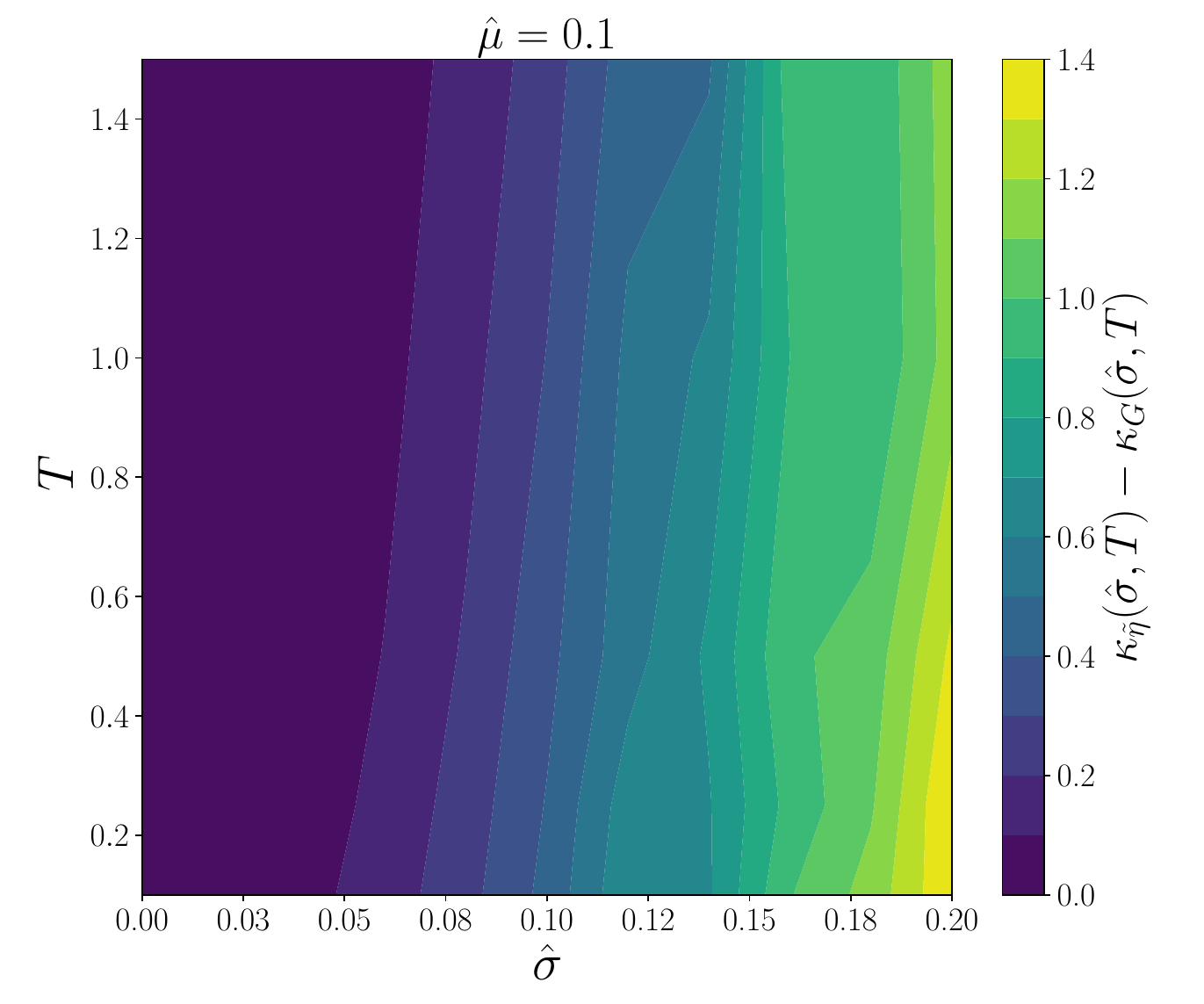}
\caption{\textbf{Deviation from Gaussianity for the species abundance for two different values of the average interactions}. The deviation is quantified as difference between the kurtosis $\kappa_{\tilde\eta}$ of the marginal distribution $\tilde\eta$ and the kurtosis $\kappa_G$ of a reference truncated Gaussian with equivalent variance and centered at the peak of $\tilde\eta$, and it is plotted as a function of $\hat{\sigma}$ and $T$ for two different values of $\hat\mu$, $\hat\mu=0.1$ and $\hat\mu=0.05$. For high $\hat{\sigma}$ it is evident that the deviation increases implying that the marginal distributions are developing strong non-Gaussian effects. The deviation from Gaussianity increases for higher $\hat\mu$. The number of species considered is $N=256$.}
\label{fig:contour-plot-etatilde}
\end{figure}

Let us precise that the practical computation of the kurtosis in Fig.~\ref{fig:contour-plot-etatilde} is slightly different from the one in Fig. \ref{fig:2d-contour}. In the main text, we computed the kurtosis of $\eta(n)$ by excluding the region too close to $n=0$, where the $1/(n+\epsilon)$ term introduces a peak around $n=0$, while retaining the entire right tail. In Fig.~\ref{fig:contour-plot-etatilde} instead, since $\tilde\eta(n)$ does not have the term $1/(n+\epsilon)$, we compute its kurtosis over the full distribution.
We want to stress that, in absence of interaction, that is, for $\hat\mu=\hat\sigma=0$, the rescaled marginal distribution $\tilde{\eta}(n)$ is a perfect Gaussian, where the variance is associated to the temperature. The new Fig. \ref{fig:contour-plot-etatilde} shows that indeed, as the interaction strength increases (through either larger $\hat\mu$ or larger $\hat\sigma$), the rescaled distribution $\tilde{\eta}(n)$ passes from a Gaussian to a strongly non-Gaussian behaviour. The reason why the interactions lead to a Gamma-like behaviour is still not clear and will be studied in subsequent works.

\section{Stability of the RS fixed point} \label{app:Stability}
In the main text we defined the transition line drawn in Figg. \ref{fig:transMUcdashed} and \ref{fig:dynamics_plots} as the line on the right of which the BP RS algorithm is no more convergent. In this section we want to show that the non-convergence of BP is not simply due to numerical errors or instability of the algorithm itself but instead it is linked to the loosing of stability of the RS fixed point, that at the transition becomes unstable, suggesting the creation of a new RSB state.
To do this, we use a well established method \cite{zdeborova2009statistical}: we numerically look at how a perturbation evolves in time under the BP iteration. To practically implement this procedure, we set the parameters of the problem in such a way that we are sure to be in the unique fixed point phase; we then let the BP algorithm reaching the unique fixed-point, $\eta^{FP}_{i\to j}(n)$. We then slightly perturb the solution in this way (even if the results are independent from the exact form of the perturbation):
\begin{equation}
\eta_{i\to j}^{pert}(n, t=0)= \frac{1}{\mathcal{Z}_{ij}}(\eta^{FP}_{i\to j}(n)\cdot w_{i,j}(n)),
\end{equation}
where $w_{i,j}(n)$ are independent random numbers extracted uniformly inside the interval $[0.95,1.05]$ for each $i, j, n$ and $\mathcal{Z}_{ij}$ is a new normalization factor ensuring the normalization of $\eta_{i\to j}^{pert}(n, t=0)$. We then let evolve $\eta_{i\to j}^{pert}(n, t=0)$ following the standard BP equations \eqref{eq-cavitymarginal} and we look at the evolution of the perturbation in time:
\begin{equation}
\varepsilon^2_{i \to j}(n,t)\equiv \left (\eta_{i\to j}^{pert}(n, t)-\eta_{i \to j}^{FP}(n)\right )^2.
\end{equation}
Defining the average square perturbation as
\begin{equation}
\varepsilon^2(t)\equiv\frac{1}{N\cdot c \cdot n_{max}}\sum_{i=1}^N \sum_{j \in \partial i}\sum_{n=0}^{n_{max}} \varepsilon^2_{i \to j}(n,t),
\end{equation}
after a first transient time it should follow the law:
\begin{equation}
\varepsilon^2(t)=\lambda \cdot\varepsilon^2(t-1),
\end{equation}
where $\lambda$ is the largest eigenvalue associated to the linearized BP square operator around the fixed-point. If $\lambda<1$ a perturbation is reabsorbed and the fixed-point is reached again: $\lambda<1$ is the necessary condition for the fixed point to be stable. When $\lambda$ reaches the value 1, a perturbation 
is no more reabsorbed, and practically it is impossible to find the fixed point, that is unstable, with the BP algorithm, that thus stops to converge.
In fig. \ref{fig:variance} we show the evolution of the perturbation $\varepsilon^2(t)$ approaching the transition in Fig. \ref{fig:transMUcdashed} for $T=0.2$, $\sigma=0$ and varying $\hat{\mu}$. In the left panel of fig. \ref{fig:variance} it is evident that, after a transient time, the variance of the perturbation evolves as $\varepsilon^2(t)=const\cdot\lambda^t(\hat\mu)$, with a value of $\lambda(\hat\mu)$ that grows approaching the point in which BP stops to converge $\hat\mu_c=0.3526$. The values of $\lambda(\hat\mu)$ are extracted with an exponential fit, and the values are reported in the right panel of fig \ref{fig:variance}, from which it is evident that $\lambda\to 1$ when $\hat\mu\to \hat\mu_c$: the non-convergence of BP is thus due to the lack of stability of the RS fixed point. The Fig. \ref{fig:variance} is done with a regularization parameter for the BP equations \eqref{eq-cavitymarginal} $\epsilon=10^{-4}$ but we have seen that the results are indistinguishable for the ones obtained for a smaller value of $\epsilon=10^{-5}$: the transition towards a multiple attractor phase does not depend on the regularization scheme.

\begin{figure}[htbp]
\includegraphics[width=0.49\columnwidth]{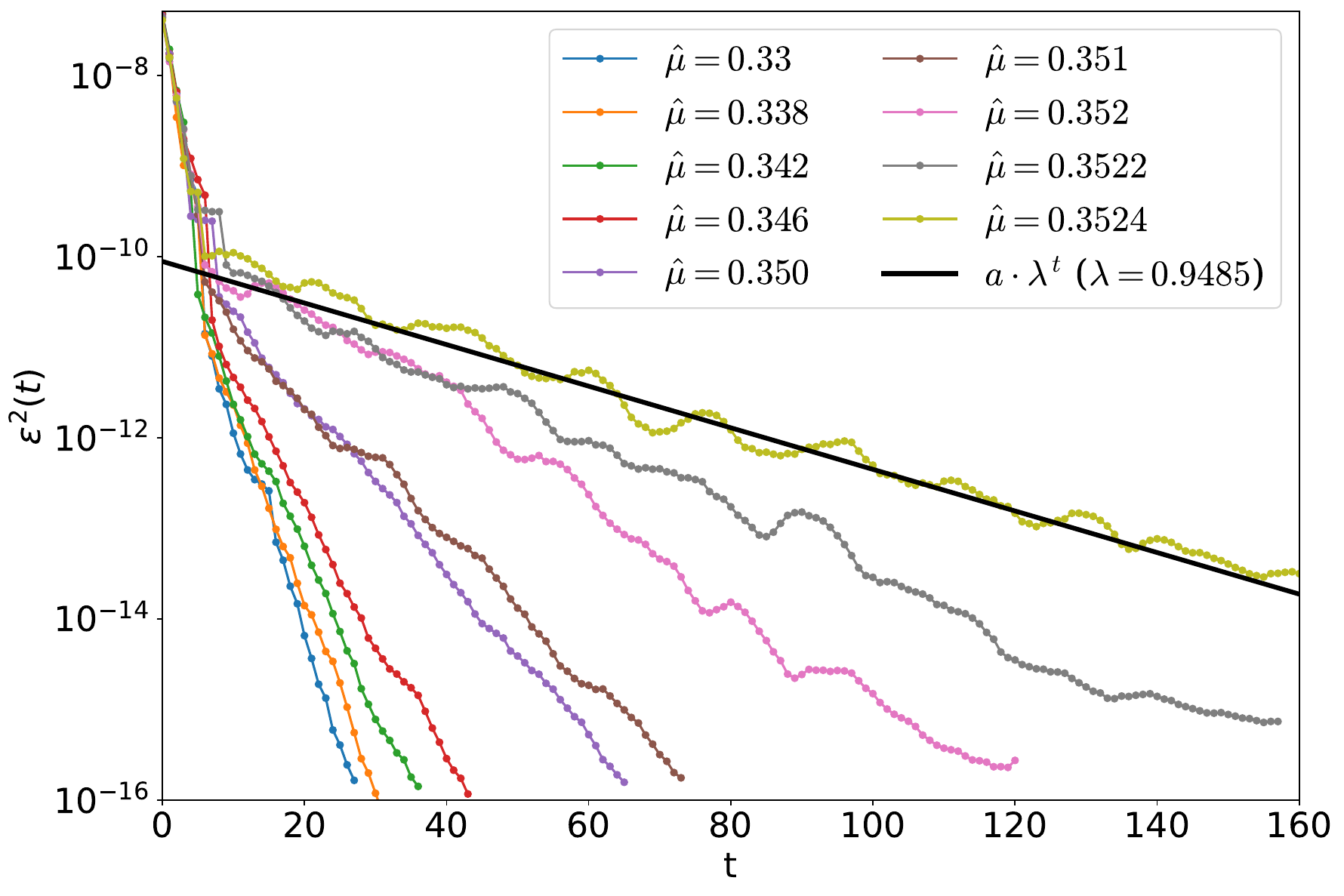}
\includegraphics[width=0.49\columnwidth]{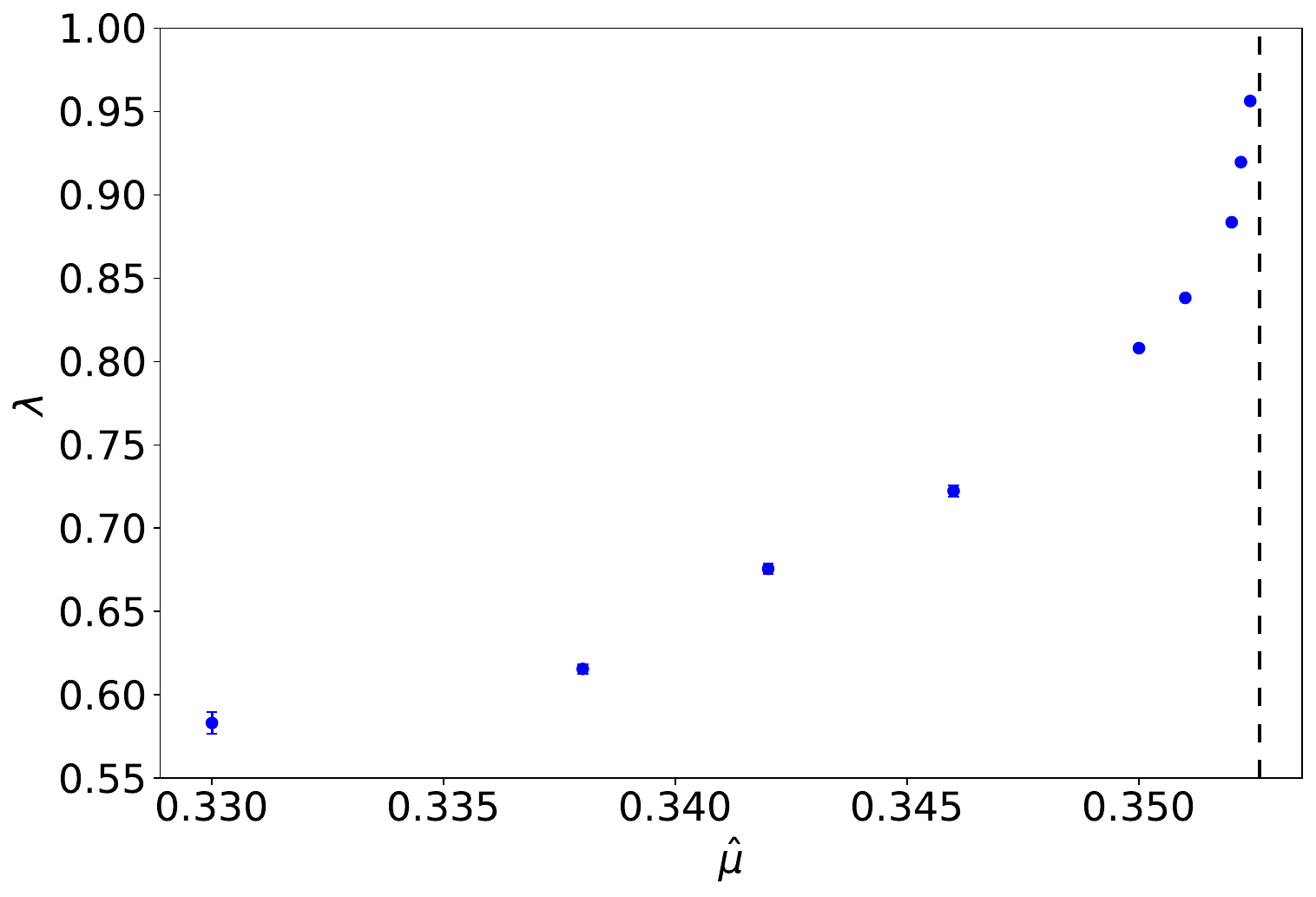}
\caption{\textbf{Left: Evolution of the variance of a perturbation around the BP RS fixed point.} After a transient time, the variance of the perturbation evolves as $\varepsilon^2(t)=const\cdot\lambda^t(\hat\mu)$, with a value of $\lambda(\hat\mu)$ that grows approaching the point in which BP stops to converge $\hat\mu_c=0.3526$. The solid line displays the best fit of the type $\varepsilon^2(t)=const\cdot\lambda^t(\hat\mu)$ for $\hat\mu=0.3524$.\\
\textbf{Right: Eigenvalue $\lambda$ extracted from the left figure as a function of $\hat\mu.$} The value of $\lambda$ approaches the value 1 as $\hat\mu$ approaches the point in which BP stops to converge $\hat\mu_c=0.3526$, signaled as a dashed line in the figure. For both figures the parameters are $T=0.2$, $\sigma=0$, $N=200$.}
\label{fig:variance}
\end{figure}

\section{Dynamics in the topological multiple attractor phase}\label{app:multiple_attractor}

\begin{figure}[htbp]
\includegraphics[width=0.45\columnwidth]{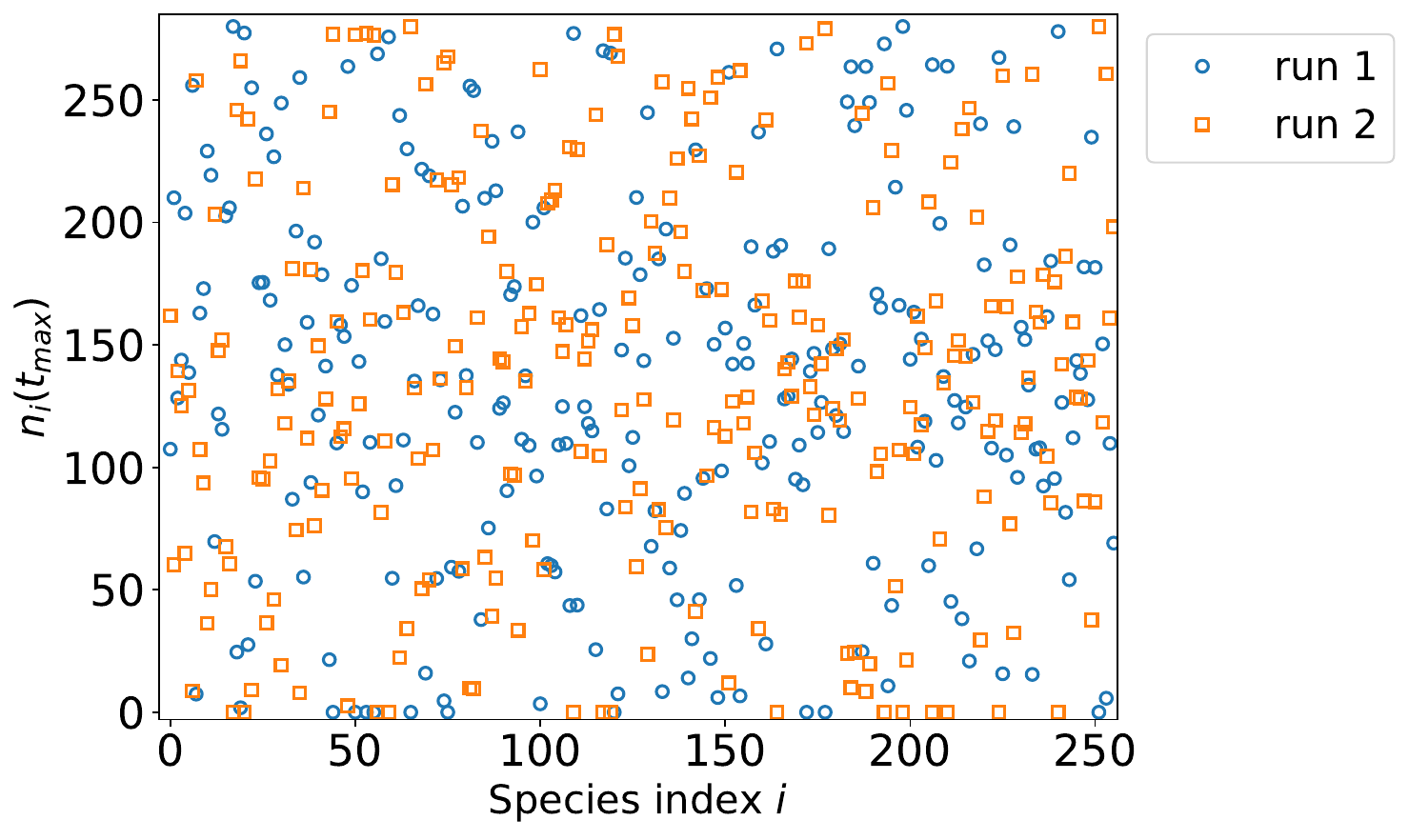}
\includegraphics[width=0.45\columnwidth]{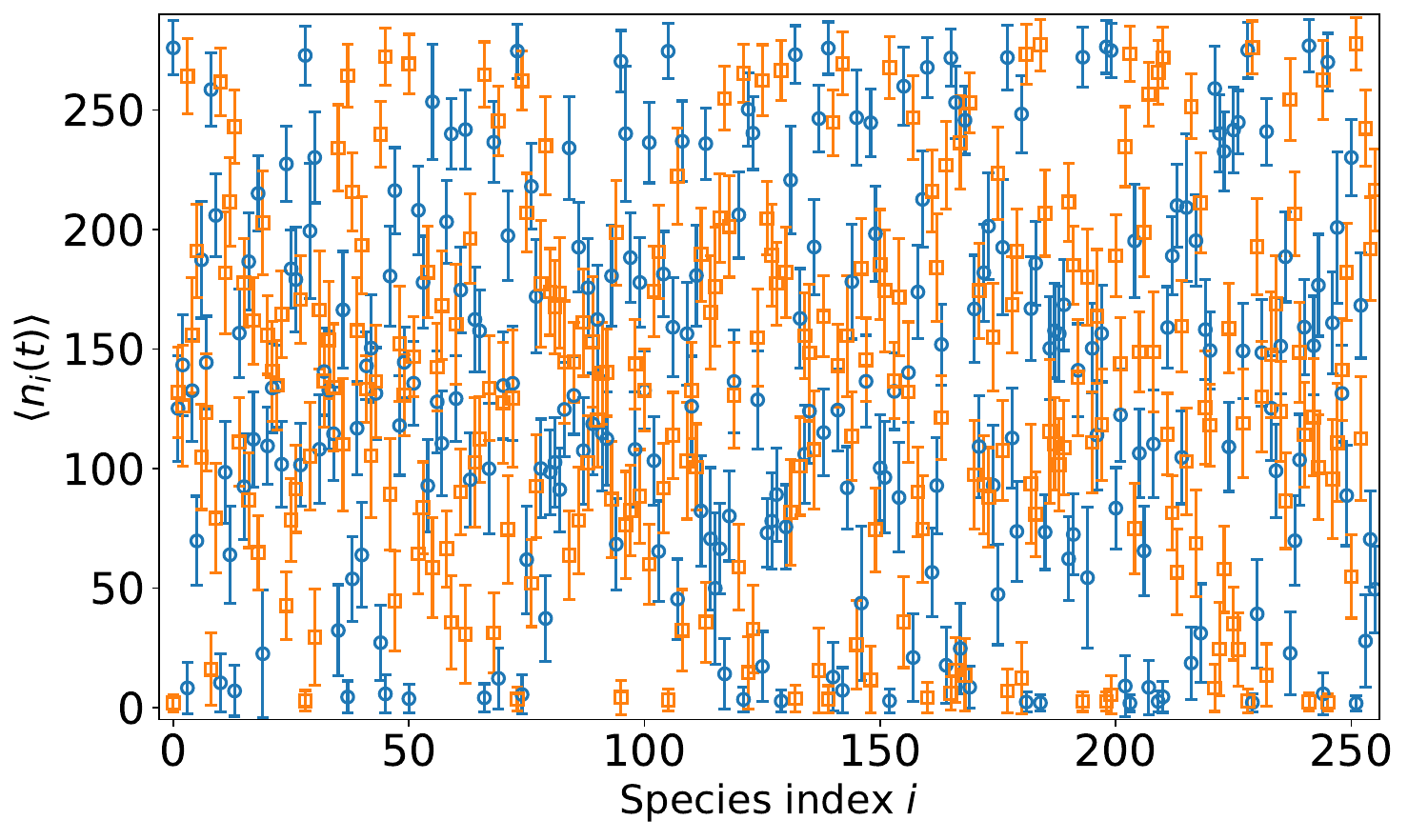}
\caption{\label{Fig:multipleattractor}
\textbf{Results from different iterations of the dynamics on the same graph in the multiple attractor phase. Left:} Final species abundance for each species $i\in[1,256]$ after the iteration of the Langevin dynamics for a time $t_{max}=4000$. The dynamics is run two times on the same graph with $\hat\mu=0.358$, $\hat\sigma=0$ and $T=0$, starting from different initial conditions, and thus reaches two different final states. \textbf{Right:} Average species abundance, and the relative standard deviation, for each species $i\in[1,256]$ during the iteration of the Langevin dynamics averaged from time $t=1000$ to time $t=4000$. The dynamics is run two times on the same graph with $\hat\mu=0.355$, $\hat\sigma=0$ and $T=0.4$, starting from different initial conditions, and thus reaches two different final states. At difference with the $T=0$ left Figure, in this case the species abundances fluctuate in time due to the thermal noise: the shown error take into account this time fluctuations, absent when $T=0$.}
\end{figure}

In the main text we show that, for $\hat\sigma=0$, when the interactions are strong enough, with $\hat\mu>\hat\mu_c(T)$, the model undergoes a transition towards what we called a \textit{topological} multiple attractor phase. We locate the exact position of the transition looking at the non convergence of the BP algorithm. In this Appendix we show the results coming from the dynamics in this new phase. We called the transition \textit{topological} because it is due to a change in the interaction network topology of surviving species. In fact, when $\hat\sigma=0$, all the species are equivalent, because there is no disorder in the model, and, as a consequence all the species should have the same marginal distribution. When the critical line $\hat\mu_c(T)$ is crossed, instead, some species go extinct: which species survive is a consequence of the randomness in the initial conditions and in the demographic noise. The extinctions introduce a new source of randomness in the model: in fact, after the extinctions not all the species have anymore 3 neighbors, and thus the marginal of the surviving species are allowed to be different. 
In Fig. \ref{Fig:multipleattractor}, we show the results from two different iterations of the dynamics in the same graph, but with different realizations of the initial conditions and of the demographic noise inside the multiple attractor phase. In the left panel we show the results for a system with parameters $\hat\mu=0.358$, $\hat\sigma=0$ and $T=0$. Because of the null temperature, after a transient time $t<1000$ the dynamics reaches a stable fixed point (see Fig. \ref{fig:dynamics_plots}): to show the multiple attractor fixed points it is thus just sufficient to look at the values that the single species abundances have after a fixed time, that we choose to be $t_{max}=4000$, after two different iterations of the dynamics. As shown in the left panel of Fig. \ref{Fig:multipleattractor}, the two different iterations lead to very different fixed points. In particular, the two sets of species that are extinct, for which $n_i(t_{max})=0$, after the two dynamics are completely different. 
In the right panel of Fig. \ref{Fig:multipleattractor} we show the results of the dynamics for parameters $\hat\mu=0.355$, $\hat\sigma=0$ and $T=0.4$. In this case, as can again be seen from Fig. \ref{fig:dynamics_plots}, after a transient time $t<1000$, the dynamics reaches a state characterized by species abundances that fluctuates in time, because of the thermal noise, but whose average and variance remains almost stable in time. We thus iterate the dynamics two times on the same graph, and we measure the average and the standard deviations of the single species abundances in both cases. Again, one can see that the two states reached are very different, because the averages of each species for the two runs are not compatible with each other inside the standard deviation. In particular, there are some species that are "extinct", characterized by both a very small mean and variance, and the two sets of extinct species for the two runs are different.

\end{widetext}

\newpage
\bibliography{biblio}

@article{may1973qualitative,
  title={Qualitative stability in model ecosystems},
  author={May, Robert M},
  journal={Ecology},
  volume={54},
  number={3},
  pages={638--641},
  year={1973},
  publisher={Wiley Online Library}
}

@article{neri2010phase,
  title={The phase diagram of L{\'e}vy spin glasses},
  author={Neri, Izaak and Metz, FL and Boll{\'e}, D},
  journal={Journal of Statistical Mechanics: Theory and Experiment},
  volume={2010},
  number={01},
  pages={P01010},
  year={2010},
  publisher={IOP Publishing}
}

@article{zdeborova2009statistical,
  title={STATISTICAL PHYSICS OF HARD OPTIMIZATION PROBLEMS},
  author={Zdeborov{\'a}, Lenka},
  journal={acta physica slovaca},
  volume={59},
  number={3},
  pages={169--303},
  year={2009}
}

@article{Crisanti2005Inverse,
  title = {Stable Solution of the Simplest Spin Model for Inverse Freezing},
  author = {Crisanti, Andrea and Leuzzi, Luca},
  journal = {Phys. Rev. Lett.},
  volume = {95},
  issue = {8},
  pages = {087201},
  numpages = {4},
  year = {2005},
  month = {Aug},
  publisher = {American Physical Society},
  doi = {10.1103/PhysRevLett.95.087201},
  url = {https://link.aps.org/doi/10.1103/PhysRevLett.95.087201}
}

@book{mezard2009information,
  title={Information, physics, and computation},
  author={Mezard, Marc and Montanari, Andrea},
  year={2009},
  publisher={Oxford University Press}
}

@article{leuzzi2011random,
  title={Random Blume-Capel model on a cubic lattice: First-order inverse freezing in a three-dimensional spin-glass system},
  author={Leuzzi, L and Paoluzzi, M and Crisanti, Andrea},
  journal={Physical Review B—Condensed Matter and Materials Physics},
  volume={83},
  number={1},
  pages={014107},
  year={2011},
  publisher={APS}
}

@article{altieri2021properties,
  title={Properties of equilibria and glassy phases of the random lotka-volterra model with demographic noise},
  author={Altieri, Ada and Roy, Felix and Cammarota, Chiara and Biroli, Giulio},
  journal={Physical Review Letters},
  volume={126},
  number={25},
  pages={258301},
  year={2021},
  publisher={APS}
}

@article{bunin2017ecological,
  title={Ecological communities with Lotka-Volterra dynamics},
  author={Bunin, Guy},
  journal={Physical Review E},
  volume={95},
  number={4},
  pages={042414},
  year={2017},
  publisher={APS}
}

@article{galla2018dynamically,
  title={Dynamically evolved community size and stability of random Lotka-Volterra ecosystems (a)},
  author={Galla, Tobias},
  journal={Europhysics Letters},
  volume={123},
  number={4},
  pages={48004},
  year={2018},
  publisher={IOP Publishing}
}

@article{diederich1989replicators,
  title={Replicators with random interactions: A solvable model},
  author={Diederich, Sigurd and Opper, Manfred},
  journal={Physical Review A},
  volume={39},
  number={8},
  pages={4333},
  year={1989},
  publisher={APS}
}

@article{suweis2024generalized,
  title={Generalized lotka-volterra systems with time correlated stochastic interactions},
  author={Suweis, Samir and Ferraro, Francesco and Grilletta, Christian and Azaele, Sandro and Maritan, Amos},
  journal={Physical Review Letters},
  volume={133},
  number={16},
  pages={167101},
  year={2024},
  publisher={APS}
}

@article{opper1992phase,
  title={Phase transition and 1/f noise in a game dynamical model},
  author={Opper, Manfred and Diederich, Sigurd},
  journal={Physical review letters},
  volume={69},
  number={10},
  pages={1616},
  year={1992},
  publisher={APS}
}

@article{BiroliBuninCammarota2018,
doi = {10.1088/1367-2630/aada58},
url = {https://dx.doi.org/10.1088/1367-2630/aada58},
year = {2018},
month = {aug},
publisher = {IOP Publishing},
volume = {20},
number = {8},
pages = {083051},
author = {Giulio Biroli and Guy Bunin and Chiara Cammarota},
title = {Marginally stable equilibria in critical ecosystems},
journal = {New Journal of Physics}
}

@article{azaele2024generalized,
  title={Generalized Dynamical Mean Field Theory for Non-Gaussian Interactions},
  author={Azaele, Sandro and Maritan, Amos},
  journal={Physical Review Letters},
  volume={133},
  number={12},
  pages={127401},
  year={2024},
  publisher={APS}
}

@article{marcus2022local,
  title={Local and collective transitions in sparsely-interacting ecological communities},
  author={Marcus, Stav and Turner, Ari M and Bunin, Guy},
  journal={PLoS computational biology},
  volume={18},
  number={7},
  pages={e1010274},
  year={2022},
  publisher={Public Library of Science San Francisco, CA USA}
}

@article{valigi2024local,
  title={Local sign stability and its implications for spectra of sparse random graphs and stability of ecosystems},
  author={Valigi, Pietro and Neri, Izaak and Cammarota, Chiara},
  journal={Journal of Physics: Complexity},
  volume={5},
  number={1},
  pages={015017},
  year={2024},
  publisher={IOP Publishing}
}

@article{poley2024interaction,
  title = {Interaction networks in persistent Lotka-Volterra communities},
  author = {Poley, Lyle and Galla, Tobias and Baron, Joseph W.},
  journal = {Phys. Rev. E},
  volume = {111},
  issue = {1},
  pages = {014318},
  numpages = {23},
  year = {2025},
  month = {Jan},
  publisher = {American Physical Society}
}

@article{ros2023generalized,
  title={Generalized lotka-volterra equations with random, nonreciprocal interactions: The typical number of equilibria},
  author={Ros, Valentina and Roy, Felix and Biroli, Giulio and Bunin, Guy and Turner, Ari M},
  journal={Physical Review Letters},
  volume={130},
  number={25},
  pages={257401},
  year={2023},
  publisher={APS}
}

@article{ros2023quenched,
  title={Quenched complexity of equilibria for asymmetric generalized lotka--volterra equations},
  author={Ros, Valentina and Roy, Felix and Biroli, Giulio and Bunin, Guy},
  journal={Journal of Physics A: Mathematical and Theoretical},
  volume={56},
  number={30},
  pages={305003},
  year={2023},
  publisher={IOP Publishing}
}

@article{lorenzana2024non,
  title={Non-reciprocal spin-glass transition and aging},
  author={Lorenzana, Giulia Garcia and Altieri, Ada and Biroli, Giulio and Fruchart, Michel and Vitelli, Vincenzo},
  journal={arXiv preprint arXiv:2408.17360},
  year={2024}
}

@book{may2007theoretical,
  title={Theoretical ecology: principles and applications},
  author={May, Robert and McLean, Angela R},
  year={2007},
  publisher={Oxford University Press}
}

@article{sole2002self,
  title={Self--organized instability in complex ecosystems},
  author={Sol{\'e}, Ricard V and Alonso, David and McKane, Alan},
  journal={Philosophical Transactions of the Royal Society of London. Series B: Biological Sciences},
  volume={357},
  number={1421},
  pages={667--681},
  year={2002},
  publisher={The Royal Society}
}

@article{azaele2016statistical,
  title={Statistical mechanics of ecological systems: Neutral theory and beyond},
  author={Azaele, Sandro and Suweis, Samir and Grilli, Jacopo and Volkov, Igor and Banavar, Jayanth R and Maritan, Amos},
  journal={Reviews of Modern Physics},
  volume={88},
  number={3},
  pages={035003},
  year={2016},
  publisher={APS}
}

@article{McGill2007SAD,
author = {McGill, Brian J. and Etienne, Rampal S. and Gray, John S. and Alonso, David and Anderson, Marti J. and Benecha, Habtamu Kassa and Dornelas, Maria and Enquist, Brian J. and Green, Jessica L. and He, Fangliang and Hurlbert, Allen H. and Magurran, Anne E. and Marquet, Pablo A. and Maurer, Brian A. and Ostling, Annette and Soykan, Candan U. and Ugland, Karl I. and White, Ethan P.},
title = {Species abundance distributions: moving beyond single prediction theories to integration within an ecological framework},
journal = {Ecology Letters},
volume = {10},
number = {10},
pages = {995-1015},
year = {2007}
}

@article{grilli2020macroecological,
  title={Macroecological laws describe variation and diversity in microbial communities},
  author={Grilli, Jacopo},
  journal={Nature communications},
  volume={11},
  number={1},
  pages={4743},
  year={2020},
  publisher={Nature Publishing Group UK London}
}

@article{garcia2022well,
  title={Well-mixed Lotka-Volterra model with random strongly competitive interactions},
  author={Garcia Lorenzana, Giulia and Altieri, Ada},
  journal={Physical Review E},
  volume={105},
  number={2},
  pages={024307},
  year={2022},
  publisher={APS}
}

@article{volkov2007patterns,
  title={Patterns of relative species abundance in rainforests and coral reefs},
  author={Volkov, Igor and Banavar, Jayanth R and Hubbell, Stephen P and Maritan, Amos},
  journal={Nature},
  volume={450},
  number={7166},
  pages={45--49},
  year={2007},
  publisher={Nature Publishing Group UK London}
}

@article{Giometto-taylor-law,
author = {Andrea Giometto  and Marco Formentin  and Andrea Rinaldo  and Joel E. Cohen  and Amos Maritan },
title = {Sample and population exponents of generalized Taylor’s law},
journal = {Proceedings of the National Academy of Sciences},
volume = {112},
number = {25},
pages = {7755-7760},
year = {2015}}

@misc{zamponi-cavity,
      title={Mean field theory of spin glasses}, 
      author={Francesco Zamponi},
      year={2014},
      eprint={1008.4844},
      archivePrefix={arXiv},
      primaryClass={cond-mat.stat-mech} 
}

@article{Anderson_1982,
year = {1982},
volume = {296},
pages = {245-248},
author = {Anderson, R. and Gordon, D. and Crawley, M. et al.},
title = {Variability in the abundance of animal and plant species},
journal = {Nature},
}

@article{Barbier2020complex,
    author = {Roy, Felix and Barbier, Matthieu and Biroli, Giulio and Bunin, Guy},
    journal = {PLOS Computational Biology},
    publisher = {Public Library of Science},
    title = {Complex interactions can create persistent fluctuations in high-diversity ecosystems},
    year = {2020},
    month = {05},
    volume = {16},
    pages = {1-14}
}

@article{Barbier2018generic,
author = {Barbier, Matthieu and Arnoldi, Jean-François and Bunin, Guy and Loreau, Michel },
title = {Generic assembly patterns in complex ecological communities},
journal = {Proceedings of the National Academy of Sciences},
volume = {115},
number = {9},
pages = {2156-2161},
year = {2018}
}
\end{document}